\documentclass[12pt,a4paper,twoside]{avh}
\usepackage{./mythesis}

\makeindex
\begin{document}

\setlength{\headheight}{15pt}

\newcommand{\thetitle}{\thesistitle}

\pagenumbering{roman}

\thispagestyle{empty}
\vspace{10ex}

{\centering
  \rule{\textwidth}{2pt}\\[5ex]
  {\Huge\sffamily{
      Entanglement and Its Applications \\[.5ex]
      in Systems with Many Degrees of Freedom}}\\[5ex]
  \rule{\textwidth}{2pt}\\
  \vspace{5ex}

  {\Large Stein Olav Skr{\o}vseth}
  \vspace{10ex}

  {Thesis submitted in partial fulfillment of the requirements\\
    for the Norwegian academic degree philosophi{\ae} doctor.}
  \vspace{20ex}


 {\Large Department of Physics \\ 
    Norwegian University of Science and Technology\\
    Trondheim, Norway\\[1.5ex]
    December, 2006\\}
}
\vfill

\includegraphics[width=.4\textwidth]{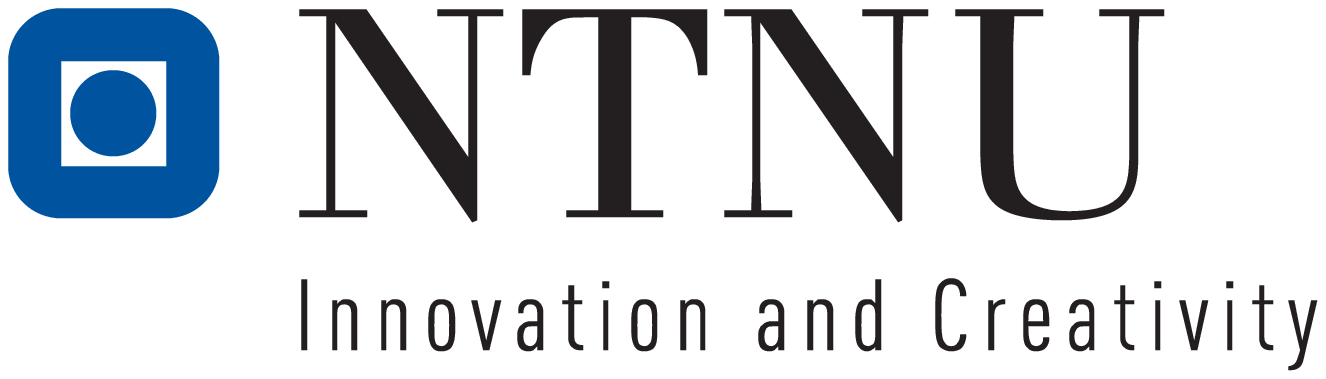}
\newpage

\chapter*{List of articles}
\begin{description}
\item[\bf Paper I \cite{SOS1}]~ \\
  Stein Olav Skr{\o}vseth and K{\aa}re Olaussen,\\
  {\it Entanglement used to identify critical systems},\\
  Phys. Rev. A, {\bf 72}, 022318 (2005) \eprint{cond-mat/0503235}
  \\[1.5ex]
  We promote use of the geometric entropy formula derived by Holzhey
  et. al. from conformal field theory, $S_\ell\sim({c}/{3})
  \log(\sin{\pi\ell}/{N})$, to identify critical regions in zero
  temperature 1D quantum systems. The method is demonstrated on a
  class of one-dimensional $XY$ and $XYZ$ spin-1/2 chains, where the
  critical regions and their correponding central charges can be
  reproduced with quite modest computational efforts. 
\item[\bf Paper II \cite{SOSbose}]~ \\
  Stein Olav Skr{\o}vseth,\\
  {\it Entanglement in bosonic systems},\\
  Phys. Rev. A, {\bf 72}, 062305 (2005)
  \eprint{quant-ph/0508160}\\[1.5ex]
  We present a technique to resolve a Gaussian density matrix and its
  time evolution through known expectation values in position and
  momentum. Further we find the full spectrum of this density matrix
  and apply the technique to a chain of harmonic oscillators to find
  agreement with conformal field theory in this domain. We also
  observe that a non-conformal state has a divergent entanglement
  entropy. 
\item[\bf Paper III \cite{SOSposter}]~ \\
  Stein Olav Skr{\o}vseth,\\
  {\it Entanglement signatures in critical quantum systems} in\\
  Proceedings of ERATO conference on Quantum Information Science
  2005, 177 (2005)\\[1.5ex]
  Conformal field theory (CFT) predicts a signature in the
  entanglement entropy of conformally invariant systems. These include 
  critical quantum systems in 1+1 dimension with local interactions,
  and we present a technique to identify criticality in such system
  through this signature. It is shown that this detection is precise
  even in systems small enough to facilitate numerical diagonalization
  of the Hamiltonian.
  \pagebreak
\item[\bf Paper IV \cite{SOSent}]~ \\
  Stein Olav Skr{\o}vseth,\\
  {\it Entanglement properties of quantum spin chains},\\
  Phys. Rev. A, {\bf 74}, 022327 (2006) \eprint{quant-ph/0602233}\\[1.5ex]
  We investigate the entanglement properties of a finite size 1+1
  dimensional Ising spin chain, and show how these properties scale
  and can be utilized to reconstruct the ground state wave
  function. Even at the critical point, few terms in a Schmidt
  decomposition contribute to the exact ground state, and to physical
  properties such as the entropy. Nevertheless the entanglement here
  is prominent due to the lower-lying states in the Schmidt
  decomposition. 
\item[\bf Paper V \cite{SOStherm}]~ \\
  Stein Olav Skr{\o}vseth,\\
  {\it Thermalization through unitary evolution of pure states},\\
  Europhys. Lett., {\bf 76}, 1179 (2006) \eprint{quant-ph/0606216}\\[1.5ex]
  The unitary time evolution of a critical quantum spin chain with an
  impurity is calculated,
  and the entanglement evolution is shown. Moreover, we show that the
  reduced density matrix of a part of the chain evolves such that the
  fidelity of its spectrum is very high with respect to a state in
  thermal   equilibrium. Hence, a thermal state occurs through
  unitary time evolution in a simple spin chain with impurity.
\end{description}

\chapter*{Acknowledgements}
\vspace{-3ex} 

The writing of a PhD thesis is a vast and lonely task, and the present
thesis is no exception in that respect. However, the work is made ever
more inspiring with colleagues and friends around for discussion and
support.

Professor K{\aa}re Olaussen has been an indispensable resource in the
work of this thesis with his great insight into numerous and varied
areas of 
physics. His sociable and friendly personality has made it a true
pleasure to work with him.

I am grateful that Professors Jos\'e Ignacio Latorre, Henrik
Johannesson, and Johannes Skaar have agreed to constitute the
evaluating committee for the present thesis and the defense to be held
on December 12, 2006.

My fellow students at NTNU have provided a great network for inspiring
coffee breaks and physics discussions. Great credit is given to Jo
Smiseth, Martin Gr{\o}nsleth, Jan Petter Morten, Kjetil B{\o}rkje,
Daniel Huertas-Hernando, and Jan {\O}ystein Haavik Bakke for their
inspiration, and for providing a great atmosphere in the physics
department. Martin, Jan Petter and Kjetil are also thanked for
proofreading parts of the thesis.

The NordForsk network on Low-dimensional Physics administered by
Susanne Viefers is acknowledged for
arranging inspiring meetings, and economic support. Stephen Bartlett
in Sydney is thanked for providing a great opportunity
to go to Australia at the end of this work.

The final years of the PhD work has been spent at the world's
northernmost university, the University of Troms{\o}. I would like to
thank Robert Jenssen, 
Eivind Brodahl, Tor Arne {\O}ig{\aa}rd, and Stian Anfinsen for making
me feel most welcome, even though their scientific interests are
at best tangential to the present work.

Parental support is indispensable, and in this case it has
always been good to know that I am welcome at my childhood home when
coming back to Trondheim, thanks to my parents and brother.

My deepest gratitude goes to my beloved Veronika Kristine
T{\o}mmer{\aa}s, who 
has always been supportive and loving, and whose company and
inspiration has been fantastic through these years.

Finally, I cannot overestimate the inspiration provided by the
creativity and great friendship of Bendik Kvale Jacobsen.

\begin{spacing}{.5}
  \tableofcontents
  \listoffigures
\end{spacing}

\fancyfoot{}
\fancyhead{}
\fancyhead[LE]{\sffamily{\thepage\hspace{5ex}\thetitle}}
\fancyhead[RO]{\sffamily{\leftmark\hspace{5ex}\thepage}}
\pagestyle{fancy}
\renewcommand{\chaptermark}[1]{%
  \markboth{#1}{}}

\newpage
\cleardoublepage
\pagenumbering{arabic}
\setcounter{page}{1}
\chapter{Introduction}
\label{Ch:Introduction}

Erwin Schr\"odinger called entanglement the characteristic trait of
quantum mechanics, the single concept that stands out not to have any
counterpart in classical physics. Richard Feynman said that there is
only one mystery in quantum mechanics, and that is entanglement and
the consequences thereof. Albert Einstein was -- along with many of
his contemporaries -- uncomfortable with what he called ``spooky
action at a distance'', and co-wrote the seminal paper \cite{EPR35}
where the authors dismissed the idea that quantum mechanics could
possibly be a complete description of Nature on the grounds that it
contained entanglement. Today entanglement is
accepted as a trait of Nature, and though many aspects are poorly
understood, the idea of a quantum computer has shifted much of the
research away from fundamental investigation to pragmatical work on
how to utilize entanglement in quantum computing and quantum
communication \cite{Nielsen&Chuang}.

Entanglement is essentially a question of correlations. Correlations
are a well known daily life phenomenon. We all know that wearing a
seat belt is correlated with our chances of avoiding serious injury in
a car accident. Other correlations are less obvious, and are often the
focus of research, e.g. whether there is a correlation between
the $\chem{CO_2}$ content in the atmosphere and global warming,
whether there is a correlation between certain diets and the risk of
acquiring cancer, or how correlated the number of years of education
is with income. In classical theories a correlation can never be
better than perfect, the best we can do is a perfect correlation,
e.g. if a certain number of years in school linearly increases the
yearly income. As all scientists know, this is not the case, but
probably there is a less than perfect, though positive
correlation. However, quantum 
mechanics opens the case that there might be better than perfect
correlations. This is the case when considering e.g. the spin of two
electrons separated by some spatial distance. We can measure the spin
of one, and it may not only be correlated with what we measure on the
other electron, but it will be more than 100\% correlated; the
electrons were {\it entangled} before the measurement. In the
simplistic case described here, this will never be evident, all we can
see in our measurements is a perfect correlation. However, John Bell
wrote down an 
inequality in 1964 \cite{Bell64} that would be satisfied for all
classical correlations, but would possibly be broken if there were
quantum correlations present. Now it is accepted as an experimental fact
that Bell's inequality may be broken, as first confidently
demonstrated by Alain Aspect \etal in 1982 \cite{Aspect82}. Thus
Nature seems to allow quantum 
correlations on top of the classical correlations. This excess
correlation is what we know as entanglement, and is evident whenever
Bell's inequality is broken.
\begin{figure}[t]
  \centering\includegraphics[width=.8\textwidth]{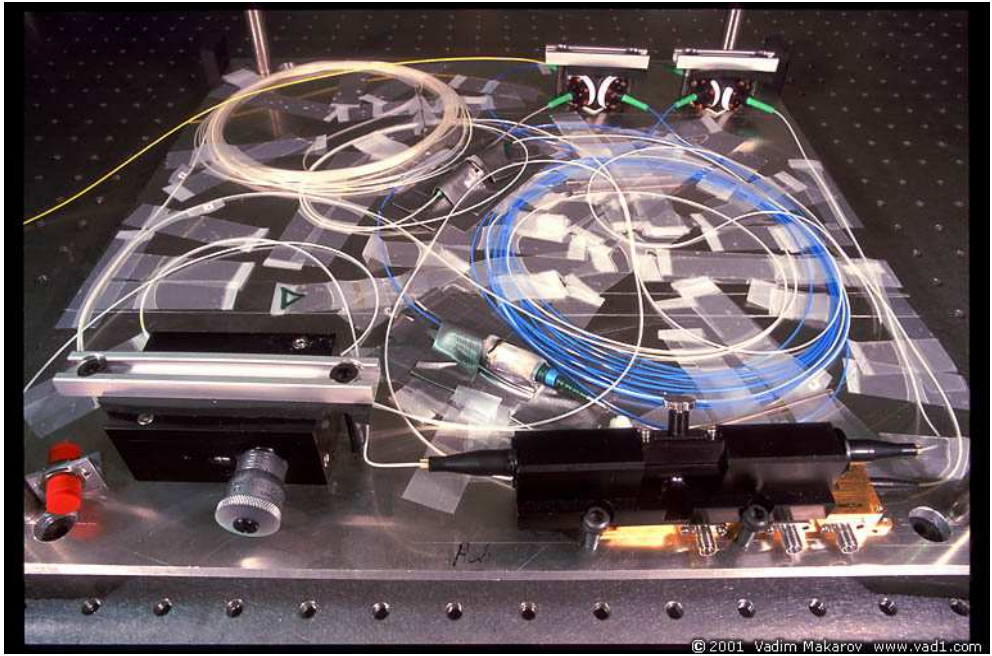}
  \caption[A physical implementation of ``Bob''.]%
  {A physical implementation of ``Bob'', one of the characters
    of quantum information theory. This is
    the setup used in an optical quantum cryptography experiment at
    the Department of Electronics and Telecommunications,
    NTNU. Image courtesy of Vadim Makarov, used with
    permission.}
  \label{fig:Bobpic}
\end{figure}

Though the connection is not obvious, entanglement is today mainly
studied in connection with quantum information theory, as it is
believed that a quantum computer will -- if ever built -- utilize
entanglement to perform certain computational tasks faster than
classically possible. Feynman pointed out in 1982 that simulating
quantum systems on a classical computer was nearly impossible due to
the much larger Hilbert space of the quantum variant
\cite{Feynman82}. Thus, he 
suggested the use of a quantum computer to simulate quantum
systems. However, it was not until Peter Shor's discovery in 1994
\cite{Shor94} that a quantum computer could factorize large numbers
exponentially faster than classically possible that the idea really
gained momentum. The quantum computer is still very much a
theoretical device, despite considerable experimental effort on small
scale systems. It has so far not emerged any single clear
candidate for 
the implementation of the quantum bits (qubits) and the operations
needed to process the quantum information.
The NMR experiment that factorized the number
15 in 2001 \cite{Vandersypen01} was the first implementation where a
quantum algorithm was proved to work, though on a relatively trivial
problem. The NMR implementation, along with 
most implementations based on atomic, molecular or optical physics
(AMO), is poorly scalable. The approach based on ion traps has
nevertheless become fashionable, and was one of the first promising
proposals for a quantum computer made in 1995 by Cirac and Zoller
\cite{Cirac95}.
Other implementations are based on condensed matter
systems, such as superconducting qubits or quantum dots. The main
problem in many of these approaches is decoherence, that is the
qubits become entangled with the environment and thus information is
lost into the environment during the computation. Recently, two
superconducting qubits were entangled \cite{Steffen06}, making it an
ever more promising candidate for quantum computing.
Also recently, a hybrid
approach has been proposed, exploiting the advantages of different
schemes \cite{Andre06}, where trapped polar molecules are controlled
by a microwave field set up by running superconductors. Despite the
extensive effort, one still doesn't
know whether the quantum computer even can be built, or if one might
encounter a fundamental reason why this is an impossible feat. Even
so, the process will learn us fundamental facts about physics and the
process cannot be said to have been in vain.

Phase transitions are also familiar phenomena of daily
life, like the freezing of water at zero temperature. Such
classical phase transitions occur when the temperature is changed
across some critical value $T_c$, and the thermal fluctuations are
exactly what drives the system from one thermodynamical state to
another. Quantum phase transitions \index{quantum phase transitions}
on the contrary, can happen at zero temperature, and therefore without
thermal fluctuations. These continuous phase transitions occur when
the
ground state changes nature through the variation of a parameter, such
as an external magnetic field. The only fluctuations present are the
quantum fluctuations that must exist due to the Heisenberg uncertainty
relation, and hence the name quantum phase transitions. The
wave function of the system at the critical point is very
complicated, as is evident from the fact that it carries more entanglement
that at off-critical points. The complexity
of the wave function at the critical point also means that more
information is encoded in the wave function, thus making it more
usable for quantum information.

This thesis is organized as follows;
Chapter \ref{Ch:Entanglement} is a brief introduction to the most
important principles in entanglement theory, with an emphasis on those
subjects which are considered in later chapters. Chapter
\ref{ch:Numerics} addresses some general issues when it comes to
modelling quantum systems on a classical computer, and the line
is drawn between 
renormalization schemes and the problem of finding the ground state
of a quantum system as addressed in \cite{SOSent}. In Chapter
\ref{Ch:ContVariable} we investigate bosonic systems, and compute the
entanglement therein. In particular, the results from \cite{SOSbose}
are connected to other results from the theory of entanglement in
continuous variable systems. Finally, Chapter \ref{Ch:QuantumCritical}
contains the bulk part of the thesis, summarizes quantum critical
systems, and applies ideas from
\cite{SOS1,SOSposter,SOSent,SOStherm}.

In all chapters, unless otherwise specified, units such that
$\hbar=c=1$ are used, where $c$ is the speed of light in vacuum. All
notation is as far as possible applied as is conventional in the
literature. In particular, the symbol $\oplus$ is used for the direct
sum of matrices, e.g. $A\oplus
B=\mathrm{diag}(A,B)$, while $\otimes$ designates 
direct product, which for matrices is defined in
e.g. \cite{Nielsen&Chuang}.

\newpage
\cleardoublepage
\chapter{Entanglement}
\label{Ch:Entanglement}
\index{entanglement|(}
In its simplest form, entanglement arises from the possibility opened
by quantum theory to write down a wave function that is non-local,
encompassing two constituents separated by arbitrarily large
distances. Consider a wave function of two
particles shared by the protagonists of quantum information theory,
Alice and Bob. Now, any
measurement on either's part implies the collapse of the wave function
at the other's side. This collapse happens instantly (in some frame of
reference), and this is the source to its
counter-intuitivity, not to mention the apparent contradiction with
special relativity. However, Alice and Bob can never exploit this
effect to communicate any information faster than the speed of light.
Due to the statistical nature of the outcomes, the parties need to
communicate classically to compare their results. Moreover, the simple
scheme here can easily be 
explained through classical means with some ``hidden variable''
interpretation. However, there are more 
complicated schemes that efficiently demonstrate the breaking of
Bell-type inequalities, thus ruling out any hidden variable
explanation \cite{Mermin85}. 
In this thesis we will restrict our investigation to bipartite
entanglement, and
ignore the case where there are more parties than Alice and Bob. Thus,
consider two systems in separate Hilbert 
spaces $\mathcal H_A$ and $\mathcal H_B$ respectively. Given a state
vector $\ket\psi_A\in\mathcal H_A$ in the first system and a state
vector $\ket\varphi_B\in\mathcal H_B$ in the second, we can define a
state in the entire Hilbert space, $\mathcal H=\mathcal
H_A\otimes\mathcal H_B$ as 
\[\ket\Psi_{\mathrm{separable}}=\ket\psi_A\otimes\ket\varphi_B.\]
As indicated, this is a separable state, and any state that can be
written in this way is separable, and thus carries no
entanglement. More generally, the state of the entire system can be
written as a Schmidt decomposition\index{Schmidt decomposition}
\begin{equation}
  \ket\Psi=\sum_{i=1}^\chi\sqrt{\lambda_i}\,\ket{\psi_i}_A\otimes\ket{\varphi_i}_B
\end{equation}
where the upper limit of the sum, $\chi$, is called the Schmidt number
of the state, and is maximally equal to the {\em smallest} of the
dimensionalities of the two Hilbert spaces in question. The states
$\ket{\psi_i}$ are orthogonal for all $i$, and the same is true for
$\ket{\varphi_i}$. The expansion is unique, and the Schmidt number
can be considered a very brute measure of entanglement, in particular
the measure $E(\ket\Psi)=\log_2\chi$ fulfills some of the requirements
for an entanglement measure as described later \cite{Vidal03b}.

Identifying an entangled state can, as opposed to what impression one
gets from the simple analysis above, be a very hard task. The problem
can be stated as finding the basis in which one can
write a given state as a separable one, or proving that no such basis
exists. It is important to realise that this task depends heavily on
the choice of partitioning of the system. The task of writing 
down this basis indeed seems so daunting that one often  resorts to
simply stating whether a state is separable or not, and if it is
non-separable, how much entanglement do we have? We will return to
this problem in Section \ref{sec:EntanglementMeasures}.

Given a Schmidt decomposition of a pure state $\ket\Psi$, one can
straightforwardly write down the density matrix of (say) $\mathcal
H_A$ as 
\[\rho^{(A)}=\sum_i\lambda_i\ket{\psi_i}\bra{\psi_i}\]
where we have omitted the index $A$ since the identity of the Hilbert
space is obvious. This density matrix contains all information
accessible to Alice. Hence, finding the Schmidt decomposition of the
pure state wave function is equivalent to diagonalizing the reduced
density matrices of the two subspaces. That is one indication of the
importance of the Schmidt decomposition.

More generally though, the state in the full Hilbert space $\mathcal
H$ may not be a pure one, but an indeterminate mixed state with
density matrix $\rho$. Still, the reduced density matrix of $\mathcal
H_A$ is well defined, but the entanglement is not, in the sense that
one cannot find a genuine measure of entanglement for the
state. Purification\index{purification} learns us
that any mixed state in a Hilbert space $\mathcal H$ can always be
interpreted as a pure state in some (imaginary) larger Hilbert space
$\mathcal H'\supset\mathcal H$. Consequently, the physical state
$\rho\in\mathcal H$ is entangled to the system of the larger Hilbert
space $\mathcal H'$. 

If Alice has access to the pure states
$\left\{\rho_i^{(A)}\right\}\in\mathcal H_A$ and Bob has the pure
states $\left\{\rho_i^{(B)}\right\}\in\mathcal H_B$, the state 
\[\rho=\sum_{ij}p_{ij}\,\rho_i^{(A)}\otimes\rho_j^{(B)}\qquad\sum_{ij}p_{ij}=1\]
carries only classical correlations, and must thus be classified as
unentangled, or separable \cite{Werner89}. A density matrix that is
non-separable 
consequently is entangled, but finding a decomposition for a given
density matrix is computationally hard. Moreover, for the case where
$\rho$ is a mixed state, it is not yet known how to measure the
entanglement of the state such that classical correlations are
eliminated. 

Assume Alice and Bob being at different locations, separated by a large
distance, and sharing a state describe by the density matrix
$\rho$. Now, each party may operate on their state by unitary
transformations or measurements, and they are even allowed to
communicate classically, e.g. through telephone, but any joint
operation on both parties are prohibited.
 This protocol is known as
LOCC\index{LOCC}, Local Operations and Classical
Communication. Through LOCC one can generate all classical
correlations, but no entanglement can ever be created this way. This
means that LOCC describes the fundamental difference between what can
be created classically and quantum mechanically and is a useful
starting point to describe quantum correlations. What
is sometimes called the fundamental law of quantum information
processing can be phrased as
\begin{impbox}
  Through LOCC alone, Alice and Bob cannot increase the total
  entanglement of the state they share.
\end{impbox}
Mathematically, local operations can be either unitary operators
$U_A\in\mathcal H_A$ and $U_B\in\mathcal H_B$ or general measurements
$\left\{\hat A_i\right\}$ and $\left\{\hat B_i\right\}$. The
generalized measurements are complete, in the sense that
\index{measurement operators}
\[\sum_i\hat A_i^\dag\hat A_i=\sum_i\hat B_i^\dag\hat B_i=\mathds 1,\]
where $\hat A_i$ are measurements performed exclusively on Alice's
system, and $\hat B_i$ are performed on Bob's. After Alice has applied
a measurement on her state $\rho^{(A)}$,the state collapses
into $\rho'=\hat A_i\rho^{(A)}\hat A_i^\dag/p(i)$ with probability
$p(i)=\trace(\hat A_i\rho^{(A)}\hat A_i^\dag)$, and correspondingly
for Bob's measurements.
LOCCs are important in the sense that it should be assumed that
entanglement cannot increase under LOCC, and should be invariant if
there are no measurements involved.

Moreover, a fundamental fact involving LOCC is that, given two states
$\ket{\psi_1}$ and $\ket{\psi_2}$, of the joint system of Alice and
Bob, one can transform the first into the latter by LOCC only if the
ordered eigenvalues $\lambda_1$ of
$\rho_1^{(A)}=\trace_B\ket{\psi_1}\bra{\psi_1}$ is majorized by the
eigenvalues $\lambda_2$ of
$\rho_2^{(A)}=\trace_B\ket{\psi_2}\bra{\psi_2}$. A decreasingly
ordered 
vector $\vec v$ majorizes\index{majorization} another
ordered sequence $\vec w$, written $v\succ w$ if
\[\sum_{i=1}^kv_i\geq\sum_{i=1}^kw_i \qquad\text{for all}\qquad
k=1,\cdots,d\]
where $d$ is the smallest of the dimensions of the two vectors. 

\section{Entanglement measures}
\label{sec:EntanglementMeasures}
\index{entanglement!measures|(}
As we have pointed out, entanglement is very much considered a
fundamental resource in nature, much like energy or information are
considered resources in other contexts. And anything called a
resource must be 
measured, so we would have some idea about how ``much'' we have of
that resource. Thus one needs to define entanglement measures, a very
non-trivial task \cite{Myhr04}. For some
state described by the density matrix $\rho$, we want the entanglement
$E(\rho)\in\mathds R$ of the state. This measure must fulfill some basic
requirements to be physically reasonable \cite{Vedral97,Horodecki00};
\begin{itemize}
\item[E1] $E(\rho)\geq0$ with equality if $\rho$ is separable.
\item[E2] $E(\rho)=E(U_A\otimes U_B\rho U_A^\dag\otimes U_B^\dag)$ where
  $U_A$ and $U_B$ are any local unitary operators on Hilbert spaces
  $\mathcal H_A$ and $\mathcal H_B$ respectively.
\item[E3] The entanglement cannot increase under LOCC, $E(\Theta \rho)\leq
  E(\rho)$ where $\Theta$ is any LOCC operator.\index{LOCC}
\item[E4] Convexity:\index{convexity}
  $\sum_ip_iE(\rho_i)\geq E\left(\sum_ip_i\rho_i\right)$.
\item[E5] Partial additivity:\index{partial additivity}
  $E(\rho^{\otimes n})=nE(\rho)$.
\item[E6] Continuity: If $\lim_{n\to\infty}\bra{\psi^{\otimes
      n}}\rho_n\ket{\psi^{\otimes n}}=1$ with $\rho_n$ a density
  matrix of $n$ pairs, then
  \[\lim_{n\to\infty}\frac 1n\left[E\left(\ket\psi\bra\psi^{\otimes n}\right)-E(\rho_n)\right]=0\]
\end{itemize}
One also often requires the normalization that the Bell
state $\ket\beta=\left(\ket{00}+\ket{11}\right)/\sqrt 2$ should have
unit entanglement. Any entanglement measure that satisfies conditions
E1-4 is 
known as an {\it entanglement monotone}\index{entanglement monotone}.

\subsection{Forming and distilling entanglement}
\label{sec:FormingDistilling}
Given some quantum state $\rho$, the question of how much entanglement
this contains is nontrivial. However, conceptually the question
can be simplified by defining entanglement of formation
\index{entanglement!of formation} and
-distillation\index{entanglement!distillation}. 
Assume that we have a large number of Bell states, and are allowed to
use only LOCC on these states, how many copies of a given bipartite
state $\rho$ can we make? The number of copies of $\rho$ divided by
the number of Bell states is known as the entanglement of formation
$E_F$ when one takes the limit of an infinite number of Bell states.
On the contrary, the entanglement of distillation is defined as
the number of Bell states we can create from a large number of copies
of $\rho$. The distillable entanglement $E_D$ is always
less than or equal to the entanglement of formation since one
inevitably loses information in the process of formation and
distillation. Moreover, it turns out that any reasonable entanglement
measure is bounded by these, \cite{Horodecki00}
\begin{equation}
  E_D(\rho)\leq E(\rho)\leq E_F(\rho).
\end{equation}

For the special case of $\rho=\ket\psi\bra\psi$ being a pure state, it
turns out that one 
can uniquely define a bipartite entanglement measure since the
entanglement of formation and -distillation are equal and equal to the
von Neumann entropy of the reduced density matrix of one of the
parties. That is,\index{von Neumann entropy}
\begin{equation}
  E_F(\ket\psi\bra\psi)=S\left(\rho'\right)=-\trace\rho'\log_2\rho'\qquad\text{with}\qquad\rho'=\trace_B\ket\psi\bra\psi
  \label{eq:defSrho}
\end{equation}
where $\rho'$ is the reduced density matrix accessible to Alice. The
entanglement must be symmetric, so it does not matter whether we trace
out Alice's or Bob's system. The von Neumann entropy is nothing but
the Shannon entropy of the 
eigenvalues of the reduced density matrix, as the Shannon entropy of
some random variable $X$ with probabilities $x_1,\ldots,x_N$ is
$H(X)=-\sum_nx_n\log_2x_n$. Note that we use logarithm base two,
meaning that the von Neumann entropy is measured in ebits
(entanglement bits), and the von Neumann entropy of a shared Bell
state is unity.

\begin{figure}[b]
  \psfrag{X}{$\lambda$}
  \psfrag{Y}{$H(\lambda)$}
  \centering
  \includegraphics[width=.5\textwidth]{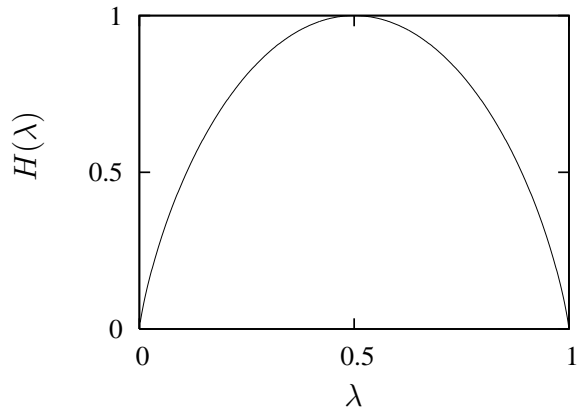}
  \caption[The binary entropy function]%
  {The binary entropy function
    $H(\lambda)=-\lambda\log_2\lambda-(1-\lambda)\log_2(1-\lambda)$,
    which is the entanglement of a single qubit whose density matrix
    has eigenvalues $\lambda$ and $1-\lambda$.} 
  \label{fig:binentropy}
\end{figure}

Consider the von Neumann entropy in the case of a single
qubit, that is the density matrix
\[\rho=\alpha\ket0\bra0+\beta\ket0\bra1+\beta^*\ket1\bra0+(1-\alpha)\ket1\bra1=
\begin{matrise}\alpha&\beta\\\beta^*&1-\alpha
\end{matrise}.\]
This has two eigenvalues $\lambda$ and $1-\lambda$, and thus the von
Neumann entropy becomes\index{binary entropy}
\begin{equation}
  S(\rho)=-\lambda\log_2\lambda-(1-\lambda)\log_2(1-\lambda)\equiv H(\lambda),
  \label{eq:binentropy}
\end{equation}
which is known as the binary entropy function, shown in Figure
\ref{fig:binentropy}. $H(\lambda)$ is zero when $\lambda=0,1$
and reaches maximum at $\lambda=1/2$ as $H(1/2)=1$. Obviously, the
former is a pure state, while the latter is a maximally entangled
state for a single qubit.

For a general mixed state, the situation is more complicated. Indeed,
it is very hard to eliminate the classical correlations in a mixed
state such that the entanglement measure contains only the quantum
correlations. The entanglement of formation
\index{entanglement!of formation} for a mixed state $\rho$ is defined
as 
\begin{equation}
  E_F(\rho)=\inf\left\{\sum_k\lambda_kS\left(\ket{\psi_k}\bra{\psi_k}\right)\,\Big|\,\rho=\sum_k\lambda_k\ket{\psi_k}\bra{\psi_k}\right\} 
\end{equation}
where the infimum is to be taken over all decompositions of $\rho$
into pure states. Here, $S(\ket\psi\bra\psi)$ indicates the von
Neumann entropy of (say) Alice's reduced density matrix
$\rho'=\trace_B\ket\psi\bra\psi$. 
The explicit computation of the entanglement of
formation is known only for a limited number of cases, such as for the
two-spin entanglement known as concurrence reviewed next, and the
two-mode Gaussian entanglement reviewed in section
\ref{sec:Twomode}.

\subsection{Concurrence}
\index{concurrence|(}
Concurrence is a measure of entanglement in a system consisting of two
spins, or any other 
two-level system that is closely related to the entanglement of
formation \cite{Hill97,Wooters98}. We consider a mixed state $\rho$,
such as the reduced density matrix of two spins in a larger spin
chain. The spin-flip operator on a qubit is defined as
$\ket{\tilde\psi}=\sigma^y\ket{\psi^*}$, which flips the spin in the
standard basis. Note that complex conjugation, like transposition, is
a basis-dependent transformation and it is thus not a physical
operation. Now, to flip both spins in the standard basis
$\{\ket{\!\uparrow\uparrow},\ket{\!\uparrow\downarrow},\ket{\!\downarrow\uparrow},\ket{\!\downarrow\downarrow}\}$
take the transformed density matrix
\[\tilde\rho=(\sigma^y\otimes\sigma^y)\rho^*(\sigma^y\otimes\sigma^y),\]
and define the Hermitian matrix $R=\sqrt{\rho^{1/2}\tilde\rho\rho^{1/2}}$
whose decreasingly ordered eigenvalues we denote
$\lambda_1\geq\lambda_2\geq\lambda_3\geq\lambda_4$, and assume that at
least two of them are non-zero. The trace of $R$ is simply
the fidelity between the spin-flipped density matrix and the
original\index{fidelity}%
\footnote{The fidelity is a distance measure between two quantum
  states, being unity if they are equal, and less than one, though
  positive, else. The fidelity is revisited in Section
  \ref{sec:timeev}.}.
Now the concurrence is defined as
\begin{equation}
  C(\rho)=\max(0,\lambda_1-\lambda_2-\lambda_3-\lambda_4).
\end{equation}
The concurrence is a valid entanglement measure in its own right, but
the entanglement of formation equals
\begin{equation}
  E_F(\rho)=H\left(\frac12\left[1+\sqrt{1-C(\rho)^2}\right]\right),
\end{equation}
where $H(x)$ is the binary entropy function (\ref{eq:binentropy}). Since
$E_F(\rho)$ is a monotonically increasing function of $C(\rho)$ and
they coincide at zero and unity, they can be considered equivalent
entanglement measures, and one usually call the concurrence simply the
entanglement of formation due to this equivalence. 

\begin{figure}[htbp]
  \centering
  \begin{pspicture}(15,5)
    \rput*(7.5,2.5){\includegraphics[width=.8\textwidth]%
      {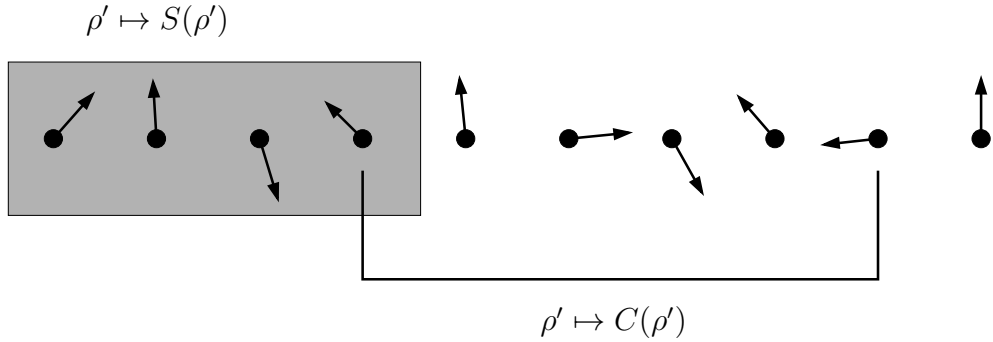}}
    \rput(3,4.5){$\rho'\mapsto S(\rho')$}
    \rput(9,.5) {$\rho'\mapsto C(\rho')$}
  \end{pspicture}
  \caption[Conceptual sketch of entropy and concurrence.]%
  {Conceptual sketch of the two entanglement measures
    entanglement entropy and concurrence in a spin chain. The former
    amounts to finding the reduced density matrix of a part of the
    spin chain and hence the entropy $S(\rho)=-\trace\rho\log_2\rho$,
    while the second considers two individual spins in the chain and
    their entanglement is measured by the concurrence.}
  \label{fig:entanglement_measures}
\end{figure}

A conceptual sketch of the different ways to measure entanglement by
concurrence and entanglement entropy is shown in Figure
\ref{fig:entanglement_measures}. The key point is that the entropy
measures the entanglement of a set of (not necessarily contiguous)
spins with the rest of the chain, where the whole chain is in a pure
state. The 
concurrence, however measures the entanglement shared between only two
spins, where these two spins together might be in a mixed state.
\index{concurrence|)}

\subsection{Other entanglement measures}
There is a number of other measures of entanglement for mixed
states that are entanglement monotones, most of which reduce to the
entanglement entropy when one considers pure states. These include
negativity\index{negativity}, which is defined on a 2-qubit state as
$\mathcal N(\rho)=2\max(0,-\lambda_{\mathrm{neg}})$ where
$\lambda_{\mathrm{neg}}$ is the sum of the negative eigenvalues of
$\rho^{\mathrm T_B}$ where one takes the partial transpose of Bob's
part of the density matrix \cite{Vidal02}. 

Another measure that has gained attention recently is {\it squashed
  entanglement}\index{squashed entanglement} or {\it conditional
  mutual information} (CMI) \cite{Christandl04}, which is defined as
\[E_{\mathrm{CMI}}(\rho_{AB})=\frac12\inf_{\rho_{ABC}}S(A:B|C)\]
where the infimum is to be taken over all tripartite states that are
such that $\rho_{AB}=\trace_C\rho_{ABC}$. The entropy $S(A:B|C)$ is
known classically as conditional mutual information and is the mutual
information of $A$ and $B$ given $C$. Mutual information is given by
\[S(A:B)=S(\rho_A)+S(\rho_B)-S(\rho_{AB}),\]
while conditional mutual information is
\[S(A:B|C)=S(\rho_C)-S(\rho_{AC})-S(\rho_{BC})+S(\rho_{ABC}).\]
That is, we introduce an artificial system $C$, and compute the
maximal information content in $A$ and $B$ given this system. Squashed
entanglement is known to fulfill many of the requirements for
entanglement measures and is a promising candidate for the final
measure of entanglement for mixed states. However, the definition
includes an extremalization that may be computationally very hard,
though little has been done on computing squashed entanglement
\cite{Christandl06}.

\subsection{R\'enyi entropy}
The entanglement entropy (\ref{eq:defSrho}) is not the only entropy
one can define given the density matrix $\rho$, whose eigenvalues are
$\lambda_i$. In particular, the von 
Neumann entropy is a special case of the R\'{e}nyi entropy,
\index{R\'enyi entropy}
\begin{equation}
  H_\alpha(\rho)=\frac1{1-\alpha}\log_2\left(\sum_i\lambda_i^\alpha\right), 
\end{equation}
since $S(\rho)=\lim_{\alpha\to1}H_\alpha(\rho)$. The R\'{e}nyi
entropy in a classical context where the eigenvalue spectrum is
replaced by a probability distribution, has applications in several
areas of physics and information theory, e.g. in signal processing
\cite{Jenssen05}, along with the Shannon entropy, which is the
classical equivalent of the von Neumann entropy. 
All R\'enyi entropies are zero if and only if $\rho$ is
a pure state and positive otherwise. Finally, in the limit
$\alpha\to\infty$ the R\'enyi entropy is simply the logarithm of the
largest eigenvalue,\index{entanglement!single-copy}
\begin{equation}
  H_\infty(\rho)=-\log_2\max_i\lambda_i.
\end{equation}
This is known as the single-copy entanglement
\cite{Eisert05},
since it measures how much entanglement can be extracted from a single
specimen of $\rho$, rather than the asymptotic notion used
to define entanglement of formation that leads to the von Neumann
entropy.
\index{entanglement!measures|)}

\section{Entanglement in quantum information}
Entanglement is an important subject in its own right, as it
challenges our understanding of Nature and how it works, and might be
at the heart of physics. Nevertheless, it is in quantum information
theory that applications of entanglement are known, and hence where
the subject has gained the most attention. Shor's
algorithm\index{Shor's algorithm} is an
example of a quantum algorithm that utilizes entanglement as a
computational resource. However, entanglement is not necessary for
quantum computation, e.g. Grover's search algorithm\index{Grover's
  algorithm} does not attain 
its power from entanglement \cite{Lloyd99}. This might be the reason
that Grover's algorithm is a mere quadratic speedup, compared to
Shor's exponential speedup relative to known classical algorithms.

Also, entanglement can be used as a resource in quantum communication,
including superdense coding\index{superdense coding}, where two
classical bits can be 
transmitted by one qubit given that Alice and Bob share an entangled
state before the communication \cite{Bennett92}. Quantum teleportation
\index{teleportation}
\cite{Bennett93} exploits a shared entangled state  to transmit an
arbitrary quantum state from Alice to Bob using only LOCC, a technique
that has been experimentally implemented in atomic systems to much
popular interest \cite{Kimble04}. The quantum cryptography scheme due
to Ekert also relies on distribution of entangled particles
\cite{Ekert91}, though the popular BB84 protocol does not use
entanglement \cite{BB84}.

\index{entanglement|)}

\chapter{Numerical Analysis}
\label{ch:Numerics}
\index{numerical analysis}
The study of quantum mechanical systems that involve more than a few
particles inevitably requires the use of classical computers to solve
the system. The size of the Hilbert space of a system of $N$ qubits
increases exponentially as $2^N$, which is a contributing factor that a
quantum computer may attain its vast computational power, but also
naturally hampers simulations of quantum systems on a classical
computer. And as long as we don't have the quantum computer, the
simulations must be done on the classical variant. Given a quantum
system of $N$ spin-$1/2$ particles, a possible encoding of the qubit, a
separable state is 
\begin{equation}
  \ket{\psi_{\mathrm{sep}}}=\bigotimes_{n=1}^N\left(\alpha_n\ket0_n+\beta_n\ket
    1_n\right)
  \label{eq:sepstate}
\end{equation}
with $\alpha_n$ and $\beta_n$ complex coefficients. This gives 
$2N$ degrees of freedom --- overall phase factors and normalization
corrected for. However, a general state is
\[\ket\psi=\sum_{\eta=0}^{2^N-1}\alpha_\eta\ket\eta\]
where $\ket\eta$ are the orthogonal basis states, and
$\sum_\eta|\alpha_\eta|^2=1$. The $\alpha$s now constitute $2^N-2$
degrees of freedom. Hence, when entanglement is not included, the
complexity in the state is dramatically reduced as opposed to the
general entangled state, although the schematic overview here is
somewhat simplified.

It is also immediately clear that if one computes the spin-spin
correlation function 
\[\Gamma_{ij}=\expt{\sigma^\alpha_i\sigma^\beta_j}-\expt{\sigma^\alpha_i}\expt{\sigma^\beta_j}\]
for the separable state (\ref{eq:sepstate}), it is identically
zero. Hence, if there are any correlations in the state, they must
arise from entanglement unless there are degeneracies or finite
temperature, which means we may not have a pure state. It is a
well-known fact from critical phenomena that the correlation function
decays exponentially with inter-spin distance,
$\Gamma_{ij}\sim\expe^{-|i-j|/\xi}$ for large distances,
$|i-j|\gg1$. $\xi$ is then known as the correlation length. However,
at critical points, the correlation function decays polynomially, thus
demonstrating the fact that entanglement is much more prevalent at
critical points than elsewhere in the parameter space.

Sometimes it is possible to map the system of interacting qubits onto
a simpler system where the degrees of freedom grows linearly as
opposed to exponentially, as we will explore in later
chapters. However, this is not always possible, and much attention has
been drawn to the subject of finding numerical recipes that can
be used to efficiently compute e.g. the ground state of a
Hamiltonian.

\section{Using symmetries}
\label{sec:symmetries}
Most models exhibit symmetries, that is operators that commute with
the Hamiltonian, and thus share a common eigenvalue set with
this. Imposing periodic boundary conditions on a model will lead to
translational invariance, and thus conservation of momentum according to
Noether's theorem. Thus, the full Hilbert space of the Hamiltonian can
be reduced into subspaces with definite momenta, and thereby somewhat
reducing the size of the matrices that one needs to diagonalize. This
is not as efficient as other methods --- by a long shot --- but it may
reduce the complexity enough to allow numerical analysis of slightly
larger systems than otherwise possible. Nevertheless, the size of the
matrices still grows exponentially and therefore
no fundamentally new window of exploration is opened.

We focus on fermionic models, that is where a Fock state wave function
can be written as a string of binary digits, and we consider
translational, reflection, and parity invariance.  
Define the operators $\mathcal T$,
$\mathcal R$, and $\mathcal P$ for these operations respectively. Each
of these symmetry operators 
$\mathcal S\in\{\mathcal T,\mathcal R,\mathcal P\}$ commute with the
Hamiltonian, $[H,\mathcal S]=0$,
and each has an order $n$ for which $\mathcal S^n=1$. Thus,
$\mathcal S$ must have $n$ eigenvalues $\lambda_k=\expe^{2\pi\imi
  k/n}$. The symmetry operators and their eigenvalues are summarized
in table \ref{tab:symops}.
\begin{table}[b]
  \caption{The symmetry operators of a generic spin chain and their
    eigenvalues and -states. $\ket\chi$ is an eigenstate of
    the Hamiltonian.} 
  \centering
  \begin{tabular}{lcccl}
    \hline
    Name & Symbol & Order & Eigenvalues & Eigenstates \\
    \hline
    Translation   & $\mathcal T$ & $N$ & $\eta_k=\exp(2\pi\imi k/N)$ &
    $\sum_{n=0}^{N-1}\expe^{2\pi\imi kn/N}\mathcal T^n\ket\chi$\\
    Reflection    & $\mathcal R$ & 2 & $r=\pm1$ & 
    $(\mathds 1\pm\mathcal R)\ket\chi$ \\
    Parity change & $\mathcal P$ & 2 & $p=\pm1$ & 
    $(\mathds 1\pm\mathcal P)\ket\chi$ \\
    \hline
  \end{tabular}
  \label{tab:symops}
\end{table}
Each of these operators will reduce the Hilbert space into their
respective subspaces, thus reducing the dimensionality
accordingly. However, they do not commute with each other,
e.g. $[\mathcal T,\mathcal R]\not=0$, so it may not be possible to
reduce the complexity with all operators simultaneously. However,
$[\mathcal P,\mathcal T]=0$, so we exploit these two symmetries
simultaneously to
block-diagonalize the Hamiltonian. The size of the matrix for a given
partition of the Hilbert space with quantum numbers $(k,p)$ is denoted
$N_{(k,p)}$.

If we look at the translation operator, this will provide the greatest
reduction of the Hilbert space since it has the largest number of
different eigenvalues, while the parity operator will split the
Hilbert space in two. Given a state $\ket\chi$, not all values of
$k$ are allowed though, e.g. look at the state $\ket\chi=\ket{0101}$;
\begin{align*}
  &\phantom{=}\left(\mathds 1+\expe^{2\pi\imi k/4}\mathcal
    T+\expe^{4\pi\imi k/4}\mathcal T^2+\expe^{6\pi\imi k/4}\mathcal
    T^3\right)\ket\chi\\
  &=\left(1+\expe^{\pi\imi k}\right)\ket{0101}+\left(\expe^{\pi\imi
      k/2}+\expe^{3\pi\imi k/2}\right)\ket{1010}\\
  &=\left(1+\expe^{\pi\imi k}\right)\left(\ket{0101}+\expe^{\pi\imi
      k/2}\ket{1010}\right).
\end{align*}
The latter expression is zero unless $k=0$ or $k=2$, and hence $k=1$
and $k=3$ do not correspond to  eigenvalues of the operator. More
generally, if $\mathcal T^s\ket\chi=\ket\chi$ for an $s\leq N$ and
some state $\ket\chi$, then $N/s$ must be an 
integer, and $k$ must be a multiple of this integer to be a valid
quantum number. The two extreme cases are if $s=N$, e.g. $\ket{0001}$,
when all $k=0,\cdots,N-1$ are good quantum numbers and if $s=1$,
e.g. $\ket{0000}$, when only $k=0$ is a good quantum number. Hence,
for a given state $\ket{\chi_i}$, there exists $s_\chi$ valid $k$-quantum
numbers, where $s_\chi$ is the number of unique states that $\ket\chi$
can be transformed into through the translation operator. We
have observed that the ground state always is in the $k=0$ segment of
the Hilbert space, at least assuming there is no spontaneous violation
of translation invariance.

\begin{figure}[htbp]
  \psfrag{X}{$N$}
  \psfrag{Y}{\hspace{-3ex}$N_{(k,p)}$}
  \centering
  \includegraphics[width=.6\textwidth]{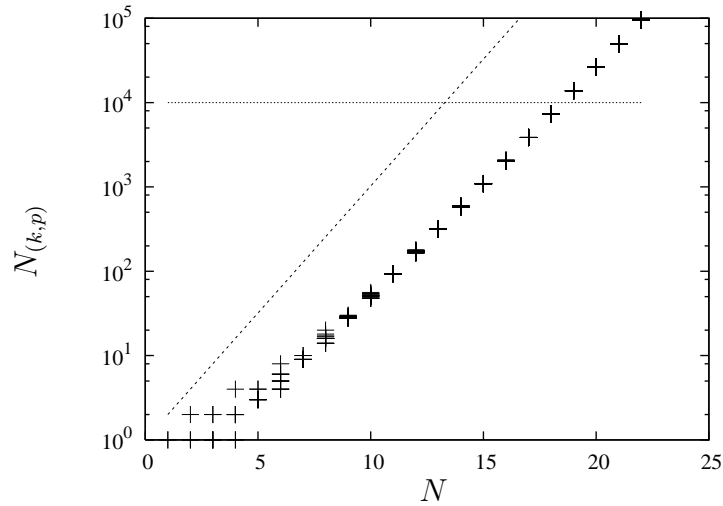}
  \caption[The reduction in complexity when utilizing symmetries.]%
  {The reduction in complexity when utilizing symmetry as measured by
    the size of the matrices needed to diagonalize. The dashed line is
    $2^N$, which is the size of the matrix if no symmetries are used,
    while the points are the size of the matrices in each
    $(k,p)$ segment of the Hilbert space. For illustration
    the horizontal line $N_{(k,p)}=10^4$ is shown as a threshold for
    how large matrices we can diagonalize, and the improvement when
    using symmetries is from $N=13$ to $N=18$.}
  \label{fig:complexred}
\end{figure}

\begin{table}[htbp]
  \caption{The groups when using translational and parity symmetries
    in a fermionic spin chain of size $N=4$.}
  \centering
  \begin{tabular}{llcc}
    \hline
    Quantum numbers & $\ket\chi$ & $s_\chi$ & $N_{(k,p)}$\\
    \hline
    \hline
    $k=0$, $p=1$ & $\ket{0000}$ & 1 & \\
    & $\ket{0011}$ & 4 \\
    & $\ket{0101}$ & 2 & \raisebox{1.5ex}[0pt]{4}\\
    & $\ket{1111}$ & 1 \\
    \hline
    $k=1$, $p=1$ & $\ket{0011}$ & 4 & 1\\
    \hline
    $k=2$, $p=1$ & $\ket{0011}$ & 4 \\
    & $\ket{0101}$ & 2 & \raisebox{1.5ex}[0pt]{2}\\
    \hline
    $k=3$, $p=1$ & $\ket{0011}$ & 4 & 1\\
    \hline
    $k=0$, $p=-1$ & $\ket{0001}$ & 4 \\
    & $\ket{0111}$ & 4 & \raisebox{1.5ex}[0pt]{2}\\
    \hline
    $k=1$, $p=-1$ & $\ket{0001}$ & 4 \\
    & $\ket{0111}$ & 4 & \raisebox{1.5ex}[0pt]{2}\\
    \hline
    $k=2$, $p=-1$ & $\ket{0001}$ & 4 \\
    & $\ket{0111}$ & 4 & \raisebox{1.5ex}[0pt]{2}\\
    \hline
    $k=3$, $p=-1$ & $\ket{0001}$ & 4 \\
    & $\ket{0111}$ & 4 & \raisebox{1.5ex}[0pt]{2}\\
    \hline
  \end{tabular}
  \label{tab:symex4}
\end{table}
To exemplify, the $N=4$ Hamiltonian is a $16\times16$ matrix. This can
be reduced according to Table \ref{tab:symex4} into groups with a
defined parity and wave number. As we see, the matrix is reduced into
one $4\times4$ and smaller matrices. In any one of these eight groups,
each generating state $\ket\chi$ defines a translationally invariant
state,
\[\ket{\psi^{(k,p)}_\chi}=\frac1{\sqrt{s_\chi}}\sum_{n=0}^{s_\chi-1}\expe^{2\pi\imi
  kn/N}\mathcal T^n\ket\chi.\]
Now each element of the new $N_{(k,p)}\times N_{(k,p)}$ Hamiltonian matrix $h$ is
\[h^{(k)}_{\chi\chi'}=\bra{\psi_\chi^{(k)}}H\ket{\psi_{\chi'}^{(k)}}=\sum_{n=0}^{s_\chi-1}\sum_{m=0}^{s_{\chi'}-1}\bra\chi\mathcal
T^nH\mathcal T^m\ket{\chi'}\expe^{2\pi\imi k(n-m)/N}\]
for two generating states $\ket\chi$ and $\ket{\chi'}$. Hence the
computational complexity is reduced somewhat. Figure
\ref{fig:complexred} shows how large the 
submatrices are for different system sizes.
The figure shows that with a computational threshold that restrict
matrix diagonalization upwards to $N=10^4$, the size of the system
that can be considered is increased from $N=13$ to $N=18$, which can
be a 
crucial improvement in certain context, such as the criticality
detection discussed in Section
\ref{ssec:determining_criticality}. Also, one may go beyond this
threshold when looking at 
individual points in the parameter space, but if scanning through a
large portion of the space one would usually apply smaller system
sizes to avoid excessive time consumption.

\section{Renormalization techniques}
Renormalization in classical condensed matter systems was conceived by
Leo P. Kadanoff in 1966 \cite{Kadanoff66}, while Wilson
\cite{Wilson75} refined the method and applied it to critical
phenomena such as the Kondo effect. The basic realization is that at
critical points, the system becomes scale invariant, that is the
thermodynamic properties can be derived from what happens on large
length scales. Hence, one can construct a mapping of a group of
lattice sites onto a ``supersite'' whose properties are derived from
the smaller group. This can now be iterated to successively larger
blocks until the thermodynamic properties can be extracted. This is
known as the renormalization group (RG)\index{renormalization group},
and it creates a 
renormalization flow in the parameter space. The technique has proved
highly successful in classical lattice models, and has become a part
of the standard theory of critical phenomena.

However, the RG method fails spectacularly for some very simple
models, such as the one dimensional particle in a box problem. The
very problem that is often used as the first example
in quantum mechanics textbooks could not be solved by this
approach. White and Noack \cite{WhiteNoack92} identified the crucial
point, that the ground state wave function which is a sine wave,
cannot be composed of the ground states of smaller partitions which
are also sine waves but with different end points. White thus proposed
a new renormalization group, the density matrix renormalization group
(DMRG), which picks out the relevant wave functions to keep in further
iterations very effectively \cite{White1992}.
\index{DMRG}
In brief, the technique amounts to dividing the full system into a
``system'', an ``environment'', and two single sites between them
that work as a connection. Now, the ground state of this so-called
superblock is found as a first approximation. The procedure is then
iterated by shrinking the system until it consists of a single site
and then growing the system until it reaches a threshold size, and so
on until the desired accuracy is reached \cite{Schollwoeck2005}. This
has proven successful for a number of models, such as the above
particle in a box and more real problems such as the Heisenberg
model, and has established itself as the predominant method for
solving low-level excitations in one dimensional systems. 

However, the DMRG fails to reproduce the algebraic decay of the
correlation functions at the critical point, which is a crucial
requirement for any such technique. It has been conjectured that this
is because, while the DMRG technique takes some of the entanglement
into consideration, it does not preserve maximal entanglement under
renormalization \cite{OsborneNielsen01}. At a critical point it
seems that the entanglement constitutes a major part of the
wave function, such that an RG technique to estimate e.g. the ground
state wave function here must take entanglement into account.

There have been a number of suggestions surfacing in the latest years,
such as the entanglement
renormalization\index{entanglement renormalization} \cite{Vidal2005},
which tries to devise a new 
renormalization scheme based on projecting out the most entangled
states. Another approach based on so-called projected entangled pair
states (PEPS) \index{PEPS}\cite{Verstraete04, Verstraete06}
enables efficient computation in two or more dimensions. 

Some issues concerning entanglement properties of the ground state in
the Ising chain are addressed in \cite{SOSent}. In particular, we see
that even though the entanglement is a highly important attribute of
the wave function, and especially so at the critical point in the
Ising chain, there are very few terms in the Schmidt decomposition
that needs to be taken into account to construct the wave
function. However, it is not yet clear how one should
proceed to identify these terms {\it a priori}, that is when the wave
function is not previously known as is the case with the Ising
chain. Nevertheless, it is enough to know the first four terms in the
Schmidt decomposition of a spin chain with $N=100$ spins to determine
the entropy with an accuracy of $10^{-4}$.

\chapter{Continuous Variables}
\label{Ch:ContVariable}
Usually quantum information theory is occupied with two-level quantum
systems, both due to its simplicity and its obvious connection to
classical information theory. In particular, the analogy that
classical bits are substituted with quantum bits in a quantum computer
is appealing, but quantum information in systems with more than two
levels, or even a continuous spectrum, is also
considered. Such systems are experimentally important since they are
easy to generate and manipulate, and quantum information based on
quantum optics is arguably where most effort is done implementing
feasible quantum systems.
Particularly, optical realizations are important in
quantum communication.
In this chapter we will study
continuous variable (CV) systems, that is where each entity can have an
infinite number of eigenvalues rather than the two-level system of a
qubit. The infinite number of eigenvalues
corresponds to the fact that any given mode of the system may contain
an arbitrary number of particles, denoted $\ket n$ with
$n=0,1,2,\cdots$. The canonical example of a CV system is the 
textbook harmonic oscillator, which has an infinite set of discrete
energy eigenvalues $E_n=\omega(n+1/2)$.
The density matrix of such a system is, in contrast to the fermionic
case, infinite
dimensional. This means a very different approach to computing
physical properties, and also that the entanglement of a state is not
bounded upwards as opposed to the fermionic case.

Generally, a CV system of $N$ particles, or modes, consists of
$N$ conjugate pairs $x_k$ and $p_k$ that satisfy the canonical
commutation relations 
\begin{equation}
  [x_k,p_l]=\imi\delta_{kl}.
\end{equation}
It is customary to define the annihilation and creation operators 
\begin{equation}
  a_n=\frac1{\sqrt2}\left(x_n+\imi p_n\right)\qquad 
  a^\dag_n=\frac1{\sqrt2}\left(x_n-\imi p_n\right),
\end{equation}
which satisfy the commutation relations
$[a_n,a_m^\dag]=\delta_{nm}$. 

The conventional way to
define the density operator is in terms of $\vkt x$ and $\vkt p$,
$\rho(\vkt x, \vkt p)$. 
We can also explicitly write down the density matrix in the position
basis, in which it is a continuous
matrix $\rho(\vkt x,\vkt x')=\sum_n\psi^*_n(\vkt x)\rho(\vkt x',\vkt
p')\psi_n(\vkt x')$ in a choice of basis with basis vectors
$\{\psi_n(\vkt x)\}$. 
This normalizes as $\int\rho(\vkt x,\vkt
x)\dd^{2N}\vkt x=1$ and the expectation value of an operator $\mathcal
O(\vkt x,\vkt p)$ is
\begin{equation}
  \expt{\mathcal O(\vkt x,\vkt p)}=\trace\mathcal O(\vkt x, \vkt
  p)\rho(\vkt x,\vkt x').
\end{equation}
However, it is not always convenient to write the states in the
explicit position representation, and we will review the equivalent
Weyl representation of the states below. 

Separability criteria for CV systems have been proven \cite{Giedke01},
and we will here recite some of the important aspects of the
continuous variable systems in terms of Gaussian states and refer to
the proofs in the literature.

\section{Weyl algebra}
\index{Weyl!algebra}
Continuous variable states as defined by the $x_k$ and $p_k$ operators
can be conveniently formulated in terms of the Weyl algebra. To this
end, we define the $2N$ operators $r_k$ as
\begin{equation}
  \vkt r=\begin{matrise}x_1, & p_1, & x_2, & p_2, & \cdots & x_N, & p_N\end{matrise}\transpose.
\end{equation}
These operators' commutation relations can be written as
\begin{equation}
  [r_k,r_l]=\imi J_{kl}
\end{equation}
where $J$ is the $2N\times 2N$ matrix
\begin{equation}
  J=\bigoplus_{k=1}^N\begin{matrise} 0 & 1\\- 1 &  0 \end{matrise}.
\end{equation}
This matrix is the defining matrix of the symplectic
group\index{symplectic!group} $\mathrm{Sp}(2N)$, which is the group
under multiplication 
of all $2N\times 2N$ matrices $M$ that satisfy
\begin{equation}
  M\transpose JM=J.
\end{equation}

Now we may define the Weyl operators\index{Weyl!operators} which are
defined on $\xi\in\mathds R^{2N}$ as
\begin{equation}
  \mathcal W(\xi)=\exp\left(-\imi \sum_k\xi_k r_k\right).
\end{equation}
Finally, defining the form $\sigma(\xi,\zeta)=\sum_{ij}\xi_iJ_{ij}\zeta_j$, the
Weyl relations\index{Weyl!relations} are
\begin{equation}
  \mathcal W(\xi)\mathcal W(\zeta)=\expe^{-\imi\sigma(\xi,\zeta)}\mathcal
    W(\zeta)\mathcal W(\xi).
\end{equation}

Given a quantum state $\rho$, the characteristic
function\index{characteristic function} $\chi(\xi)$ of the state
is simply the expectation value of the Weyl operators,
\begin{equation}
  \chi(\xi)=\trace\rho\mathcal W(\xi).
\end{equation}
Inversely, the density matrix can be written in terms of the
characteristic function and the Weyl operators,
\begin{equation}
  \rho=\frac1{(2\pi)^N}\int\chi(\xi)\mathcal W(-\xi)\,\dd^{2N}\xi.
\end{equation}
All expectation values of polynomials of the canonical operators $r_k$
can be expressed in terms of differentiations of the characteristic
functions. Indeed, the two first momenta are
\begin{subequations}
  \begin{align}
    \expt{r_k}&=\imi\frac\partial{\partial
      t}\left.\chi(t\hat e_k)\right|_{t=0}\\
    \expt{r_kr_l}&=-\frac{\partial^2}{\partial t_1\partial
      t_2}\left.\expe^{-\frac\imi2 t_1t_2J_{kl}}\chi(t_1\hat e_k+t_2\hat e_l)\right|_{t_1=t_2=0},
  \end{align}
  \label{Expt12momenta}
\end{subequations}
where $\hat e_k$ is the unit vector of element $k$.

According to the above, the formulation of the theory with Weyl
operators is equivalent to the formulation with canonical
operators. Hence, which one is applied is a matter of taste and
computational convenience. A general state in the form of a density
matrix is completely determined by all $n$th order momenta. However,
we will focus on the important class of Gaussian
states\index{Gaussian states}. These are experimentally important due
to the simplicity involved in creating and manipulating them. Also,
they are simple and elegant to compute analytically but nevertheless
involves a rich structure. Further, for the Gaussian
states only the first and second order momenta are necessary to
completely determine the state. This is of course a convenient
situation where both experimental and theoretical simplicity meets and
Gaussian states are therefore by far the most studied continuous
variable states in both camps.

\section{Gaussian states}
\index{Gaussian states}
A state $\rho$ is called Gaussian if its characteristic function is
a Gaussian,
\begin{equation}
  \chi(\xi)=\exp\left[-\frac14\sum_{ij}\xi_i\Gamma_{ij}\xi_j+\imi\sum_i\delta_i\xi_i\right].
\end{equation}
The $2N\times 2N$ matrix $\Gamma$ must be positive and symmetric to
give rise to finite expectation values. $\Gamma$ is known as the
correlation matrix and $\delta\in\mathds R^{2N}$ as the displacement
vector. Equivalently, and according to the convention used in
\cite{SOSbose}, a Gaussian state can be written in the position basis
as
\begin{align}
  \rho(\vkt{q},\vkt{q}')=&
  \sqrt{\frac{\det\left(A'-C'\right)}{\pi^D}}\exp\left[-d'_i\left(A'_{ij}-C'_{ij}\right)^{-1}d'_j\right]\nonumber\\
  &\times\exp\Big[-\frac{1}{2}\left(q_iA_{ij}q_j+q'_iA^*_{ij}q'_j\right)+q_iC_{ij}q'_j+d_iq_i+d^*_iq'_i\Big]. 
  \label{rhoqq}
\end{align}
Here $A'=\real A$, $C'=\real C$, and $d'=\real d$ with $A$ and
$C$ $N\times N$ matrices and $d$ an $N$ dimensional vector.
The equivalence of the two formulations is evident when we compute the
first and second order momenta from
Eqs. (\ref{Expt12momenta}) and from direct calculation on the density
matrix. To simplify the equations we consider the variational matrices
\begin{subequations}
  \begin{align}
    Q_{kl}&=\expt{q_kq_l}-\expt{q_k}\expt{q_l}\\
    P_{kl}&=\expt{p_kp_l}-\expt{p_k}\expt{p_l}\\
    S_{kl}&=\frac12\expt{q_kp_l+p_lq_k}-\expt{q_k}\expt{p_l}.
  \end{align}
\end{subequations}
Computing these for both formulations, we find
\begin{subequations}
  \begin{alignat}{2}
    Q_{kl}&=\Gamma_{2k-1,2l-1}&=&\,\frac12\left\{\left(\real A-\real C\right)^{-1}\right\}_{kl}\\
    P_{kl}&=\Gamma_{2k,2l}&\,=&\,A_{kl}-(A_{ki}-C_{ki})Q_{ij}(A_{jl}-C_{jl})\transpose\\
    S_{kl}&=\frac12\Gamma_{2k-1,2l}&=&-Q_{ki}(\imag A_{il}+\imag C_{il})\\
    \expt{q_k}&=\xi_{2k-1}&=&\,2Q_{ki}\,\real d_i\\
    \expt{p_k}&=\xi_{2k}&=&-2(\imag A_{ki}-\imag C_{ki})Q_{ij}\,\real d_j+\imag d_k.
  \end{alignat}
\end{subequations}
This shows the mapping between the two formulations, and though they
are equivalent, the mapping is complicated. However, the position
representation gives a more physical view that counters the
mathematical convenience of the Weyl representation.

As shown in \cite{SOSbose}, using the position representation one can
perform a rotation, scaling and another rotation of the coordinate
system to separate the density matrix into single particle density
matrices for $N$ virtual particles. Having found this transformation,
it is easy to find the eigenvalues of the single particle density
matrices and thus the eigenvalues of the full density matrix. Also, we
show that the massless Klein-Gordon 
field possesses conformal symmetries (see Section
\ref{sec:conformal_symmetries}), and that this can be identified by the
entanglement signatures in the model.

It is easy to realize that a state with $C=0$ is separable with
$\rho(\vkt q,\vkt q')=\rho_1(\vkt q)\rho_2(\vkt q')$, and thus
a pure state. Indeed, this is a necessary criterion, and for the Weyl
formulation this is equivalent to $\det\Gamma=1$.

\section{The Klein-Gordon field}
\index{Klein-Gordon field}
As our main example of a bosonic field we use the simplest possible,
namely the free Klein-Gordon field placed on a $1+1$ dimensional chain
with periodic boundary conditions. That is, we use the Lagrangian
\begin{equation}
  \mathcal
  L=\sum_n\left[\dot\varphi_n^2-\left(\nabla\varphi_n\right)^2-\kappa^2\varphi_n^2\right] 
\end{equation}
summed over all lattice points. The Euler-Lagrange equations gives the
equations of motion and the dispersion relation
\begin{equation}
  \omega_k^2=\frac4{a^2}\sin^2(k/2)+\kappa^2
\end{equation}
where $a$ is the lattice constant. To maintain a conformal invariance
of the model, one must keep $Na$ constant when rescaling the
theory. Now follow the expectation value matrices for the ground state
of the field,
\begin{align}
  Q_{mn}&=\frac1{2N}\sum_k\frac1{\omega_k}\,\expe^{\imi k(m-n)}\\
  P_{mn}&=\frac1{2N}\sum_k\omega_k\,\expe^{\imi k(m-n)}\\
  S_{mn}&=0.
\end{align}
The sums over $k$ are over all wave numbers $2\pi n/N$ with
$n=\{0,\cdots,N-1\}$, discretized through the boundary conditions.

In the massless limit, the field is expected to be conformal, and thus
follows the conformal signature derived by Holzhey, Larsen, and
Wilczek \cite{Holzhey94,SOSbose} (see Section
\ref{sec:conformal_symmetries}). Massless here means roughly 
$\kappa\leq0.1$. The true massless system $\kappa=0$ is not accessible
as $Q$ diverges due to the $k=0$ term. A massive theory in the limit
$\kappa\to\infty$ carries no entanglement.

One can also impose anti-periodic boundary conditions on the field,
thereby explicitly violating the translational invariance in the
target space, and hence the
conformal symmetry. The entanglement diverges as $N\to\infty$, like in
the periodic case. However, in the limit $\kappa\to 0$ the periodic
wave function leads to diverging entanglement, while the anti-periodic
one does not. This is an example of entanglement
divergence even though we do not have conformal invariance. Usually,
in non-critical systems, the entanglement saturates at some point.

\section{Two-mode entanglement}
\label{sec:Twomode}
The reduced density matrix of two sites in the lattice is generally a
mixed state, and generally the entanglement in this state is
ill-defined. However, for Gaussian CV states, it is known how to find
the entanglement of formation for this state \cite{Wolf04}. That is,
the so-called Gaussian entanglement of formation $E_G$ of a state
$\rho_\Gamma$ with correlation matrix $\Gamma$ is
\begin{equation}
  E_G(\rho_\Gamma)=\inf_{\Gamma_p}\left\{S(\rho_{\Gamma_p})\big|\Gamma_p\leq\Gamma\right\}
\end{equation}
where $\Gamma_p$ are correlation matrices of all pure states and
$S(\rho_{\Gamma_p})$ is the von Neumann entropy of the state. For the
case of Gaussian states this is determined by the correlation matrix'
symplectic eigenvalues\index{symplectic!eigenvalues}. Given a
symmetric, positive matrix $\Gamma$, one can find a symplectic matrix
$M\in\mathrm{Sp}(2N)$ and a diagonal matrix $D\in \mathds R^{N\times N}$
such that
\begin{equation}
  M\Gamma M\transpose=D\oplus D.
\end{equation}
This is known as the symplectic
diagonalization\index{symplectic!diagonalization} of $\Gamma$, and it
is unique up to permutations. The elements of
$D=\mathrm{diag}(d_1,\cdots,d_N)$ are known as the symplectic
eigenvalues of $\Gamma$. Now, the von Neumann entropy of the pure
state $\Gamma_p$ is \cite{Wolf04}
\begin{gather}
  S(\rho_{\Gamma_p})=\sum_k \zeta(\alpha_k)
\end{gather}
where
\begin{gather}
  \zeta(x)=\cosh^2(x)\log_2(\cosh^2x)-\sinh^2(x)\log_2(\sinh^2x)
  \label{zetaxdef}\\
  d_k=\cosh\alpha_k,\qquad\alpha_k\geq0.\nonumber
\end{gather}

Given a 1+1$D$ chain of bosonic systems, one can write the correlation
matrix for two sites, known as the two-mode correlation matrix, as
\begin{equation}
  \Gamma=
  \begin{matrise}
    q_0 & 0   & q_1 & 0   \\
    0   & p_0 & 0   & p_1 \\
    q_1 & 0   & q_0 & 0   \\
    0   & p_1 & 0   & p_0 
  \end{matrise}
  =\Gamma_q\otimes
  \begin{matrise}1&0\\0&0\end{matrise}+
  \Gamma_p\otimes
  \begin{matrise}0&0\\0&1\end{matrise}
\end{equation}
where
\[\Gamma_q=\begin{matrise}q_0&q_1\\q_1&q_0\end{matrise}\qquad\text{and}
\qquad\Gamma_p=\begin{matrise}p_0&p_1\\p_1&p_0\end{matrise}.\]
Here $q_0=Q_{00}$, $q_1=Q_{0n}$, $p_0=P_{00}$, and $p_1=P_{0n}$ with $n$
the distance between the two sites under consideration.
Generally, any two-mode correlation matrix can under local unitary
transformations be written in the so-called standard form 
\begin{equation}
  \tilde\Gamma=
  \begin{matrise}
    n_a & 0   & k_q & 0   \\
    0   & n_a & 0   & k_p \\
    k_q & 0   & n_b & 0   \\
    0   & k_p & 0   & n_b 
  \end{matrise}
\end{equation}
with $k_q\geq|k_p|$.
For the special case that $n_a=n_b=n$, the state is known as a
symmetric state \cite{Giedke03} and a necessary condition for this to be
a correlation matrix is that $n^2-k_q^2\geq1/4$. Also, the state is
entangled if and only if $(n-k_q)(n+k_p)<1$ \cite{Giedke01}.
Symmetric states
arise for example as two light beams from a parametric down converter
is sent through optical fibres or when a squeezed state is sent
through two identical lossy fibres.  
The entanglement of the
symmetric Gaussian state is particularly simple,
\begin{equation}
  E_G(\rho_\Gamma)=\zeta(p),\qquad p=\frac12\ln\left[(n-k_q)(n+k_p)\right]
\end{equation}
with $\zeta(x)$ defined in (\ref{zetaxdef}).
In terms of the correlation matrix this can be
written as a symmetric state through the symplectic transformation
\[S=\mathrm{diag}\left(\sqrt\eta, 1/\sqrt\eta, \sqrt\eta, 1/\sqrt\eta\right)\]
where $\eta$ is the squeezing factor $\eta=\sqrt{p_0/q_0}$.
This is a local, unitary Gaussian operation. Hence
\begin{equation}
  n=\sqrt{q_0p_0}\qquad k_q=q_1\sqrt{\frac{p_0}{q_0}}\qquad k_p=p_1\sqrt{\frac{q_0}{p_0}},
\end{equation}
and the entanglement follows from this. 

\begin{figure}[htbp]
  \psfrag{X}{$N$}
  \psfrag{Y}{$E_G$}
  \includegraphics[width=.5\textwidth]{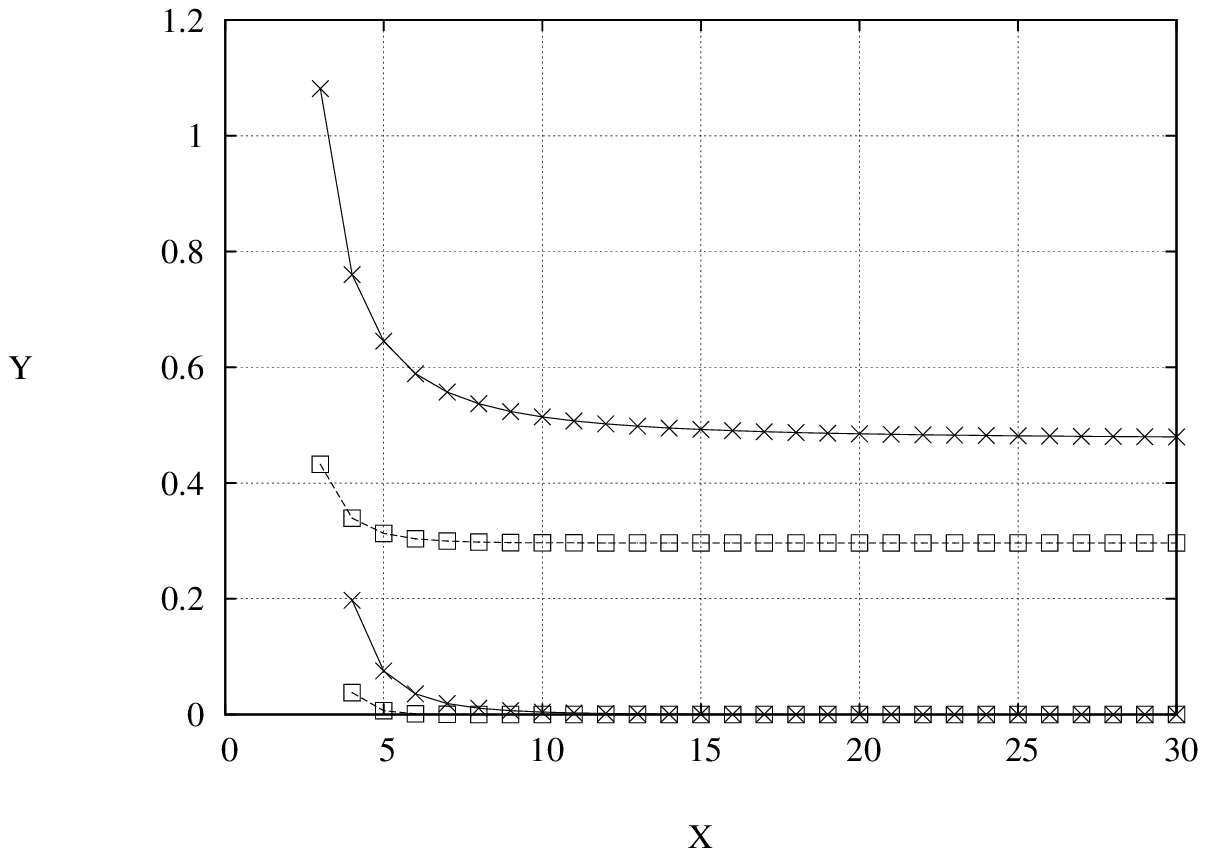}
  \psfrag{X}{$\kappa$}
  \includegraphics[width=.5\textwidth]{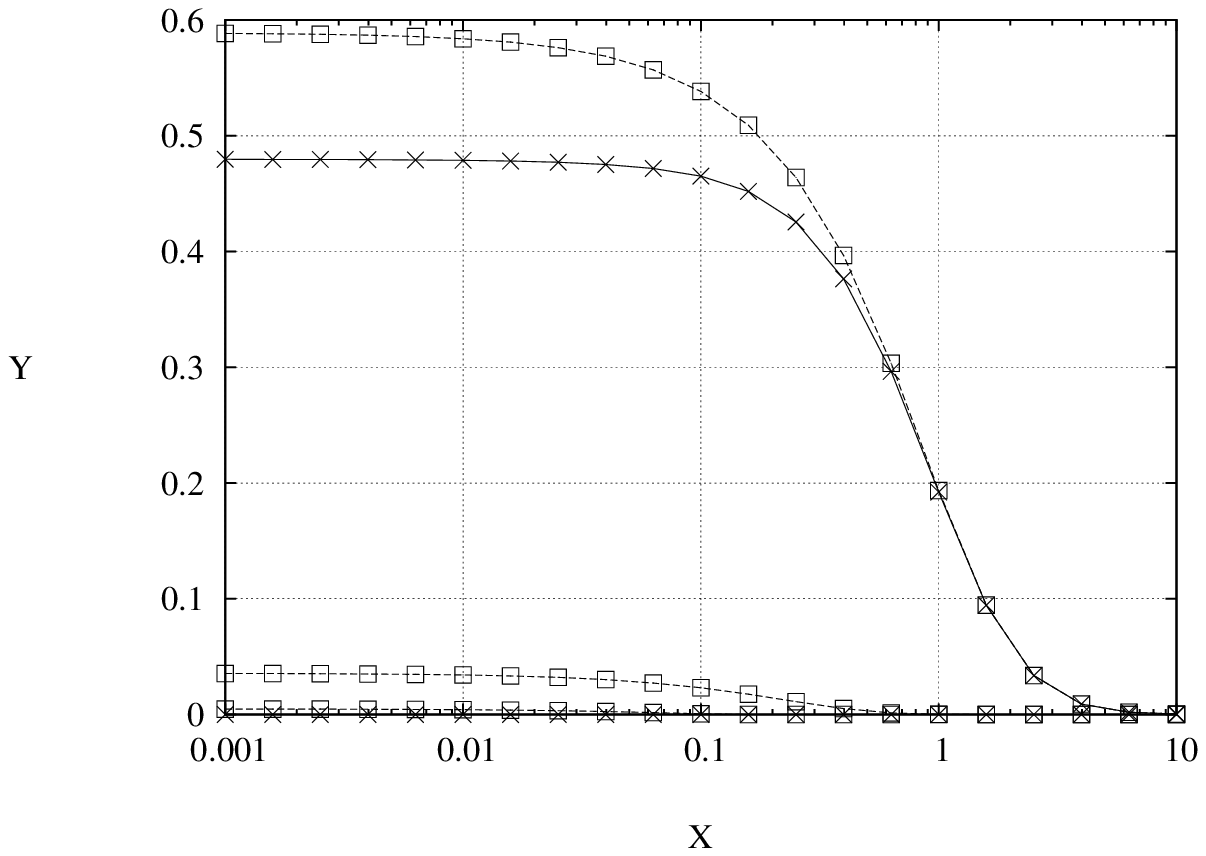}
  \caption[Two-mode Gaussian entanglement in the Klein-Gordon field]%
  {The two-mode Gaussian entanglement between two sites in the 1D
    Klein-Gordon field with periodic boundary conditions.\\
    {\it Left:} $E_G$ for the nearest neighbor and next-nearest
    neighbor as function of total system size $N$. The data are for a
    massless ($\kappa=10^{-3}$, $\times$) and a massive
    ($\kappa=10^{-0.2}$, $\Box$) system. The next-neighbor
    entanglement vanishes for large systems, while the 3rd neighbor
    entanglement is non-zero only for $N=6$.\\
    {\it Right:} $E_G$ for the three nearest neighbor pairs as
    function of $\kappa$ for a large ($N=30$, $\times$) and a small
    ($N=6$, $\Box$) system.
  }
  \label{fig:Entanglement_KG}
\end{figure}

Some results for the Klein-Gordon field is shown in
Fig. \ref{fig:Entanglement_KG}. We see that the entanglement in the
massless field vanishes quickly with distance, indeed for large
systems the entanglement extends only to the nearest neighbor. For
small systems it may extend to the third neighbor, due to
the periodic boundary conditions. Also, all entanglement vanish for a
massive system of $\kappa\gtrsim 1$, as is the case for the
entanglement entropy \cite{SOSbose}. However, there is no divergent
entanglement for the two-mode case as opposed to the entanglement
entropy, but the entropy saturates for $N\to\infty$, $\kappa\to0$ at
$E_G\approx0.48$. 

\section{State evolution}
The massless Klein-Gordon field, or the bosonic vacuum investigated in
\cite{SOSbose} gives directly rise to a conformal symmetry and the
properties that can be deduced from that. However, the time evolution
of a state in 
vacuum is trivial, and to provoke a non-trivial time evolution of the
system, we introduce an impurity and consider the toy model
with Lagrangian density
\begin{equation}
  \mathcal L_n=\frac12\left[\dot\varphi^2_n-(\nabla\varphi_n)^2-\kappa^2\varphi_n^2\right]+\varepsilon\delta_{nl}.
\end{equation}
This is the usual Klein-Gordon field bracketed and an impurity
of strength $\varepsilon$ at a site $l$ to provoke a non-trivial time
evolution of the state. The system is still linear, and the discussion
of this system is an example of how linear systems can still
provide interesting behaviour. We assume the initial condition that
the initial state is that of the unperturbed case $\varepsilon=0$.

We apply 
periodic boundary conditions, and Fourier expanding the field in the 
Heisenberg picture gives
\begin{equation}
  \varphi_n(t)=\sum_k\frac1{\sqrt{2N\omega_k}}\left[a_k(t)\expe^{\imi
      kn}+a_k^\dag(t)\expe^{-\imi kn}\right]
\end{equation}
with the initial condition $a_k(0)=a_k$.
The time evolution of the operators $a$ and $a^\dag$ can now be computed by the
Euler-Lagrange equations to be
\begin{equation}
  a_k(t)=a_k\expe^{-\imi\omega_kt}+\frac{h_k}{\omega_k^2}\left(1-\expe^{-\imi\omega_kt}\right),
\end{equation}
where 
\[\omega^2_k=\frac4{a^2}\sin^2(k/2)+\kappa^2\qquad\text{and}\qquad
h_k=\sqrt{\frac{\omega_k}{2N}}\,\varepsilon\,\expe^{-\imi kl}.\]
Now, we consider the vacuum state $\ket\Omega$,
that is the ground state of the unperturbed system where
$a_k\ket\Omega=0$. Then the correlation matrix is unchanged, but the
expectation values of the field become
\begin{equation}
  \expt{\varphi_n(t)}=\frac\varepsilon
  N\sum_k\frac1{\omega_k^2}\left[\cos
    k(n-l)-\cos\left(\omega_kt-k(n-l)\right)\right],
  \label{varphiexpt}
\end{equation}
and, if we consider the site of the impurity, $n=l$ we find the
equation of motion
\begin{equation}
  \expt{\varphi_n(t)}=\frac\varepsilon
  N\sum_k\frac1{\omega_k^2}\left(1-\cos\omega_kt\right).
\end{equation}
When $N\to\infty$ we can write the constant term as
\begin{equation}
  \frac\varepsilon N\sum_k\frac1{\omega_k^2}=\frac\varepsilon\pi\int_0^\pi\frac1{\omega_k^2}=\frac{\pi a}{\kappa\sqrt{4+\kappa^2a^2}}.
\end{equation}

\begin{figure}[b]
  \psfrag{X}{$t$}
  \psfrag{Yphi}{$\expt{\varphi_l(t)}$}
  \psfrag{Yphi2}{$\expt{\varphi_l(t)}/1000$}
  \psfrag{Ypi}{$\expt{\pi_l(t)}$}
  \centering\includegraphics[width=.9\textwidth]{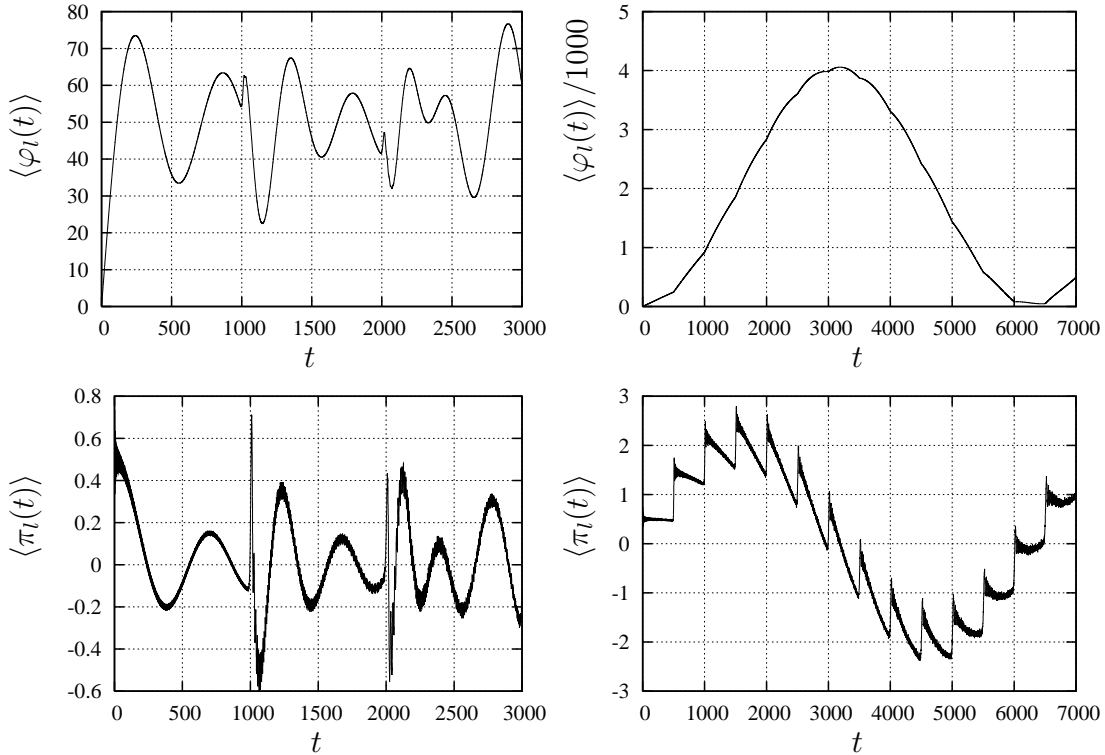}
  \caption[Field expectation values with time in a bosonic field with impurity]%
  {The field expectation values $\expt{\varphi_l(t)}$ and
    $\expt{\pi_l(t)}$ at the perturbation site $l$ with time. To the
    left $\kappa=10^{-2}$ and $N=1000$ ($\kappa N=10 > 2\pi$), and to
    the right $\kappa=10^{-3}$ and $N=500$ ($\kappa N=0.5 <
    2\pi$). Note that a perturbation hits every $N$ time units and that
    the predominant oscillations have period $2\pi\kappa^{-1}$.}
  \label{fig:timeevolution}
\end{figure}
Figures \ref{fig:timeevolution} show the time evolution of the field
expectation value $\expt{\varphi_n(t)}$ and its derivative, the
expectation value of the conjugate field
$\expt{\pi_n(t)}=\expt{\dot\varphi_n(t)}$. The field has a roughly
cosine shape, governed by the main contributing $k=0$ term
(``the zero-mode'') of the sum \ref{varphiexpt}. Hence the period is
$2\pi\kappa^{-1}$. Also, given that we consider the impurity term, a
new perturbation hits this site every time a perturbation has propagated
through the system to the site. Thus, with units where the
perturbation speed in the system equals the lattice constant per time
unit, there 
will be a new perturbation every $N$ time units. If we consider another
site $\delta$ sites away from the perturbation site, the perturbation will
hit every $\delta+Nn$ and $N-\delta+Nn$ time units with integer $n$.
In the following we consider the impurity points and
units where the perturbation speed is one lattice point per time unit
and an invariant lattice constant $a=1$. 
Now, we have two different regimes dependent on $\kappa
N\lessgtr2\pi$. For large systems where $\kappa N>2\pi$, the main
contribution is the oscillations of the point itself. The field tends
to relax its oscillation over time until a perturbation hits which
excites the field again, followed by another relaxation period. For small
systems with $\kappa N<2\pi$ the relaxation never occurs since the
perturbation hits the point too quickly, and thus the perturbations
are the main contribution to the time evolution. Figures
\ref{fig:timeevolution} illustrate these two regimes. In both cases the leading
oscillatory term is with frequency $2\pi\kappa^{-1}$, even though the
main oscillations have different origins. Also, the expectation value
$\expt{\pi_l(t)}$ has small oscillations of frequency $2\pi$, which in
the figure appears as a smearing of the graph.

\chapter{Quantum Critical Systems}
\label{Ch:QuantumCritical}
Classical phase transitions occur under the change of temperature where
at some critical temperature $T_c$ the system has an abrupt
thermodynamical change, the hallmark example being the liquid-solid
transition in e.g. water. Quantum phase transitions \cite{Sachdev}
occur at zero temperature, and thus quantum effects can automatically
be assumed to be profound. Indeed, the thermal fluctuation that take a 
classical state across a phase transition are non-existent, and the
only fluctuations present are the quantum fluctuations determined by
Heisenberg's uncertainty principle.

The general setup is to consider a quantum
system under the change of some external parameter $\gamma$, which
causes a transition at some critical value $\gamma_c$, where a
previously 
excited state now becomes the ground state. This crossover is called
the quantum critical point. At this critical point there are, just as
in classical phase transitions, scaling laws. For example the energy
gap $\Delta$, which necessarily is zero at the transition itself, will
scale as
\[\Delta\sim J|\gamma-\gamma_c|^{z\nu}\]
where $z\nu$ is known as the critical exponent\index{critical
  exponents} of the phase transition and $J$ is some energy
  scale. Moreover, the correlation length 
$\xi$ will diverge (in an infinite system) at the critical point, and
scale according to $\xi\sim L|\gamma-\gamma_c|^\nu$. This also means
  that we have a scaling law between the correlation length and the
  energy gap, \index{correlation length}
\[\Delta\sim\xi^{-z}.\]
Since the correlation length diverges -- or extends to the entire
system in 
the case of a finite system size -- the system is scale invariant
\index{scale invariance} at the critical point, and the 
properties of the system becomes independent of the scale on which we
view the system.

Any classical phase transition invariably has temperature fluctuations
which drive the phase transition. However, as the quantum phase
transitions occur at $T=0$, there can be no classical fluctuations in
these models. Nevertheless there are always quantum fluctuations due to
Heisenberg's uncertainty principle, and these fluctuations have to
drive the quantum transition, in the sense that they are necessary to
explore the low-lying excitations of the system. Moreover, the
classical correlations 
present in the classical models can be partially replaced by the
quantum correlations which present themselves as entanglement. 

\begin{figure}
  \begin{pspicture}(15,5)
  \psset{labels=none,ticks=none}
  \psaxes{->}(1,1)(7,5)
  \psaxes{->}(9,1)(15,5)
  \rput(.5,5){$T$}
  \rput(8.5,5){$T$}
  \rput(7,.5){$\gamma$}
  \rput(15,.5){$\gamma$}
  \pscircle*(4,1){.1}
  \rput(4,.5){$\gamma_c$}
  \pscircle*(12,1){.1}
  \rput(12,.5){$\gamma_c$}
  \definecolor{shade}{rgb}{.7,.7,.7}
  \pscustom[linecolor=white,fillstyle=solid,fillcolor=shade,linewidth=0pt]{%
    \pscurve{-}(9.2,4.5)(11,3.4)(12,1.4)(12,1)(12,1.1)(11,2.7)(9.2,3.5)
  }
  \pscurve[linewidth=1pt]{-}(12,1)(12,1.2)(11,3)(9.2,4)
\end{pspicture}
  \caption[Phase diagram for a quantum phase transition]%
  {Two scenarios for a quantum phase transition in which a
    parameter $\gamma$ is taken across the critical value
    $\gamma_c$. To the left there is no classical phase transition
    close to the quantum transition, while to the right there is a
    classical phase transition across the line, and in the shaded
    region the theory of classical phase transitions can be
    applied. \cite{Sachdev}} 
  \label{fig:quantumclassicalPT}
\end{figure}
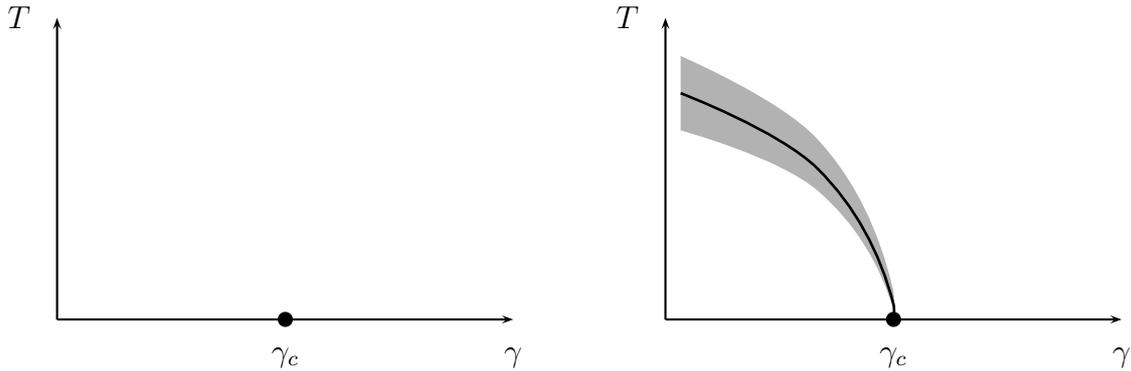

The quantum phase transition may or
may not have a classical counterpart in the $T>0$ regime, see
Fig. \ref{fig:quantumclassicalPT}, in which case there will be a
region where there are competing classical and quantum fluctuations
across the boundary. However, there are also cases where the
thermodynamic variables are analytical close to the $T=0$ line, and
thus no classical phase transition takes place.

In the study of quantum statistical mechanics, we use the density
operator, which at finite temperatures take the form
\[\rho=\expe^{-\beta H},\]
where $\beta$ is the inverse temperature and $H$ the
Hamiltonian. However, in the zero-temperature limit $\beta\to\infty$
only the ground state will contribute to 
this expression, and assuming the Hamiltonian is diagonal in some
basis $H=\sum_nE_n\ket n\bra n$ with lowest energy level
$E_0=0$, the
density operator becomes 
\[\rho(T=0)=\expe^{-\beta\sum_n E_n\ket n\bra n}=\expe^{-\beta E_0\ket
  0\bra0}=\ket 0\bra 0.\]
Hence, when computing properties of a quantum critical system, we need
to identify the ground state and hence follows the density matrix and
all thermodynamic properties.

A phase transition can, strictly speaking, only occur in the
thermodynamic limit of $\sim10^{23}$ particles, which is naturally
inaccessible when implementing the system on a computer. However, one
may see traces of what may become a phase transition at smaller
systems, even with as little as $\sim10$ particles
\cite{SOS1}. However, it is important to emphasize that no actual
phase transition can occur in this system, and that when we speak
about parameters where we get a phase transition, it should be
interpreted as parameters where a phase transition occurs
in the thermodynamic limit.

\section{Ground state entropy}
When the ground state of a quantum system changes macroscopic features
across a quantum phase transition, the details of the ground state is
vastly more complicated at the phase transition point than at
non-critical places in the parameter space. This is indicated by the
increase of entanglement at the critical point as
measured both by the concurrence \cite{Osterloh:2002} and the
entanglement entropy \cite{Vidal:2002rm,Latorre:2003kg}. As the
entropy involves tracing out some part of the system, it will have a
dependence on the size of the system traced out, as well as the
overall system size. However, it should not depend on the small-scale
details of the models since these are irrelevant at critical points
according to renormalization theory.

Specifically, at the critical point in theories where one has
conformal invariance in $1+1$ dimension, the entropy diverges
logarithmically 
with system size, as opposed to off-critical points where the entropy
saturates \cite{Latorre:2003kg}. However, close to the critical point,
the system size where the entropy saturates is very large, and thus
this is not a very efficient indicator of whether or not a system is
critical due to the computational resources required.

\section{Consequences of conformal symmetry}
\label{sec:conformal_symmetries}
\index{conformal symmetries}
Since Polyakov \cite{Polyakov70} it has been known that local scale
invariance implies conformal invariance and that this fact can be used
to investigate criticality in statistical mechanical systems. Belavin,
Ployakov and Zamolodchikov \cite{BPZ84b} showed how the rich structure
of two-dimensional conformal theories can be used to extract
information on the critical properties of a quantum field theory.

A quantum field theory becomes conformal if it fulfills the
following requirements \cite{Francesco97, Ginsparg1988};
\begin{itemize}
\item{Translation and rotation invariance}
\item{Local interactions}
\item{Lorentz invariance}
\item{Scale invariance}
\end{itemize}
The last condition is the crucial one, since it is
only fulfilled at critical points. Thus we expect conformal
symmetry on the critical point, but not elsewhere in the parameter
space. The conformal group in two dimensions is infinite as
opposed to larger dimensions\footnote{The one dimensional case is
  trivial, since any continuous transformation is conformal.}. That is,
any holomorphic transformation is conformal, and to specify a
holomorphic transformation means to know all the coefficients of its
Laurent series, which are infinitely many. This means that 
conformal symmetries in two dimensions are easier to analyze and has a
richer structure than those in larger dimensionalities, and this is
the reason that we will exclusively focus on this case.

We define the real coordinates $(z^0, z^1)$ on the
plane, and introduce complex coordinates,
\[z=z^0+\imi z^1\qquad \bar z=z^0-\imi z^1,\]
with differentiation rules
\[\partial\equiv\partial_z=\frac12(\partial_0-\imi\partial_1)\qquad\bar\partial\equiv\partial_{\bar
  z}=\frac12(\partial_0+\imi\partial_1).\]
Under a holomorphic transformation $z\mapsto f(z)$, a line element $\dd
s^2=\dd z\dd\bar z$ transforms as
\[\dd s^2\mapsto\left(\frac{\partial f}{\partial
    z}\right)\left(\frac{\partial\bar f}{\partial\bar z}\right)\dd
    s^2.\]
This transformation law is generalized under the assumption that
$\Phi(z,\bar z)\dd z^h\dd\bar z^{\bar h}$ is invariant under the
transformation. Here $\Phi(z,\bar z)$ is a field and $h$ and $\bar h$
are real parameters. Then the field transforms as
\begin{equation}
  \Phi(z,\bar z)\mapsto\left(\frac{\partial f}{\partial
      z}\right)^h\left(\frac{\partial\bar f}{\partial\bar
      z}\right)^{\bar h}\Phi\left(f(z),\bar f(\bar z)\right).
\end{equation}
This defines a primary field\index{primary field} of conformal
weight\index{conformal weight}
$(h,\bar h)$. Non-primary fields are known as secondary fields. 

A key ingredient when computing correlators in conformal symmetry is
the operator product expansion (OPE) \index{operator product
  expansion}. This is the assumption that the product of two local
fields $\phi_i(x)$ and $\phi_j(y)$ can be written as a linear
combination of local operators,
\[\phi_i(x)\phi_j(y)=\sum_kC^k_{ij}(x-y)\phi_k(y)\]
where the coefficients $C^k_{ij}$ are c-numbers. Now, one is usually
interested in the singular behavior as $x\to y$, and we use $\sim$ to
indicate that regular terms are discarded. For a primary field
$\varphi$ with conformal dimension $h$, the OPE of this with the
energy momentum tensor is
\[T(z)\varphi(w,\bar w)\sim\frac h{(z-w)^2}\varphi(w,\bar w)+
\frac1{z-w}\partial_w\varphi(w,\bar w)\]
and correspondingly for the anti-holomorphic fields. For the
energy-momentum tensor with itself the OPE is
\index{energy-momentum tensor}
\begin{equation}
  T(z)T(w)\sim\frac{c/2}{(z-w)^4}+\frac{2T(w)}{(z-w)^2}+\frac{\partial
    T(w)}{z-w}. 
  \label{eq:OPE_TT}
\end{equation}
The constant $c$ is called the {\it central charge} \index{central
  charge} of the specific model in question, or sometimes denoted the
conformal anomaly. Apart from this anomaly, the
energy-momentum tensor is simply a conformal field of conformal
dimension $h=2$. Of course, the relation (\ref{eq:OPE_TT}) holds also
for the anti-holomorphic parts of the theory, thereby defining an
independent anti-holomorphic central charge $\bar c$. When the theory
has a Lorentz-invariant and conserved two-point function
$\expt{T_{\mu\nu}(z)T_{\alpha\beta}(-z)}$, one must have $c=\bar c$
\cite{Ginsparg1988}. This holds for all relevant quantum statistical
systems.

Central in conformal field theories is the study of correlation
functions, a 
subject with obvious connections to critical systems. In particular,
the two-point correlation function of two fields $\phi_1$ and $\phi_2$
is nonzero if and 
only if they have the same conformal weights $h_1=h_2=h$ and $\bar
h_1=\bar h_2=\bar h$. Then,
\[\expt{\phi_1(z_1,\bar z_1)\phi_2(z_2,\bar
  z_2)}=\frac{C_{12}}{(z_1-z_2)^{2h}(\bar z_1-\bar z_2)^{2\bar h}}.\]
Notably, the correlation function depends only on the distances
$z_1-z_2$ and $\bar z_1-\bar z_2$, a statement that is true also for
the three- and four-point correlation functions.

The Laurent expansion of the energy-momentum tensor can be written
\begin{equation}
  T(z)=\sum_{n=-\infty}^\infty z^{-z-2}L_n\qquad\text{and}\qquad
  \bar T(\bar z)=\sum_{n=-\infty}^\infty \bar z^{-z-2}\bar L_n,
\end{equation}
and the inverse relations are
\begin{equation}
  L_n=\oint\frac{\dd z}{2\pi\imi}\,z^{n+1}T(z)\qquad\text{and}\qquad
  \bar L_n=\oint\frac{\dd\bar z}{2\pi\imi}\,\bar z^{n+1}\bar T(\bar z).
\end{equation}
Now, one can compute the commutators of $L_n$ by performing the
integrals and expressing the OPE of the energy-momentum tensor, and
the results are
\begin{align*}
  [L_n,L_m]&=(n-m)L_{n+m}+\frac c{12}(n^3-n)\delta_{n,-m}\\
  [\bar L_n,\bar L_m]&=(n-m)\bar L_{n+m}
  +\frac{\bar c}{12}(n^3-n)\delta_{n,-m}\\
  [L_n,\bar L_m]&=0.
\end{align*}
Thus we have two independent algebras on $L_n$ and $\bar L_n$ of
infinite dimension. This is known as the Virasoro algebra\index{Virasoro
  algebra} of the central charge $c$. If $c=0$, the algebra defined on
$L_n$ is the classical algebra of the generators for infinitesimal
conformal transformations in the plane. 

\subsection{The free boson}
The free massless boson in two dimensions has an Euclidean action
\cite{Francesco97}
\begin{equation}
  S[\varphi]=\frac12
  g\int\dd^2x\,\partial_\mu\varphi\partial^\mu\varphi,
\end{equation}
where $g$ is a normalization constant. The two-point correlator of
this field is 
\[\expt{\varphi(x)\varphi(y)}=-\frac1{4\pi
  g}\ln(x-y)^2+\mathrm{const}.\]
When introducing the complex coordinates $z$ and $\bar z$, this
becomes
\[\expt{\varphi(z,\bar z)\varphi(w,\bar w)}=-\frac1{4\pi
  g}\left[\ln(z-w)+\ln(\bar z-\bar w)\right]+\mathrm{const},\]
or when separating the holomorphic and anti-holomorphic parts through
a differentiation, the operator product expansion of the holomorphic
field $\partial\varphi=\partial_z\varphi$ becomes
\begin{align*}
  \expt{\partial_z\varphi(z,\bar z)\partial_w\varphi(w,\bar
    w)}&=-\frac1{4\pi g}\frac1{(z-w)^2}\\
  \partial\varphi(z)\partial\varphi(w)&\sim-\frac1{4\pi
    g}\frac1{(z-w)^2}.
\end{align*}
The bosonic symmetry property of this OPE is immediately obvious. Now,
the energy momentum tensor in complex coordinates is
\begin{equation}
  T(z)=-2\pi g:\partial\varphi(z)\partial\varphi(z):
  \label{eq:Tdef}
\end{equation}
Here $:\cdot:$ denotes normal ordering to ensure that vacuum
expectation values vanish. Wick's theorem allows us to compute the OPE
of the energy-momentum tensor with itself;
\begin{align*}
  T(z)T(w)&=4\pi^2g^2:\partial\varphi(z)\partial\varphi(z)::\partial\varphi(w)\partial\varphi(w):\\
  &=8\pi^2g^2\wick{21}{:\partial<1\varphi(z)\partial<2\varphi(z)::\partial>2\varphi(w)\partial>1\varphi(w):}\\
  &\phantom{=}+16\pi^2g^2\wick{1}{:\partial<1\varphi(z)\partial\varphi(z)::\partial>1\varphi(w)\partial\varphi(w):}\\
  &\sim\frac{1/2}{(z-w)^4}-4\pi
  g\frac{:\partial\varphi(z)\partial\varphi(w):}{(z-w)^2}\\
  &\sim\frac{1/2}{(z-w)^4}+\frac{2T(w)}{(z-w)^2}+\frac{\partial T(w)}{(z-w)},
\end{align*}
the last equation arises by a linear expansion around $w$,
$\partial\varphi(z)=\partial\varphi(w)+(z-w)\partial^2\varphi(w)$.
Thus, the  
OPE has an anomalous term $(1/2)/(z-w)^4$ which fixes the central
charge of the model at $c=1$, and the energy-momentum tensor is not a
primary field.

\subsection{The free fermion}
The free fermion in two dimensions is described by the two-component
spinor $\Psi=(\psi,\bar\psi)$,
and the action is
\begin{equation}
  S(\Psi)=g\int\dd^2x\,(\bar\psi\partial\bar\psi+\psi\bar\partial\psi).
\end{equation}
The propagator of this field becomes
\[\expt{\psi(z,\bar z)\psi(w,\bar w)}=\frac1{2\pi g}\frac1{z-w},\]
and thus the OPE follows trivially
\[\psi(z)\psi(w)\sim\frac1{2\pi g}\frac1{z-w},\]
which of course has fermionic symmetry properties under particle
exchange. The energy-momentum tensor is now
\[T(z)=-\pi g:\psi(z)\partial\psi(z):,\]
and by Wick's theorem we arrive at the OPE for $T(z)$ with itself;
\begin{align*}
  T(z)T(w)&=\pi^2g^2:\psi(z)\partial\psi(z)::\psi(w)\partial\psi(w):\\
  &\sim\frac{1/4}{(z-w)^4}+\frac{\pi g}{2}\frac1{(z-w)^2}
  \left(:\psi(z)\psi(w):-:\psi(z)\partial\psi(w):+\partial\psi(z)\psi(w):\right)\\
  &\phantom{==}-\frac{\pi g}2\frac1{z-w}:\partial\psi(z)\partial\psi(w):\\ 
  &\sim\frac{1/4}{(z-w)^4}+\frac{2T(w)}{(z-w)^2}+\frac{\partial T(w)}{z-w},
\end{align*}
thus proving that the free fermion has a central charge $c=1/2$.

\subsection{Entanglement entropy in a $1+1$ dimensional strip}
After Hawking's discovery of black hole radiation in 1975
\cite{Hawking75}, the study 
of quantum effects and information exchange in black holes became a
topic of intense study, and thus followed investigations on entropy in
conformal field theory that have resurfaced for use in quantum
information theory. Srednicki \cite{Srednicki93} computed numerically
the entropy of both $3+1$ and $1+1$ dimensional bosonic, massless
models and identified the logarithmic divergence of the latter. Later,
Callan and Wilzcek \cite{Callan94} came up with the concept of
geometric entropy on a $1+1$ dimensional conformal strip, which was
pursued by Holzhey, Larsen and Wilczek \cite{Holzhey94}. 
The results have been generalized by Calabrese and Cardy to other
geometries and finite temperature \cite{Calabrese04, Korepin2004}, and
also the time evolution of the entanglement has been investigated
\cite{calabrese-2005-0504}.

\begin{figure}[htb]
  \psfrag{A}{$\tau=-\infty$}  
  \psfrag{B}{$\tau=\infty$}
  \psfrag{z}{$z$}
  \psfrag{x}{$x$}
  \psfrag{l}{$\ell$}
  \psfrag{0}{$0$}
  \psfrag{zeta}{$\zeta$}
  \psfrag{tau}{$\tau$}
  \psfrag{w}{$w$}
  \psfrag{w0}{$0$}
  \psfrag{Rw}{$\real w$}
  \psfrag{Iw}{$\imag w$}
  \psfrag{Sp}{$\mathcal S'$}
  \includegraphics[width=\textwidth]{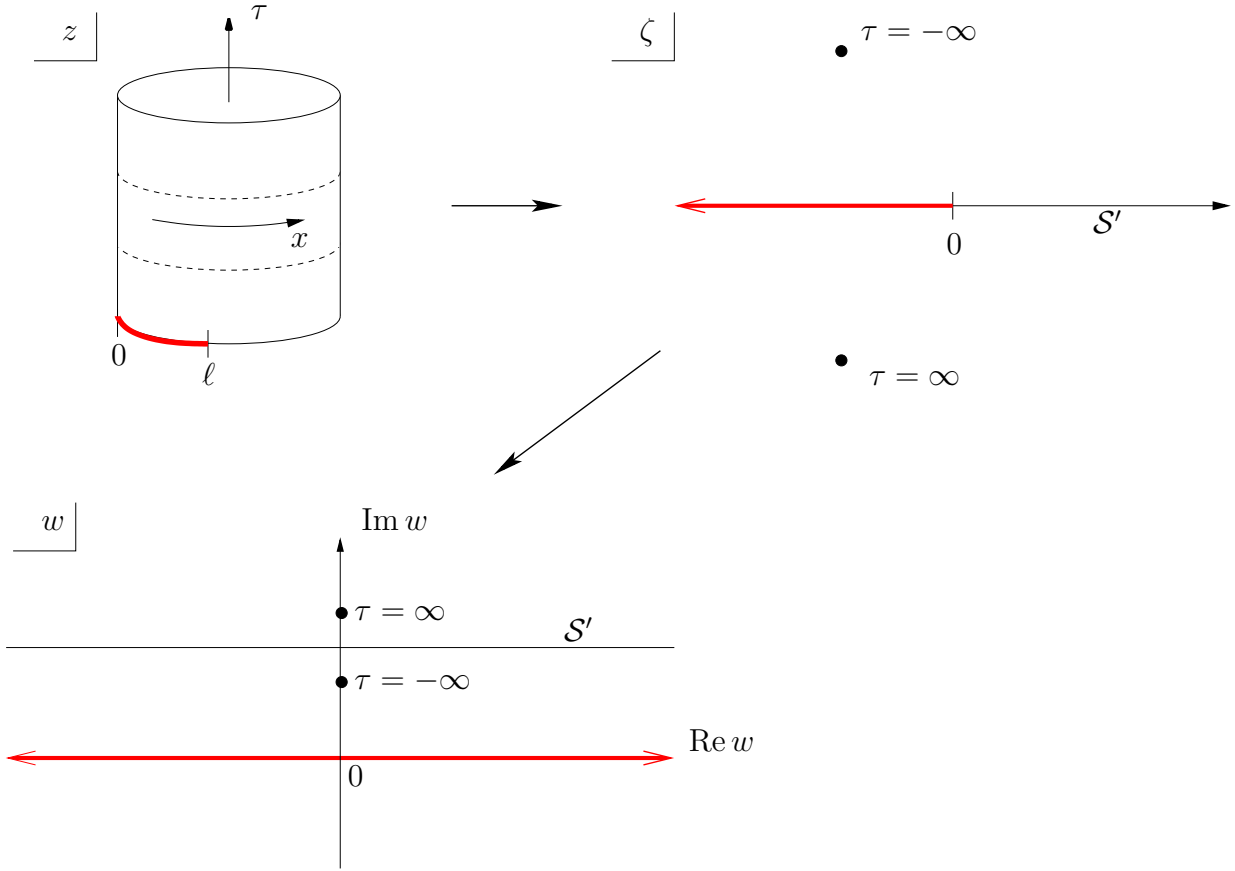}
  \caption[Conformal transformations in $1+1$ dimensions.]%
  {The conformal transformations described in the text, $z\to\zeta$
    defined by (\ref{eq:z_to_zeta}) and $\zeta\to
    w=\frac1\pi\ln\zeta$. The red arc indicates the system, $\mathcal
    S$.} 
  \label{fig:conformal_transf}
\end{figure}

The analysis starts with a series of conformal transformations in the
plane. Assume that the initial coordinate is $z=x+\imi\tau$, where $x$
is the spatial coordinate and $\tau$ is the time. The system is of
length $L$, such that $0<x<L,$ and the subsystem we're interested in
is $0<x<\ell$. We denote the surfaces at $\tau=0$ as $\mathcal S$ for
the subsystem and $\mathcal C$ for the entire system. Now, periodic
boundary 
condition in $x$ make this model an infinitely long cylinder of
circumference $L$, the length of the system. The first conformal
transformation is \cite{Holzhey94}
\begin{equation}
  \zeta=-\frac{\sin\left[\frac\pi L\left(z-\ell\right)\right]}
  {\sin\left(\frac{\pi z}L\right)}.
  \label{eq:z_to_zeta}
\end{equation}
This maps $\mathcal C$ onto the real axis, with $\mathcal S$ as the
negative half-axis and the rest $\mathcal S'=\mathcal C\setminus\mathcal S$
as the positive axis. Moreover, the 
infinite past is a point in the upper half-plane,
$\zeta(\tau=-\infty)=-\expe^{-\imi\pi\ell/L}$, and the infinite
future is at the
point $\zeta(\tau=\infty)=-\expe^{\imi\pi\ell/L}$ in the lower
half-plane. Next, we employ the transformation
$w=\frac1{\pi}\ln\zeta$, which maps $\mathcal S$ onto the real axis
with $x\to0$ corresponding to $w\to\infty$, and $x\to\ell^-$
corresponding to $w\to-\infty$. The remainder of the system is the
line $w=\imi$ with $z\to\ell^+$ becoming $w\to-\infty+\imi$, and
$z\to L$ becoming $w\to\infty+\imi$. The infinite past and future maps
to the points $w(\tau=\pm\infty)=\imi(1\pm\ell/L)$. Both 
transformations $z\to\zeta\to w$ are sketched in Figure
\ref{fig:conformal_transf}. 

Now, we can use the so-called replica trick to make the identification 
\begin{equation}
  S=-\trace\rho\log\rho=-\lim_{n\to 1}\frac{\dd}{\dd n}\trace\rho^n.
\end{equation}
This assumes that integer values of $n$ can be made into an analytical
continuation for non-integer $n$. Thus the problem is reduced into
computing $\rho^n$ on the surface defined by $w$. The functional
integral defining $\rho$ is 
\[\rho(\phi,\phi')=\int\mathcal
D\chi\ket{\phi,\chi}\bra{\phi',\chi},\]
where $\chi$ designates the fields on the upper line, while $\phi$ are
the fields on $\mathcal S$, that is the real axis. This integral is
equivalent to cloning the strips $\mathcal C$, with the intersection
along $\mathcal S'$, with boundary conditions specifying $\phi$ and
$\phi'$. Thus one gets a strip of width $2$ to integrate
along. Using functional integral representation of the wave function
$\ket{\phi,\chi}$ one finds that
\[\trace\rho^n=\frac{\mathcal Z(n)}{\mathcal Z(1)^n},\]
where $\mathcal Z(n)$ is the path integral of the strip $\mathcal C$
cloned $n$ times such that one gets a structure of width $2n$ and
boundary conditions restricting the fields on the extremal
surfaces. From this one finds the entropy (or entanglement entropy) to
be \cite{Holzhey94,Calabrese04} 
\begin{equation}
  S\sim\frac{c+\bar
    c}6\log\left[L\sin\left(\frac{\pi\ell}L\right)\right]
  \label{eq:Holzhey}
\end{equation}
where $\sim$ indicates up to an additive constant independent of
$\ell$ and $L$. The derivation is valid in 
the limit $\ell,L-\ell\gg a$ with $a$ the lattice spacing of the
model. The entropy is symmetric around $\ell=L/2$, at which point the
entropy is largest. From this it is easy to see that
for an (semi-)infinite system, $L\to\infty$, the entropy diverges
logarithmically $S\sim\frac{c+\bar c}6\log\ell$. Also, when
considering systems of different sizes $L$, with a constant fraction
$\ell/L$, one also has a logarithmic divergence, $S\sim\frac{c+\bar
  c}6\log L$.

\subsection{Conformal invariance in QPT}
We expect conformal invariance at quantum critical points
in $1+1$ dimensional models due to the scale invariance that emerges
here. It is 
reasonable to demand nearest-neighbor (or perhaps next
nearest-neighbor) interaction to keep interactions local. Thus it is
possible to identify a central 
charge to the quantum phase transition (QPT), and the central charge
defines the corresponding Virasoro algebra. Hence it is, in principle,
possible to identify the critical exponents of the model according to
the scaling dimensions of the fields. Friedan \etal \cite{Friedan84}
proved that in 
models where the Virasoro algebra has unitary representation%
\footnote{Such models are usually called {\it unitary models}, and are
  those where the representation of the Virasoro algebra contains no
  negative norm states. All such representations with $c>1$ (and
  $h>0$) are unitary.},
\index{unitary models}
and $c\leq 1$, the central charge is restricted to the values
\begin{equation}
  c=1-\frac6{m(m+1)}\qquad m=3,4,\cdots,
  \label{Kacvalues}
\end{equation}
and for each value of $c$ there are $m(m-1)/2$ allowed values of $h$
given by
\[h_{r,s}(m)=\frac{[(m+1)r-ms]^2-1}{4m(m+1)}\qquad 
r=1,\cdots,m-1\quad\text{and}\quad s=1,\cdots,p.\]
Moreover, each such value of $c$ corresponds to a universality class
(or a set of such), and for the first few values the mapping is shown
in Table \ref{tab:c_universalityclass}.\index{universality class}
\begin{table}[htbp]
  \setlength{\extrarowheight}{1ex}
  \caption{The universality classes corresponding to the first allowed
  central charges $c\leq1$. \cite{Friedan84,Ginsparg1988}}
  \centering
  \begin{tabular}{ccl}
    \hline
    $m$ & $c$          & Universality class\\
    \hline
    $3$ & $\frac12$    & Ising, free fermion\\
    $4$ & $\frac7{10}$ & Tricritical Ising\\
    $5$ & $\frac45$    & 3-state Potts\\
    $6$ & $\frac67$    & Tricritical 3-state Potts\\
    \vdots & \vdots & \\
    $\infty$ & $1$     & Free boson\\
    \hline
  \end{tabular}
  \label{tab:c_universalityclass}
\end{table}

It is remarkable that the conformal
invariance at the critical point gives information about
which universality class the model belongs to. There is a deep
connection between the conformal invariance of the underlying
structure and the statistical mechanics of the model, a feature that
we will exploit in later chapters.

\section{Quantum models}
We will in this section present a limited number of quantum models
which exhibits quantum phase transitions and describe some
properties. In general, a spin-$1/2$ quantum chain with periodic
boundary conditions has a Hamiltonian
\begin{equation}
   H = -\sum_{n=1}^{N} \left[\sum_{\alpha=x,y,z} f_\alpha\, \sigma^\alpha_n\, \sigma^\alpha_{n+1}
     +\vkt{g}\cdot\left(\vkt{\sigma}_n \times\vkt{\sigma}_{n+1}\right)
     + \vkt{\lambda}\cdot\vkt{\sigma}_n\right],
   \label{General_spin12}
\end{equation}
where $\vkt f$, $\vkt g$, and $\vkt \lambda$ are real 3-vectors, and
$\vkt\sigma_{N+1}\equiv\vkt\sigma_1$. Further,
$\vkt\sigma_n=(\sigma^x_n,\sigma^y_n,\sigma^z_n)$ are the conventional
Pauli matrices. Fixing the coordinate system,
this leaves us with seven degrees of freedom in the Hamiltonian, which
is a huge parameter space to explore for phase transitions. The models
with various parameters zero, or equal are usually known under a
plethora of names. We will always consider $\vkt g=0$, and 
assume $\vkt\lambda$ to point in the positive $z$-direction,
$\vkt\lambda=(0,0,\lambda)$. Still 
the model can be adjusted by a constant, so we can write down a
general three parameter model, which we will denote the $XYZ$ model;
\index{XYZ@$XYZ$ model}
\begin{equation}
  H_{XYZ}=-\sum_n\left[\sum_{\alpha} f_\alpha\, \sigma^\alpha_n\,
    \sigma^\alpha_{n+1}+\lambda\sigma_n^z\right]\qquad\text{with}\qquad f_x+f_y=1.
\end{equation}
When $f_y=f_z=0$ this is known as the quantum Ising model, and when
$f_y\geq0$ we get the $XY$ model, both of which will be discussed in
detail shortly. In particular the critical $XY$ model with
$f_x=f_y$ and $f_z=\lambda=0$ is sometimes known as the $XX$ model
\cite{Wang01}.\index{XX@$XX$ model} When 
$f_x=f_y=f_z\geq0$ and $\lambda\geq 0$ we have the $XXX$ model or
simply the 1D Heisenberg model \cite{Arnesen01}.
When $\lambda=0$ and $\vkt f>0$ we get what is also sometimes known as the
$XYZ$ 
model, a model that in terms of transfer matrices can be considered
equivalent to the classical eight-vertex 
model \cite{Sutherland70}\index{vertex models}. In the same
way, the case $\lambda=0$, $f_x=f_y\geq0$ and $f_z\geq0$ is known as the
$XXZ$ model, or in terms of the
transfer matrix equivalent to the classical six-vertex model
\index{vertex models} 
\cite{Lieb67,Wang01b}, which in its quantum version has been suggested
to describe the $d$-density wave competing with superconductivity in
high-$T_C$ superconductors 
\cite{Chakravarty02,Syljuasen06}. Curiously,
the model was in its early days considered for explaining
``residual entropy'' in ice at low temperatures \cite{Lieb67}, which
suggests the wide applicability of these models, not restricted to
quantum mechanics. All these models investigated and some solved
exactly, 
which makes them useful as benchmark tools. In particular, the Ising
model is simple and intuitively easy to assess.
The various models are not only mathematical playgrounds -- though
they are useful as such -- but are also efficient models to simulate
real life systems in addition to the examples already
mentioned. E.g. the
vertex models were originally conceived to model a second order
polarization transition at 122 K in potassium dihydrogen phosphate,
$\chem{KH_2PO_4}$ \cite{Slater41}. 

\subsection{The Ising model}
\index{Ising model|(}
The Ising model is perhaps the quantum model which is studied
in greatest detail, and where most is known. Hence it makes for a good
starting point for further studies. We investigate the quantum Ising
model in one spatial dimension, whose Hamiltonian is
\begin{equation}
  H_{\mathrm{Ising}}=-\sum_{n=1}^N\left(\sigma_n^x\sigma_{n+1}^x+\lambda\sigma_n^z\right).
  \label{HIsing}
\end{equation}
The $\sigma$s are the conventional Pauli spin matrices while $\lambda$ is our
external parameter, a magnetic field in the $z$-direction. Na\"{i}vely,
one expects in the limit $\lambda\to\infty$ all spins aligned in the
$z$-direction, which is a product state and thus has no
entanglement. However, in the zero field limit, the ground state would
have to be a Schr\"odinger cat state\index{Schr\"odinger cat state} with a
superposition between the 
spins in positive and negative $x$-direction respectively.

This 1D quantum model can be mapped onto the 2D classical Ising model,
which Onsager famously solved in 1944 \cite{Onsager44}. The classical
model has a continuous phase 
transition at a critical temperature $T=T_c$, which in the quantum
case is mapped onto a critical magnetic field $\lambda_c=1$.

In the $\lambda\to0$ limit we have unit entropy due to the one-qubit
entanglement in the ground state, which is a Schr\"odinger cat state,
while the large $\lambda$ limit  
the product state means that the entropy falls to zero. However, we
see that with larger system sizes there appears what seems to be a
divergence at $\lambda=1$. Even in small systems the entropy here is
larger than elsewhere, but for larger systems the effect is more
pronounced. Note that we need a large system (roughly $N>50$) to say
with any confidence that there is a phase transition at $\lambda=1$,
which indeed is the thermodynamic solution to this system.

\begin{figure}[b]
  \psfrag{X}{$\lambda$}
  \psfrag{Y}{$\Delta E$}
  \centering
  \includegraphics[width=.7\textwidth]{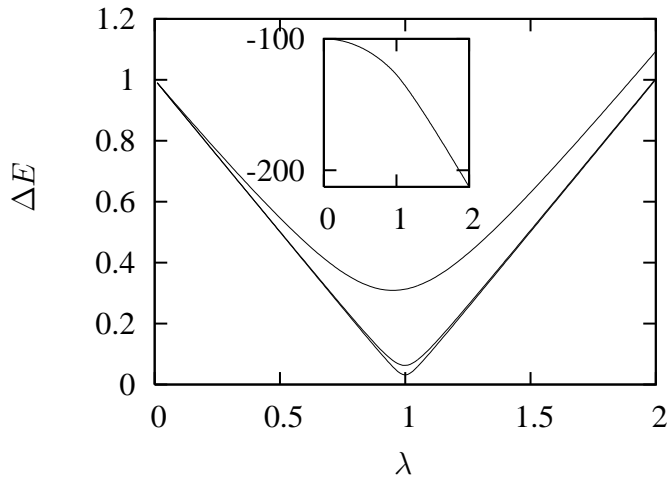}
  \caption[The gap in the Ising model.]%
  {The gap $\Delta E=\min_n\bar\omega_n$ between the ground state and first excited state
    energies 
    in the Ising model with periodic boundary conditions. From top,
    the system sizes 
    shown are $N=10,50,100$. We see clearly that the
    system approaches a gapless state at the critical point
    $\lambda=1$ as the system size increases. The inset shows the
    ground state energy for the $N=100$ system.}
  \label{fig:Gap_Ising}
\end{figure}

In Figure \ref{fig:Gap_Ising} the gap and ground state energy of the
Ising model is shown, and the approach towards a gapless state as
$N\to\infty$ is clearly shown. Note also that the $N=10$ system is far
from gapless, and one should therefore not anticipate much sign of the
phase transition in this system, contrary to what shown in
\cite{SOS1} and Section \ref{ssec:determining_criticality}. Indeed, it
is remarkable that even though a system of size of order 10 is far
from gapless, and shows no critical properties, the
conformal signature can still be very reliably used to identify the
critical parameters. The gap is
here defined as $\Delta E=\min_n\bar\omega_n$, within the ground state
parity segment.
\index{Ising model|)}

\subsection{The $XX$ model}
\index{XX@$XX$ model|(}
The $XX$ model in an external magnetic field is defined by
the Hamiltonian
\begin{equation}
  H_{XX}=-\sum_{n=1}^N\left(\sigma^x_n\sigma^x_{n+1}+\sigma^y_n\sigma^y_{n+1}+\lambda\sigma^z_n\right).
\end{equation}
This model has a continuous $U(1)$ symmetry in the $xy$ plane, which
by the Mermin-Wagner theorem \cite{MerminWagner66}\index{Mermin-Wagner
theorem} cannot be
spontaneously broken at non-zero temperatures in two dimensions in
classical systems. Thus, in quantum models in one dimension at zero
temperature, no spontaneous symmetry breaking can occur. Nevertheless,
polynomial decay of the correlation function may exist due to vortex
excitations such as the Kosterlitz-Thouless transition
\index{Kosterlitz-Thouless transition} in the
classical $XY$ model \cite{KosterlitzThouless73}. The parity
operator\index{parity} 
$\mathcal P=\bigotimes_k\sigma^z_k$ commutes with the Hamiltonian, and thus
its eigenvalue, the parity $\mathcal P=\pm1$ is a good quantum number, and the
solution space is split into two subspaces, as demonstrated in Section
\ref{sec:symmetries}. In this model the generator of rotation,
$\mathcal L=\sum_n\sigma^z_n$ also commutes with the Hamiltonian, and
its eigenvalue $\hat l$ is thus also a good quantum number. In terms 
of the spin flip operators 
$\sigma^\pm=\left(\sigma^x\pm\imi\sigma^y\right)/\sqrt2$, the Hamiltonian
can be rewritten
\[H_{XX}=-\sum_n\left(\sigma^+_n\sigma^-_{n+1}+\sigma^-_n\sigma^+_{n+1}\right)-\lambda\mathcal
L.\]
Obviously, the product states $\ket{N}=\ket1^{\otimes N}$ and
$\ket{-N}=\ket0^{\otimes N}$ (labelled by their $\hat l$ values) are
the exact ground states in the limits $\lambda\to\pm\infty$
respectively. More precisely, applying the Hamiltonian to these states
yield $H_{XX}\ket N=-\lambda N\ket N$ and $H_{XX}\ket{-N}=\lambda
N\ket N$, showing that the state $\ket N$ has lower energy of these
two when $\lambda>0$. However, it is not obvious that these states are
the ground states for all $\lambda$. The parities of the states are
$p(\ket N)=1$ and $p(\ket{-N})=(-1)^N$. 

Consider the simplest ``excitations'' to the state $\ket N$, the $N$
states
\[\ket{N-2,m}=\ket1^{\otimes(m-1)}\otimes\ket0\otimes\ket1^{\otimes(N-m)}\qquad
m=1,\cdots,N.\]
Applying the Hamiltonian to this state gives
\[H_{XX}\ket{2-N,m}=-\left[\ket{N-2,m-1}+\ket{N-2,m+1}+\lambda(N-2)\ket{2-N,m}\right].\]
Thus, it is not an eigenstate of the Hamiltonian, but we can define
a translationally invariant state, or a spin-wave state
\[\cet{N-2,k}=\frac1{\sqrt N}\sum_m\expe^{2\pi\imi km/N}\ket{N-2,m},
\qquad k=1,\cdots,N,\]
which diagonalizes the Hamiltonian. That is,
\begin{equation}
  H_{XX}\cet{N-2,k}=\left[-\lambda N+2\left(\lambda-\cos\frac{2\pi
        k}N\right)\right]\cet{N-2,k}. 
\end{equation}
The excitation energy of the spin-wave state compared to the product
state $\ket N$ is $\Delta E=+2\left(\lambda-\cos\frac{2\pi
    k}N\right)$, which may be negative when $\lambda<1$. Likewise, we
can define the simplest excitation to the state $\ket{-N}$ as
\[\ket{-N+2,m}=\ket0^{\otimes(m-1)}\otimes\ket1\otimes\ket0^{\otimes(N-m)}\qquad
m=1,\cdots,N,\]
and the corresponding spin-wave state $\cet{-N+2,k}$, which
diagonalizes the Hamiltonian and has energy eigenvalues
\begin{equation}
  H_{XX}\cet{-N+2,k}=\left[\lambda N-2\left(\lambda+\cos\frac{2\pi
        k}N\right)\right]\cet{-N+2,k}.
\end{equation}
Here, the excitation energy is $\Delta E=-2\left(\lambda+\cos\frac{2\pi
    k}N\right)$, which can be negative when $\lambda>-1$. Thus, the
ground state is shown to be $\ket{N}$ when $\lambda>1$ and $\ket{-N}$
when $\lambda<-1$, while the ground state in the intermediate region
is undetermined. 
\index{XX@$XX$ model|)}

\section{The $XY$ model}
\label{sec:XY-model}
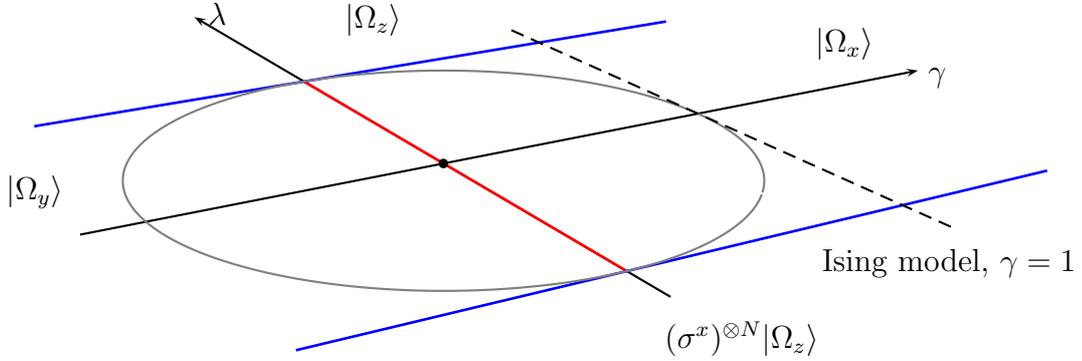
\begin{figure}[htpb]
  \psset{THETA=-120,PHI=20,Dobs=300,Decran=25}
\begin{pspicture}(-7,-3)(5,4)
  \pNodeThreeD(0,0,0){Or}
  \pNodeThreeD(80,-50,0){PM}
  \pNodeThreeD(100,0,0){PZc}
  \pNodeThreeD(80,0,0){PZ}
  \pNodeThreeD(80,50,0){PP}
  \pNodeThreeD(-50,-50,0){MM}
  \pNodeThreeD(-60,0,0){MZc}
  \pNodeThreeD(-50,0,0){MZ}
  \pNodeThreeD(-50,50,0){MP}
  \pNodeThreeD(0,100,0){ZP}
  \pNodeThreeD(0,-60,0){ZM}
  \pNodeThreeD(0,-50,0){G0M}
  \pNodeThreeD(0,50,0){G0P}
  \pNodeThreeD(50,-60,0){Ising1}
  \pNodeThreeD(50,60,0){Ising2}
  \pNodeThreeD(100,-5,0){xlabel}
  \pNodeThreeD(5,100,0){ylabel}
  \pNodeThreeD(40,-70,0){Isinglabel}
  \pNodeThreeD(100,20,0){Oxlabel}
  \pNodeThreeD(-60,20,0){Oylabel}
  \pNodeThreeD(30,80,0){Ozlabel}
  \pNodeThreeD(0,-75,0){nOzlabel}
  \psline{->}(ZM)(ZP)
  \psline{->}(MZc)(PZc)
  \psline[linecolor=red,linewidth=1pt](G0M)(G0P)
  \psline[linecolor=blue,linewidth=1pt](MM)(PM)
  \psline[linecolor=blue,linewidth=1pt](MP)(PP)
  \psset{normaleLatitude=90}
  \CircleThreeD[linecolor=gray](0,0,0){50} 
  \psline[linestyle=dashed](Ising1)(Ising2)
  \psdots(O)
  \rput(xlabel){$\gamma$}
  \rput(ylabel){$\lambda$}
  \rput(Isinglabel){Ising model, $\gamma=1$}
  \rput(Oxlabel){$\ket{\Omega_x}$}
  \rput(Oylabel){$\ket{\Omega_y}$}
  \rput(Ozlabel){$\ket{\Omega_z}$}
  \rput(nOzlabel){$(\sigma^x)^{\otimes N}\ket{\Omega_z}$}
\end{pspicture}
  \caption[Phase diagram of the $XY$ model]{The phase diagram of the
    $XY$ model as defined in (\ref{HXY}). The blue lines indicates the
    phase transition between the
    pure state with spins aligned in the $z$ direction and the
    Schr\"odinger cat states. The red separates the Schr\"odinger cat
    state in $\pm x$ direction from $\pm y$ direction. The circle
    indicates the BM-circle.}
  \label{fig:XY_phasediag}
\end{figure}

\index{XY@$XY$ model|(}
The $XY$ model is a simple generalization of the Ising and $XX$ models,
conventionally written
\begin{equation}
  H_{XY}=-\sum_n\left(\frac{1+\gamma}2\sigma_n^x\sigma_{n+1}^x+\frac{1-\gamma}2\sigma_n^y\sigma_{n+1}^y+\lambda\sigma_n^z\right),
  \label{HXY}
\end{equation}
where $\gamma=1$ is the Ising model and $\gamma=0$ the $XX$ model
already discussed. First properly solved by
Barouch and McCoy \cite{BarouchMccoy71} this is now a thoroughly
investigated model \cite{Ortega05, Osborne:2002} which is useful for
benchmarking due to its 
simplicity and still rich structure. A schematic phase diagram of the
model is presented in Figure \ref{fig:XY_phasediag}.
A simple assessment of the model shows that the ground state in the
large field limit where $\lambda\to\infty$ is the pure state
$\ket{\Omega_z}=\ket{\!\uparrow\,}^{\otimes N}$. In the low field, when
$\gamma>0$, the 
ground state will be the Schr\"odinger cat state with positive and
negative $x$ direction being equivalent,
$\ket{\Omega_x}=\left(\ket{\rightarrow}^{\otimes
    N}+\ket\leftarrow^{\otimes N}\right)/\sqrt2$. Finally, for negative
$\gamma$ the ground state will be the Schr\"odinger cat state with
superpositions in the $y$ direction,
$\ket{\Omega_y}=\left(\ket\nearrow^{\otimes
    N}+\ket\swarrow^{\otimes N}\right)/\sqrt2$. Here vertical arrows indicate
eigenstates of $\sigma^z$, horizontal arrows eigenstates of $\sigma^x$
and diagonal arrows eigenstates of $\sigma^y$. Strictly speaking these
assessments are only valid in the extreme limits, but one can assume
that they are reasonably close to the actual states also in some
region away from these limits. In the thermodynamic limit one can
assume this to be the case arbitrarily close to the phase transition.

Barouch and McCoy also acknowledged the existence of the unit circle
$\gamma^2+\lambda^2=1$ as the division line between what they denoted
the oscillatory region inside the circle and monotonic region
outside due to the behavior of the $xx$ correlation function in the
large distance limit \cite{BarouchMccoy71}. However, this is not
considered a true phase transition, but rather a boundary region. The
three phases of the model have been denoted ordered oscillatory
($\gamma^2+\lambda^2<1$), ordered ferromagnetic
($\gamma^2+\lambda^2>1$ and $|\lambda|<1$) and paramagnetic
($|\lambda|>1$) \cite{Wei05}. At this so-called BM-circle after its
discoverers, the entanglement entropy is always unity \cite{latorre05}.

\subsection{Fermionization}
\index{fermionization}
One of the reasons for the feasibility of the $XY$ model is that it
can be mapped onto a string of spinless fermions, using the technique
of a Jordan-Wigner transform. This reduces the dimensionality of the
problem from the $2^N\times2^N$ matrices of the spin formulation to
$2N\times2N$ sized matrices. The technique can be applied more
generally, but we will demonstrate it on the $XY$ model here.

We apply a Jordan-Wigner transform\index{Jordan-Wigner transform}
that introduces fermionic operators $a_n$ and $a_n^\dag$ 
\begin{align}
    a_n&=\frac12\left(\bigotimes_{k=1}^{n-1}\sigma_k^z\right)\otimes\left(\sigma_n^x+\imi\sigma_n^y\right)\qquad
    a^\dag_n=\frac12\left(\bigotimes_{k=1}^{n-1}\sigma_k^z\right)\otimes\left(\sigma_n^x-\imi\sigma_n^y\right),
\end{align}
which ensures the anti-commutation of these operators,
\[\left\{a_n,a_m^\dag\right\}=\delta_{nm}\qquad\left\{a_n,a_m\right\}=0.\]
The spin operators can be expressed in terms of these fermion
operators as
\begin{align}
  \sigma_n^z&=2a_na_n^\dag-\mathds 1\\
  \begin{bmatrix}
    \sigma_n^x\sigma_{n+1}^x\\
    \sigma_n^x\sigma_{n+1}^y\\
    \sigma_n^y\sigma_{n+1}^x\\
    \sigma_n^y\sigma_{n+1}^y
  \end{bmatrix}&=
  \begin{bmatrix}
    -1&-1&1&1\\
    \imi&-\imi&-\imi&\imi\\
    \imi&\imi&\imi&\imi\\
    1&-1&1&-1
  \end{bmatrix}
  \begin{bmatrix}
    a_na_{n+1}\\a_na_{n+1}^\dag\\a_n^\dag a_{n+1}\\a_n^\dag
    a_{n+1}^\dag
  \end{bmatrix}.
\end{align}
In turn, this means that the Hamiltonian can be written as a sum of
quadratic fermion operators, which in general is
\[H=A_{mn}(a_m^\dag a_n-a_ma_n^\dag)+B_{mn}a_m^\dag
a_n^\dag-B_{mn}^*a_ma_n\]
with $A$ and $B$ being $N\times N$ matrices where $A$ is real and
symmetric. In the special case of the $XY$ model these parameter
matrices become
\begin{align}
  A=\begin{bmatrix}
    \lambda&-\frac12&0&&\\
    -\frac12&\lambda&-\frac12&0&\\
    0&-\frac12&\lambda&-\frac12&\\
    &0&-\frac12&\lambda&\\
    &&&&\ddots&\\
  \end{bmatrix}\qquad
  B=\frac12\gamma\begin{bmatrix}
    0&-1&0&&\\
    1&0&-1&0&\\
    0&1&0&-1&\\
    &0&1&0&\\
    &&&&\ddots&\\
  \end{bmatrix}.
  \label{ABopen}
\end{align}

The above holds for the case of open boundary conditions, where the
sum in (\ref{HXY}) goes from 1 to $N-1$. However, for the case of
periodic boundary conditions, the upper limit of the sum is $N$, and
by definition $\sigma_{N+1}\equiv\sigma_1$. For this case we must
utilize the parity operator $\mathcal P=\bigotimes_{k=1}^N\sigma_k^z$,
\index{parity}
which commutes with the Hamiltonian, and
$a_Na_1=-\sigma_N^+\sigma_1^-\mathcal P$. \footnote{%
  To see this, use direct computation and the fact that
  $\sigma^\pm=\mp\sigma^\pm\sigma^z$.}
Also, for generality one can assume that the chain is
non-homogeneous, such that the $\lambda$ and $\gamma$s are local,
$\{\lambda_n\}$ and $\{\gamma_n\}$. For the case of the $A$ and $B$
matrices above, they now become 
\begin{align}
  A=\begin{bmatrix}
    \lambda_1&-\frac12&0&\cdots&&\frac12\mathcal P\\
    -\frac12&\lambda_2&-\frac12&&&0\\
    0&-\frac12&\lambda_3&&&0\\
    \vdots&&&\ddots&&\vdots\\
    &&&&&-\frac12\\
    \frac12\mathcal P&0&0&\cdots&-\frac12&\lambda_N\\
  \end{bmatrix}\qquad
  B=\frac12\begin{bmatrix}
    0&-\gamma_1&0&\cdots&0&-\gamma_N\mathcal P\\
    \gamma_1&0&-\gamma_2&&&0\\
    0&\gamma_2&0&&&0\\
    \vdots&&&\ddots&&\vdots\\
    0&&&&&-\gamma_{N-1}\\
    \gamma_N\mathcal P&0&0&\cdots&\gamma_{N-1}&0\\    
  \end{bmatrix}.
  \label{ABgeneral}
\end{align}
Replacing homogeneous parameters and a virtual parity $\mathcal P=0$,
we recover Eq. (\ref{ABopen}). Hence, the open boundary conditions can
easily be recovered from the equations for periodic boundary
conditions by setting $\mathcal P=0$.

\begin{figure}[htbp]  
  \centering
  \begin{pspicture}(15,5)
    \psfrag{X}{$\alpha_k$}
    \psfrag{Y}{$\Omega_k$}
    \psfrag{A}{\small \,$\mathcal Q_1$}
    \psfrag{B}{\small \,$\mathcal Q_2$}
    \psfrag{C}{\small \,$\mathcal Q_3$}
    \psfrag{D}{\small \,$\mathcal Q_4$}
    \rput(3.5,2.5){\includegraphics{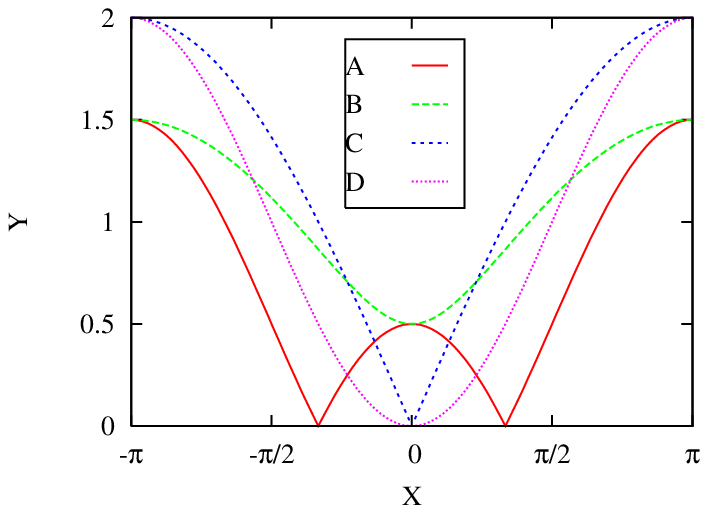}}
    \psline{->}(9,0)(9,5)
    \psline{->}(8,1)(13,1)
    \psline[linecolor=red,linewidth=2pt]{-}(9,0)(9,4)
    \psline[linecolor=blue,linewidth=2pt]{-}(8,4)(13,4)
    \rput*(13.5,1){$\gamma$}
    \rput*(8.5,5){$\lambda$}
    \psline[linestyle=dashed]{-}(12,0)(12,5)
    \rput(9,2.5){\pscircle*{.1}}
    \rput(12,2.5){\pscircle*{.1}}
    \rput(12,4){\pscircle*{.1}}
    \rput(9,4){\pscircle*{.1}}
    \rput(9.5,2.2){$\mathcal Q_1$}
    \rput(12.5,2.2){$\mathcal Q_2$}
    \rput(12.5,4.3){$\mathcal Q_3$}
    \rput(9.5,4.3){$\mathcal Q_4$}
  \end{pspicture}
  \caption[The energy modes for various places in the $XY$ model.]%
  {{\it Left:} The energy modes $\Omega_k$ are shown for the four points in the
  phase space $(\gamma,\lambda)$; 
  $\mathcal Q_1=(0,0.5)$,
  $\mathcal Q_2=(1,0.5)$,
  $\mathcal Q_3=(1,1)$, and
  $\mathcal Q_4=(0,1)$.\\
  {\it Right:} The points shown designated in the phase space of the
  $XY$ model.}
  \label{fig:Diagmodes}
\end{figure}

For the homogeneous, closed chain we thus obtain
\begin{equation}
  H_{XY}=\frac12\sum_{n=1}^N\left[a_na^\dag_{n+1}-a_n^\dag a_{n+1}+
    \gamma\left(a_na_{n+1}-a_n^\dag a_{n+1}^\dag\right)+
    \lambda\left(a_n^\dag a_n-a_na_n^\dag\right)\right]
\end{equation}
where we must interpret $a_Na_{N+1}\equiv-a_Na_1\mathcal P$. Now,
considering the positive parity subspace, $\mathcal P=1$, the boundary
condition in terms of the fermions become anti-periodic, and we can
expand the field in an anti periodic Fourier series of new fermionic
operators $b_k$ and $b^\dag_k$;
\[a_n=\frac1{\sqrt N}\sum_{k=1}^Nb_k\expe^{\pi\imi(2k+1)n/N},\qquad
a_n^\dag=\frac1{\sqrt
  N}\sum_{k=1}^Nb^\dag_k\expe^{-\pi\imi(2k+1)n/N}.\]
In terms of these, the positive parity Hamiltonian $H_{XY}^{(+)}$
becomes
\[H_{XY}^{(+)}=\sum_{k=1}^N\left[\left(\cos\frac{\pi(2k+1)}N-\lambda\right) 
  \left(b_kb_k^\dag-b_k^\dag b_k\right)+\imi\gamma\sin\frac{\pi(2k+1)}N
  \left(b_{-k}b_k-b_k^\dag b_{-k}^\dag\right)\right].\]
We have used the convention that $-k\equiv N-k-1$. Next, we use a
Bogoliubov transformation \index{Bogoliubov transformation}
 to diagonalize the Hamiltonian,
\[\begin{matrise}\imi b_k\\-\imi b_{-k}^\dag\end{matrise}=
\begin{matrise}\cos\phi_k&-\sin\phi_k\\\sin\phi_k&\cos\phi_k\end{matrise}
\begin{matrise}c_k\\c_{-k}^\dag\end{matrise}\]
where the parameter $\phi_k$ is assumed to fulfill $\phi_{-k}=-\phi_k$
and $c_k$,$c_k^\dag$ are new fermion operators. For brevity, we define
$\alpha_k=\frac{\pi(2k+1)}N$, and get
\begin{align*}
  H_{XY}^{(+)}=\sum_k\Big\{&\left[\left(\cos\alpha_k-\lambda\right)\cos2\phi_k-
    \gamma\sin\alpha_k\sin2\phi_k\right]\left(c_kc_k^\dag-c_k^\dag c_k\right)\\
  &+\left[\left(\cos\alpha_k-\lambda\right)\sin2\phi_k+
    \gamma\sin\alpha_k\cos2\phi_k\right]
  \left(c_{-k}c_k+c_k^\dag c_{-k}^\dag\right)\Big\}.
\end{align*}
This becomes diagonal upon choosing
\[\cos2\phi_k=\frac{\lambda-\cos\alpha_k}{\mathcal N_k}\qquad
\text{and}\qquad\sin2\phi_k=\frac{\gamma\sin\alpha_k}{\mathcal N_k},\]
with
\[\mathcal N_k^2=\left(\lambda-\cos\alpha_k\right)^2+\gamma^2\sin^2\alpha_k.\]
Note that the condition $\phi_{-k}=-\phi_k$ is fulfilled. This finally
means that the Hamiltonian becomes
\begin{equation}
  H_{XY}^{(+)}=\sum_{k=1}^N\underbrace{\sqrt{\left(\lambda-\cos\alpha_k\right)^2+\gamma^2\sin^2\alpha_k}}_{\Omega_k}
  \left(2c_k^\dag c_k-1\right).
  \label{eq:H_XY_pos}
\end{equation}
The same procedure can be applied to the negative parity subspace,
only now with periodic boundary conditions and periodic Fourier
series,
\[a_n=\frac1{\sqrt N}\sum_{k=1}^N\hat b_k\expe^{2\pi\imi kn/N},\]
and we arrive at (\ref{eq:H_XY_pos}) with $\alpha_k=2\pi k/N$ and
corresponding new fermion operators.
In Figure \ref{fig:Diagmodes} the energy eigenvalues is plotted with
$\alpha_k$, for various parameters, and we see clearly that the
critical points correspond to zeros in the energy gap.

\begin{figure}[htbp]
  \psfrag{X}{$\gamma$}
  \psfrag{Y}{$\lambda$}
  \centering
  \begin{pspicture}(15,8)
    \rput(7.5,4){\includegraphics[width=.6\textwidth]{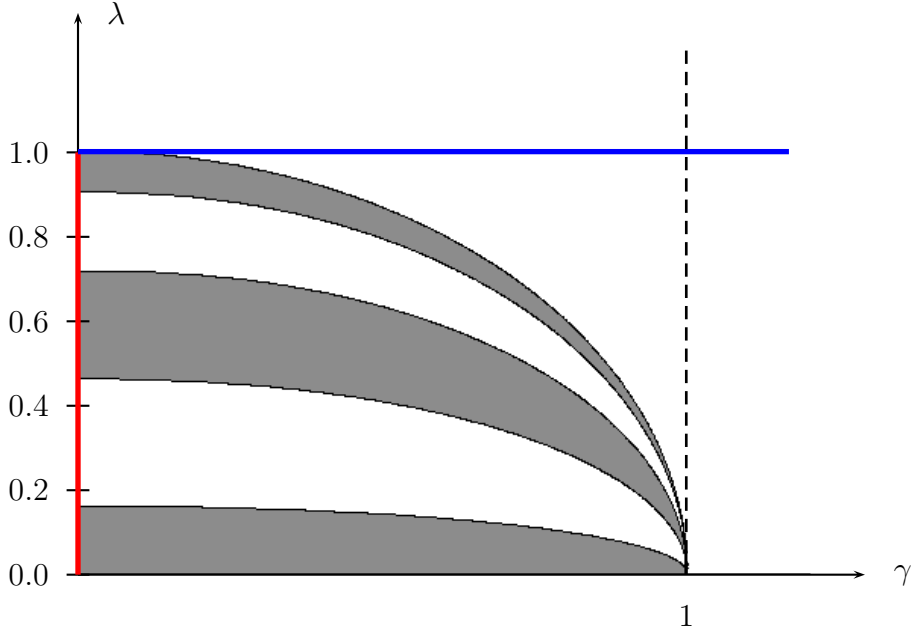}}
    \rput*(13.5,.55){$\gamma$}
    \rput*(3.15,8){$\lambda$}
    \rput*(10.65,0){$1$}
    \multido{\n=.55+1.12,\nl=0.0+0.2}{6}{
      \psline{-}(2.5,\n)(2.8,\n)
      \rput*(2,\n){\nl}}
    \psline{->}(2.65,0.55)(2.65,8)
    \psline{->}(2.65,.55)(13,.55)
    \psline[linewidth=2pt,linecolor=red]{-}(2.65,.55)(2.65,6.15)
    \psline[linewidth=2pt,linecolor=blue]{-}(2.65,6.16)(12,6.16)
    \psline[linewidth=1pt,linestyle=dashed]{-}(10.65,.55)(10.65,7.5)
  \end{pspicture}
  \caption[The parity of the ground state in the $XY$ model.]%
  {The areas in the first quadrant of the parameter space of the $XY$
    model where the ground 
    state has negative parity are shaded for an $N=10$ system. The
    outermost line coincides precisely with the BM circle
    $\gamma^2+\lambda^2=1$. The critical lines and the Ising line
    (dashed) is drawn for reference.} 
  \label{fig:GSparity}
\end{figure}

The parity of the ground state is important when investigating small
systems --- for large systems the difference is minute --- in order to
know which subspace to look for the ground state. This is not a
trivial question, and we resort to direct computation of the two
subspaces, and check which yields the lowest ground state energy. 
Brute computation shows that when $\gamma^2+\lambda^2>1$, that is
outside the BM-circle, the ground state parity is always positive%
\footnote{When $N$ is odd, it will be negative outside this circle for
negative $\lambda$. We consider only even $N$ in the following.}. 
Inside the BM-circle, the nature of the ground state is oscillatory,
and the parity of the ground state in a $N=10$ system is shown in
Figure \ref{fig:GSparity}. The BM-circle limits the outer region of
the circle, and the oscillatory nature of the ground state inside the
circle is shown. When the system size increases the difference between
the ground state in the different parity subspaces become
negligible.
\index{parity}

\subsection{Diagonalization in terms of Majorana fermions}
Having established the fermionic shape of the Hamiltonian in question,
which in this case is the $XY$ model, and which take on the form of
the two matrices $A$ and $B$, we must diagonalize the Hamiltonian. To
this end, define the $2N$ Majorana fermions in terms of the operators
\index{Majorana fermions}
\begin{align}
  \check\gamma_{2i-1}=\frac1{\imi\sqrt2}(a_i-a_i^\dag)\qquad\check\gamma_{2i}=\frac1{\sqrt2}(a_i+a_i^\dag).
  \label{eq:Majoranadef}
\end{align}
These operators satisfy the Majorana anti-commutation relations
$\{\check\gamma_i,\check\gamma_j\}=\delta_{ij}$. Moreover, in terms of
these operators the Hamiltonian becomes
\[H=\mathcal C_{ij}\check\gamma_i\check\gamma_j,\]
where the Hermitian matrix $\mathcal C$ is
\[\mathcal C=\imi\begin{bmatrix}
  \Xi_{11}&\Xi_{12}&\cdots\\
  \Xi_{21}&\Xi_{22}&\\
  \vdots& &\ddots
\end{bmatrix}\qquad
\Xi_{mn}\equiv\begin{bmatrix}
  -\imag B_{mn} & -(A_{mn}+\real B_{mn})\\
  A_{mn}-\real B_{mn} & \imag B_{mn}
\end{bmatrix}.\]
$\mathcal C$ being imaginary and antisymmetric means that its
eigenvalues are real and come in positive/negative pairs, which we
denote $\pm\bar\omega_n$. The eigenvectors corresponding to such a
pair are complex conjugates. In turn we now diagonalize $\mathcal C^2$
with the orthogonal and real matrix $O$,
\[O\transpose\mathcal
C^2O=\bigoplus_{n=1}^N\begin{bmatrix}\bar\omega_n^2&0\\0&\bar\omega_n^2\end{bmatrix}.\]
If we are careful in permuting the columns of $O$, we can 
block diagonalize $\mathcal C$ such that
\[O\transpose\mathcal
CO=\bigoplus_{n=1}^N\begin{bmatrix}0&-\imi\bar\omega_n\\\imi\bar\omega_n&0\end{bmatrix}.\]
Now we can return to fermions again, defining the $N$ fermionic operators
\begin{align}
  \hat a_n&=\frac1{\sqrt2}\sum_{i=1}^{2N}\check\gamma_i(O_{i,2n-1}+\imi
  O_{i,2n})\\
  \hat a_n^\dag&=\frac1{\sqrt2}\sum_{i=1}^{2N}\check\gamma_i(O_{i,2n-1}-\imi
  O_{i,2n}).
\end{align}
These diagonalize the Hamiltonian, such that
\begin{equation}
  H=\sum_{n=1}^N\bar\omega_n\left(\hat a_n^\dag\hat a_n-\hat a_n\hat
    a_n^\dag\right).
  \label{HXY_diagonal}
\end{equation}
Hence it is clear that the eigenstates of the Hamiltonian will be
states with a defined number of $\hat a$-fermions. How this maps back
to the real space fermions --- never mind the spins --- is far from
trivial though. Nevertheless, it is clear that the ground state is the
one with zero $\hat a$ fermions. Any state in this basis we denote
$\ket\eta=\ket{n_1,n_2,\cdots,n_N}$ where $\eta$ is the set of
fermionic occupation numbers and $n_k=0,1$.

This shows how the orthogonal matrix $O$ plays a crucial role in
diagonalizing the Hamiltonian, and that the new Hamiltonian is
diagonal in terms of quasi-fermions related to the original fermions
through the Bogoliubov transformation
\index{Bogoliubov transformation}
\begin{align}
  \begin{split}
    \hat a_m=\Sigma_{mn}a_n+\Delta_{mn}a_n^\dag&\qquad
    \hat a_m^\dag=\Delta_{mn}^*a_n+\Sigma_{mn}^*a_m^\dag
    \label{Bogoliubov}
  \end{split}
\end{align}
where the transformation matrices are
\begin{align*}
  \Sigma_{mn}&=\frac12\left[O_{2n,2m-1}+O_{2n-1,2m}+\imi\left(O_{2n,2m}-O_{2n-1,2m-1}\right)\right]\\
  \Delta_{mn}&=\frac12\left[O_{2n,2m-1}-O_{2n-1,2m}+\imi\left(O_{2n,2m}+O_{2n-1,2m-1}\right)\right].
\end{align*}
Preservation of commutation relations and unitarity of the transformation
ensure the following properties of the
matrices $\matr\Sigma=\{\Sigma_{mn}\}$ and $\matr\Delta=\{\Delta_{mn}\}$,
\begin{subequations}
  \label{unitarityconds}
  \begin{align}
    \label{UC:a}
    \matr\Sigma\matr\Sigma^\dag+\matr\Delta\matr\Delta^\dag&=\matr 1\\
    \label{UC:b}
    \matr\Sigma^\dag\matr\Sigma+\matr\Delta\transpose\matr\Delta^*&=\matr 1\\
    \label{UC:c}
    \matr\Sigma\matr\Delta\transpose+\matr\Delta\matr\Sigma\transpose&=\matr 0\\
    \label{UC:d}
    \matr\Sigma^\dag\matr\Delta+\matr\Delta\transpose\matr\Sigma^*&=\matr 0
  \end{align}
\end{subequations}

Hence, when transforming from real fermions to the quasi-fermions, we
apply the columns of $O$, where for each particle labeled $n$, two
columns of $O$ contribute, and every second term in this column
contributes at a time. There is as we see a complex structure of the
orthogonal matrix that creates the quasi-fermions. In general, the
ground state of the Hamiltonian (\ref{HXY_diagonal}) will be
$\ket{000\cdots0}$ in terms of the quasi-fermions, while it is not
obvious how this can be mapped back to the real
fermions. Nevertheless, this provides us with an easy manner in which
we can find the properties of the ground state, such as the entropy.

The following argument ensures that we are able to diagonalize the
basis we are working in, based on a matrix notation of
(\ref{Bogoliubov}),
\[\begin{bmatrix}\hat{\matr a}\\\hat{\matr a}^\dag\end{bmatrix}
=\begin{bmatrix}\matr{\Sigma} & \matr\Delta\\\matr\Delta^*&\matr\Sigma^*\end{bmatrix}
\begin{bmatrix}\matr a\\\matr a^\dag\end{bmatrix}.\]
Here, $\matr a$ is a column vector consisting of the $a_n$'s, while
$\matr a^\dag$ is a column vector consisting of $a_n^\dag$, not
the adjoint of $\matr a$. We can diagonalize the matrices
$\matr\Sigma\matr\Sigma^\dag$ and  $\matr\Delta\matr\Delta^\dag$
simultaneously, since they commute by equation (\ref{UC:a}). Hence, we
can write 
\[\matr\Sigma\matr\Sigma^\dag=\matr U\matr D_\Sigma^2\matr U^\dag\qquad\text{and}\qquad
\matr\Delta\matr\Delta^\dag=\matr U\matr D_\Delta^2\matr U^\dag,\]
$\matr U$ being unitary and $\matr D_{\Sigma|\Delta}^2$ are
diagonal. Precisely likewise, we can do the same for the matrices
involved in (\ref{UC:b}),
\[\matr\Sigma^\dag\matr\Sigma=\matr V\matr D_\Sigma^2\matr V^\dag\qquad\text{and}\qquad
\matr\Delta\transpose\matr\Delta^*=\left(\matr\Delta^\dag\matr\Delta\right)^*=\matr V\matr D_\Delta^2\matr
V^\dag,\]
$\matr V$ being another unitary matrix. The eigenvalues are however
the same. Now, in order to preserve the two latter conditions in
\ref{unitarityconds} we must have
\[\left\{\matr D_\Delta,\matr D_\Sigma\right\}=0.\]
This can be accomplished since we have double degeneracy in both
matrices by making these matrices $2\times2$ block 
diagonal with two different Pauli matrices along the diagonal,
\[\matr{D}_\Sigma=\bigoplus_{k=1}^{N/2}\cos\vartheta_k\,\sigma^z\qquad\text{and}\qquad 
\matr{D}_\Delta=\bigoplus_{k=1}^{N/2}\sin\vartheta_k\,\expe^{\imi\chi_k}\,\sigma^y.\]
These will be diagonal when squared, and thus comply with the
conditions above.
We have assumed that the two degenerate terms are coincidental,
if not the matrices can be rearranged to conform with this.
If $N$ is odd, one will have to add a single term along the diagonal
consisting of a 0 (1) for the $\matr D_\Sigma$ ($\matr
D_\Delta$). This procedure will ensure that the Bogoliubov
transformation can be written
\begin{equation}
  \begin{bmatrix}\hat b_{2n-1}\\\hat b^\dag_{2n}\end{bmatrix}=
  \begin{bmatrix}\cos\vartheta_n &
    -\imi\sin\vartheta_n\expe^{\imi\chi_n}\\
    -\imi\sin\vartheta_n\expe^{-\imi\chi_n} & -\cos\vartheta_n
  \end{bmatrix}
  \begin{bmatrix}b_{2n-1}\\b^\dag_{2n}\end{bmatrix}  
  \label{Bogoliubov_b}
\end{equation}
where the operators are unitarily transformed, 
\begin{equation}
  \hat b_n=\sum_k(U^{-1})_{nk}\hat a_k\qquad b_n=\sum_kV_{nk}a_k.
  \label{Unitary_b}
\end{equation}
Hence we have shown that the diagonalization can be accomplished
through the unitary transformation (\ref{Unitary_b}) and Bogoliubov
transformation (\ref{Bogoliubov_b}).

\subsection{Entropy in the $XY$ model}
Having defined the $XY$ model and how to diagonalize it, we now proceed
to find the entropy of the ground state as described in section
\ref{sec:EntanglementMeasures}. As mentioned, the ground state is
$\ket{00\cdots0}$ in terms of the delocalized quasi-fermions. We
define the 
correlation matrix of the Majorana fermions
\begin{equation}
  \Gamma_{ij}=\expt{\check\gamma_i\check\gamma_j-\check\gamma_j\check\gamma_i}.
  \label{Gamma_def}
\end{equation}
Note that the Majorana fermions have a two-to-one correspondence to
the real fermions, so tracing out a real fermion would amount to
tracing out two neighbouring Majorana fermions from this matrix. In
the $\hat a$-basis the correlator becomes
\[\Gamma_{ij}=\frac\imi2\sum_{k=1}^N\left(O_{i,2k}O_{j,2k-1}-O_{i,2k-1}O_{j,2k}\right).\]

\begin{figure}[htb]
  \centering
  \includegraphics[width=.9\textwidth]{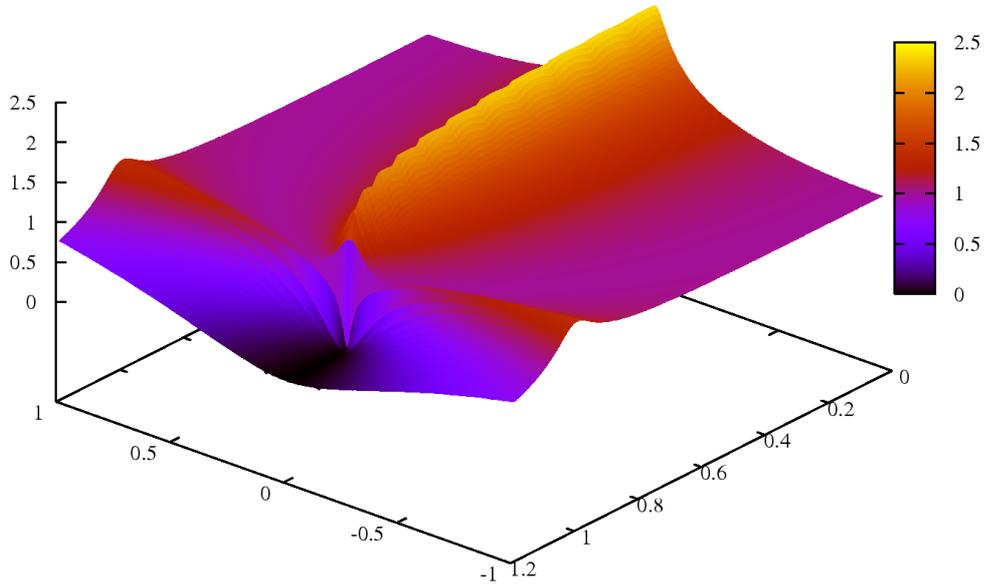}
  \caption[Entropy of the $XY$ model.]%
  {Entropy of the $XY$ model when half of the spins are traced
    out. The critical lines $\gamma=0$ and $\lambda=1$ are seen as
    areas of increased entanglement. Here $N=50$.} 
  \label{fig:XY_entropy}
\end{figure}

So far we have considered the entire system, which is in a pure state,
and the entropy is obviously zero unless we trace out some of the
system leaving us with $N'\leq N$ particles in the
subsystem. Formally, this amounts to tracing out the two columns of
$\Gamma$ corresponding 
to these particles. Now, the density matrix is assumed diagonal in some
quasi-fermion basis denoted $\hat a$. Note that this basis is
different depending on the size (or shape) of the system traced out,
but when nothing is traced out, it coincides with the basis that
diagonalizes the Hamiltonian. Assume that in this basis the density
operator can be written
\begin{equation}
  \rho=\sum_\eta\Lambda_\eta\ket\eta\bra\eta\qquad\Lambda_\eta=\prod_{k=1}^{N'}\left[(1-n_k)\lambda_k+n_k(1-\lambda_k)\right]
\end{equation}
where $\eta$ denotes the set of fermionic occupation numbers
$\{n_1,n_2,\cdots,n_{N'}\}$ of the quasi-fermions, and $\lambda_k\in[0,1]$ are
real coefficients. The sum is over all possible occupation
numbers, $\sum_\eta=\sum_{n_1=0}^1\cdots\sum_{N_{N'}=0}^1$. Moreover,
the ground state $\rho=\ket0\bra0$ is identified by all
$\lambda_k=1$. Computing the 
expectation value matrix (\ref{Gamma_def}) in the $\hat a$ basis we
find that the matrix can be written
\[\Gamma_{ij}=\sum_\eta\Lambda_\eta\bra\eta[\check\gamma_i,\check\gamma_j]\ket\eta=\imi\sum_{k=1}^{N'}\left(O_{i,2k}O_{j,2k-1}-O_{i,2k-1}O_{j,2k}\right)\left(\lambda_k-\frac12\right),\]
and in turn this means that the transformation $O$ block-diagonalizes
this matrix;
\[O\transpose\Gamma
O=\imi\bigoplus_{k=1}^{N'}\begin{bmatrix}0&-\lambda_k+\frac12\\\lambda_k-\frac12&0\end{bmatrix}.\]
This shows that the eigenvalues of $\Gamma$ are $\pm\left(\lambda_k-\frac12\right)$. The
idea of this exercise is that given the diagonalization of the
Hamiltonian, the matrix $\Gamma$ is known, and the eigenvalues can be
numerically extracted efficiently even after tracing out some
particles. Knowing the $\lambda_k$'s means 
that we know the density matrix, and can compute the entropy.
The entropy becomes
\begin{align}
  S&=-\sum_\eta\Lambda_\eta\log_2\Lambda_\eta
  =\sum_{k=1}^{N'}H\left(\lambda_k\right)
\end{align}
where $H(x)$ is the binary entropy\index{binary entropy}
(\ref{eq:binentropy}). 
This gives us an efficient way to compute the entropy of any block of
real spins through the transformation into quasi-fermions. Also the
formulas here gives an easy relation between the {\it classical}
entropy of the eigenvalues of the density matrix in the reduced
$\hat a$ basis and the entropy of the state itself. Figure
\ref{fig:XY_entropy} shows the entropy of the $XY$ model when half-size
is traced out. The entropy clearly has maxima along critical lines,
while the actual height of the maximum is larger along the
$\gamma=0$-line than the $\lambda=1$-line due to the larger central
charge of this transition.

\subsection{Other correlators}
\begin{figure}[htbp]
  \psfrag{X}{$\lambda$}
  \psfrag{Y}{$\xi$}
  \centering\includegraphics[width=.6\textwidth]{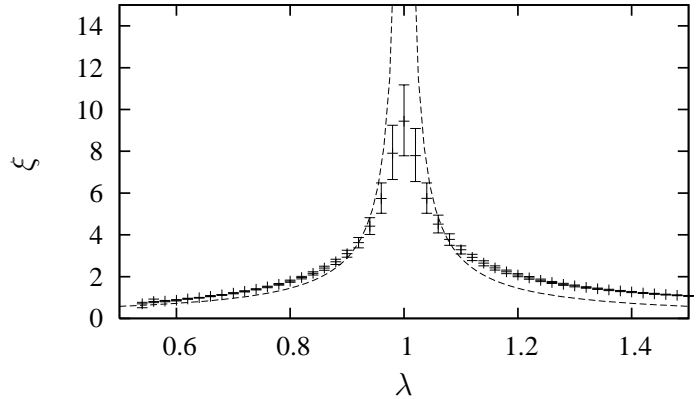}
  \caption[Correlation length in the Ising model]%
  {An estimate of the correlation
    length in the Ising model with magnetic field. The phase
    transition at $\lambda=1$ is clearly shown as a divergence in the
    correlation length. The error bars indicate the variation of the
    estimate of the correlation length, thus small error bars indicate
    a good approximation to exponential decay, while the large error
    bars close to the critical point reflect the polynomial decay
    here. The dashed line is the thermodynamical
    approximation $\xi\sim|\lambda-\lambda_c|^{-1}$. Data for $N=100$.}
  \label{fig:Ising_correlation}
\end{figure}

When we have established the matrices $\matr\Sigma$ and $\matr\Delta$
in the Bogoliubov transformation (\ref{Bogoliubov}), we can compute
expectation values and correlators of the model in
question. Specifically, the second moment of annihilation and creation
operators is
\[\bra\eta
a_na_n^\dag\ket\eta=\sum_{k=1}^{N'}\left[n_k|\Delta_{kn}|^2+(1-n_k)|\Sigma_{kn}|^2\right],\]
and hence follows e.g. the spin expectation value in the $z$-direction;
\[\expt{\sigma^z_n}=\sum_\eta\Lambda_\eta\left(2\bra\eta
  a_na_n^\dag\ket\eta-1\right)=\sum_{k=1}^{N'}\left[\lambda_k|\Sigma_{kn}|^2+(1-\lambda_k)|\Delta_{kn}|^2\right].\]
Some more effort gives
the fourth momenta, which enables us to compute the spin-spin
correlation function,
\[s(k,l)\equiv\expt{\sigma_k^z\sigma_l^z}-\expt{\sigma_k^z}\expt{\sigma_l^z}.\]
This correlation will, when the distance $|k-l|$ is
sufficiently large, fall off exponentially, and the correlation
length\index{correlation length} $\xi$ is the cutoff of this
exponential, $s(k,l)\sim\expe^{-|k-l|/\xi}$.
Hence we find the correlation
length in the Ising model as shown in figure
\ref{fig:Ising_correlation}.
The correlation function follows the expected path nicely, though
there obviously is no true divergence since the system size is
finite.

The framework provided so far gives a good overview of the methods
that can be utilized to find specific information about a large class
of models. Nevertheless, there is an even larger class of models that
do not conform to this framework, and where different approaches to
finding the properties of the model has to be found. 
\index{XY@$XY$ model|)}

\subsection{Determining criticality}
\label{ssec:determining_criticality}
In \cite{SOS1} we describe how to use the entanglement entropy and
conformal field theory to determine the critical surfaces of a
model. To this end, 
define the entropy of a block of $\ell$ spins in a system of size $N$
and inspired by the result (\ref{eq:Holzhey}) define the critical
signature of the entropy, \index{critical signature}
\begin{equation}
  s_\ell(c)=\frac c3\log_2\sin\left(\frac{\pi\ell}N\right).
  \label{eq:s_ell}
\end{equation}
This signature should fit to the actual numerical values, subtracted
the value at $\ell=N/2$ since $s_{1/2}(c)=0$.
We define the error between the entropy data
for a given model, $\hat s_\ell$, and the critical signature for a
given central charge as
\begin{equation}
  \varepsilon_c=\frac1M\sum_\ell\left(\hat s_\ell-s_\ell\right)^2.
  \label{eq:errfunct}
\end{equation}
Thus, the natural requirement is that this error is minimal. It is
obvious that the error would be infinite if we included the endpoints
in the error, since the critical signature diverges when $\ell=0$ or
$\ell=N$, while the actual entropy would be zero in those cases (since
the system is empty).

 Moreover, we define an estimated central charge
$c_{\mathrm{est}}$ as the central charge that minimizes this error,
\begin{equation}
  c_{\mathrm{est}}=\argmin_c\left\{\frac1M\sum_\ell\left(\hat
      s_\ell-s_\ell\right)^2\right\} 
\end{equation} 
where $\hat s_\ell$ are the measured entanglement entropies. $M$ is
the number of lattice sites that are within the chosen cutoff. A
convenient choice is to sum over values of $\ell$ satisfying
$0.2N<\ell<0.8N$, and $M$ follows subsequently. 

\begin{figure}[htbp]
  \begin{pspicture}(15,11)
    \psfrag{X}{$\ell$}
    \psfrag{Y}{$\hat s_\ell$}
    \rput*(4,8.5){\includegraphics{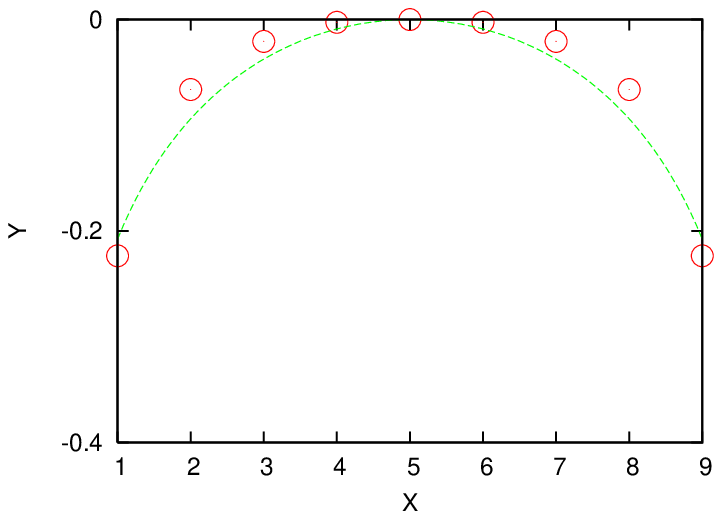}}
    \rput*(12,8.5){\includegraphics{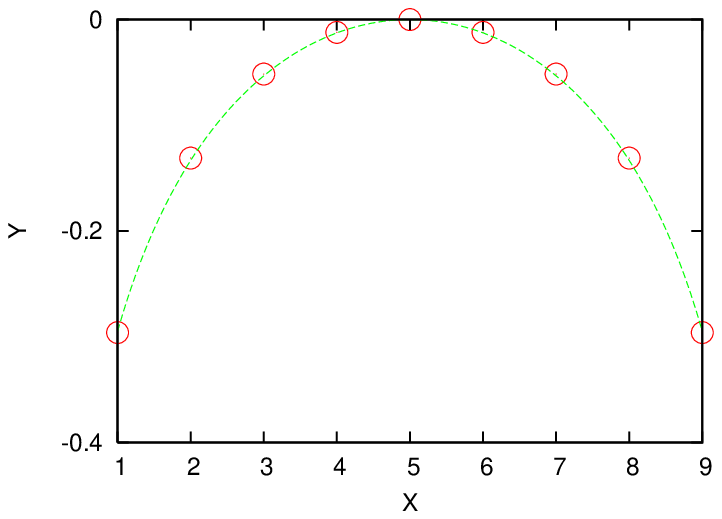}}
    \psfrag{X}{$\lambda$}
    \psfrag{Y}{$c_{\mathrm{est}}$}
    \rput*(4,3){\includegraphics{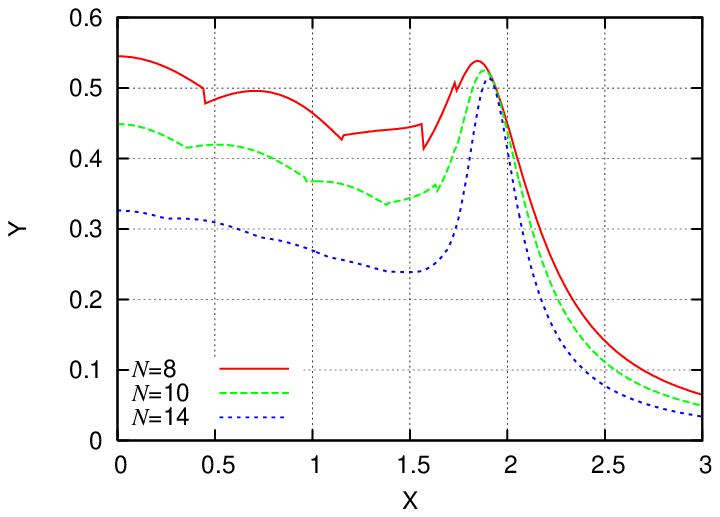}}
    \psfrag{Y}{$\log\varepsilon_c$}
    \rput*(12,3){\includegraphics{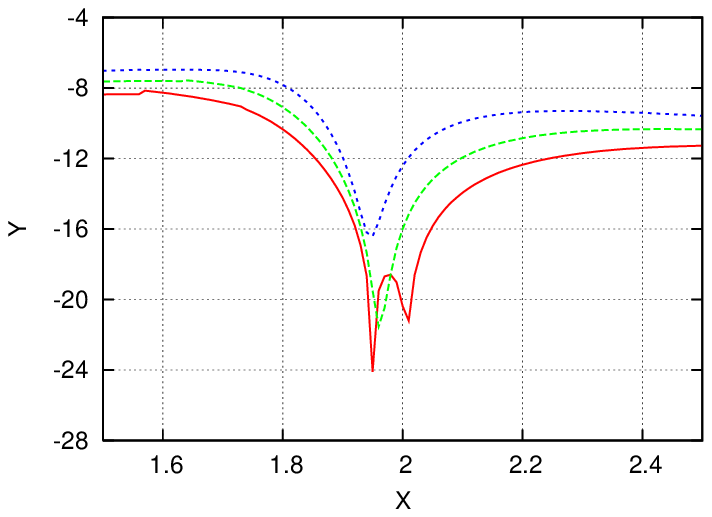}}
  \end{pspicture}
  \caption[Illustration of the technique to find a critical line.]%
  {Illustration of the technique to find a critical line as described
    in the text. We traverse the parameter line in the $XYZ$ model
    with $\Delta=-1/2$, $\gamma=1$ and $\lambda>0$. Top are
    $\hat s_\ell$ and the critical entropy function (\ref{eq:s_ell})
    for off-critical ($\lambda=1$, left  with
    $c_{\mathrm{est}}\approx0.368$) and near-critical 
    ($\lambda=1.9$, right with $c_{\mathrm{est}}\approx0.512$) 
    systems for $N=10$. At bottom the estimated central charge and the
    error function (\ref{eq:errfunct})
    are shown when crossing the phase
    transition. We see the clear identification of the phase
    transition at $\lambda_{\mathrm c}=1.9\pm0.05$. The error
    function's minimum is at a slightly higher $\lambda$ than the
    estimated central charge's maximum and the latter overshoots the
    assumed true value $c=0.5$ slightly. All this is in accordance
    with the benchmarking done in \cite{SOS1}, and one may conclude
    that the critical point is at $\lambda\approx1.9$ with the Ising
    central charge $c=1/2$.}
  \label{fig:detcrit}
\end{figure}

It turns out that the error is very small also in near-critical
systems, and a minimum in the error when adjusting parameters across a
critical line not always corresponds very good to the actual
phase transition. However, the if one looks at $c_{\mathrm{est}}$ when
traversing a critical region, this almost always has a maximum at the
critical point, except perhaps at confluence points%
\index{confluence point}%
\footnote{There are less well-defined critical regions,
  such as near the point $\gamma=0$, $\lambda=1$ in the $XY$ model
  where two critical lines with different central charges
  intersect. Here the central charge is ill-defined, and the method
  inevitably fails. We denote these points in the parameter space
  confluence points.}. 
Using the parameter value of that maximum and comparing to the
possible values of the central charge defined in (\ref{Kacvalues}),
one can very often determine 
the central charge and critical parameter values with great
certainty. The technique is illustrated for a critical line in the
$XYZ$ model in Figure \ref{fig:detcrit}.

\section{Criticality in the $XYZ$ model}
We extend the $XY$ model into what we call the $XYZ$ model, with
Hamiltonian
\index{XYZ@$XYZ$ model}
\begin{equation}
  H_{XYZ}=H_{XY}-\Delta\sum_n\sigma_n^z\sigma_{n+1}^z.
\end{equation}
Despite the simple extension, this model cannot be fermionized in the
way done with the $XY$ model. Nevertheless, we can still find the
critical regions and the 
corresponding central charges as described in Section
\ref{ssec:determining_criticality} and Refs. \cite{SOS1,
  SOSposter}. When scanning a large portion of the parameter space,
using $N=10$ (with symmetries) is useful, since the computation is
rather reliable and the time consumption is reasonable. Near
confluence points one can use larger systems for small areas of the
space. All data in the next section are taken with $N=10$ to
demonstrate the power of the method using such a small system.

\subsection{Critical surfaces}
\begin{figure}[b]
  \psfrag{X}{$\lambda$}
  \psfrag{Y}{$\Delta$}
  \centering\includegraphics[width=.7\textwidth]{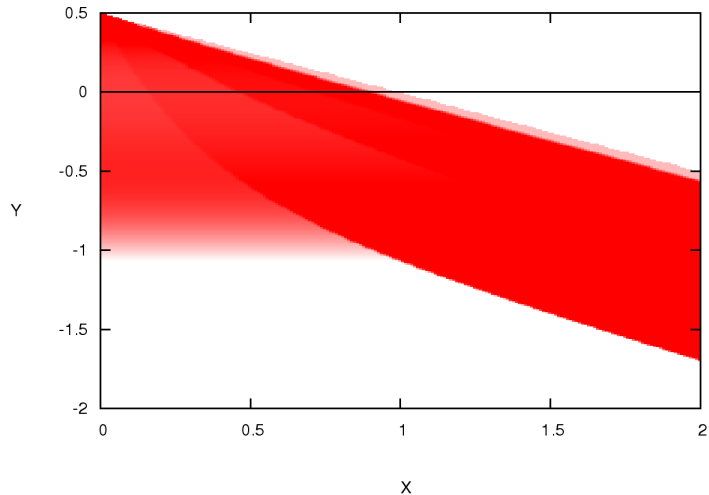}
  \caption[The critical region of the $XYZ$ model with $\gamma=0$.]%
  {The critical region of the $XYZ$ model with $\gamma=0$. The
    critical surface has central charge $c=1$, and is faded where the
    critical nature of the model becomes indeterminate for this
    determination which is done with $N=10$. The black line
    indicates the intersection with the $XY$ model. The model is
    symmetric under $\lambda\to-\lambda$.}
  \label{fig:XYZ_g0}
\end{figure}

We investigate the $XYZ$ model in the range $-2<\Delta<1/2$,
$\lambda>0$, and $\gamma>0$. 
First, consider the case
$\gamma=0$, which is critical in the region shown in Figure \ref{fig:XYZ_g0}.
That is, in the region where $\lambda<1-2\Delta$ and $\lambda$ larger
than a lower limiting line as indicated in the figure. In the
region $\Delta\approx-1$ and 
$|\lambda|\lesssim0.9$ the critical region ends in the sense that for
smaller $\Delta$ the model is massive, but the end of the criticality
is indeterminate. This means that the estimated central charge falls
below one, and the minimizing error gradually becomes larger, though
there is no sharp transition to distinguish the critical from the
non-critical region. The two lines mentioned above, however, have very
sharp transitions between critical and non-critical regions and the
central charge is readily identified at $c=1$. The $c=1$ critical
surface exclusively belongs to the regime $\gamma=0$, except perhaps
for $\Delta<-1$, where this surface seems to spread out into regions
of non-zero $\gamma$. The details here are not investigated in detail,
though.

\begin{figure}[htb]
  \psfrag{X}{$\lambda$}
  \psfrag{Y}{$\Delta$}
  \psfrag{A}{$\gamma=0.5$}
  \psfrag{B}{$\gamma=1$}
  \psfrag{C}{$\gamma=1.5$}
  \centering
  \includegraphics[width=.6\textwidth]{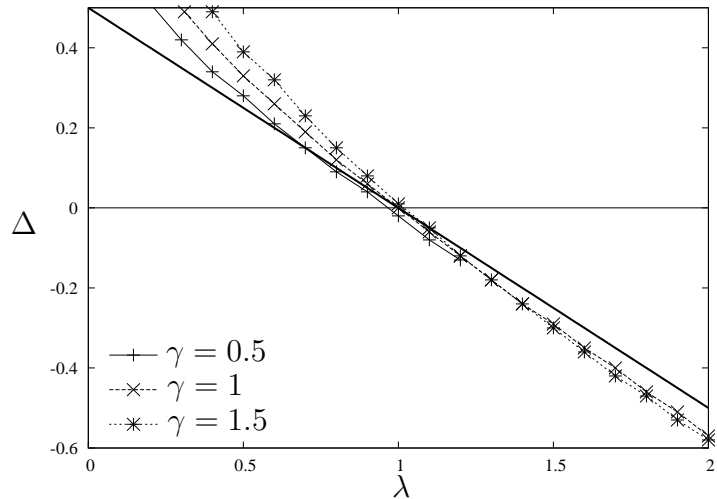}
  \caption[Critical lines of the $XYZ$ model with $c=1/2$.]%
  {Intersections of the critical surface with $c=1/2$ in the $XYZ$
    model for three values of $\gamma$ seen in the $\Delta$-$\lambda$
    plane. There is an uncertainty in the $\Delta$ value at each point
    of about $\pm10^{-2}$ due to the inherent uncertainty in the
    method. The lines are computed based on the maximal estimated
    central charge for $N=10$. The horizontal line is the
    intersection with the $XY$ model. Note that it is possible to draw
    the line for $\gamma=0.5$ only for $\lambda<1$, since it becomes
    indeterminate for this system size for larger $\lambda$. This
    apparent indeterminacy is due to the proximity to the $c=1$
    critical line, and larger system sizes will make it possible to
    draw this line further. The thick line is $\lambda=1-2\Delta$, the
    end point of the $c=1$ surface, and presumed starting point for
    the $c=1/2$ surface for $\gamma\to0$.}
  \label{fig:c.5_line}
\end{figure}

In addition to the $c=1$ surface, there is a $c=1/2$ surface in the
parameter space, which in the intersection with the $XY$ model at
$\Delta=0$ is at $|\lambda|=1$ for all $\gamma$s. When $\Delta$ is
larger than $1/2$, the $c=1$ critical surface vanish for $\gamma=0$,
while the $c=1/2$-line becomes indeterminate. However for
$\Delta<1/2$, the $c=1/2$ transition is illustrated in Figure
\ref{fig:c.5_line}. It has a very weak dependency on $\gamma$, and only for
positive $\Delta$.

\section{Time evolution}
\label{sec:timeev}
Given the formalism of the fermionization developed in section
\ref{sec:XY-model}, it is easy to find the time evolution of the state
\cite{SOStherm}. In the Heisenberg picture, the Majorana fermions
defined in (\ref{eq:Majoranadef}) has a time evolution according to
\begin{equation}
  \frac{\dd}{\dd
    t}\check\gamma_k=\imi[H,\check\gamma_k]=-2\imi\sum_i\mathcal
  C_{ki}\check\gamma_i.
\end{equation}
This is a linear first order system of differential equations, whose
solution is 
\begin{equation}
  \check\gamma_k(t)=\mathds T_{kl}\check\gamma_l,
\end{equation}
where $\check\gamma_k=\check\gamma_k(t=0)$. The time evolution matrix
$\mathds T$ is given by
\begin{equation}
  \mathds T_{kl}(t)=\sum_iS_{ki}S_{li}^*\expe^{-2\imi\xi_it}
\end{equation}
when $S$ is the eigenvector matrix of $\mathcal C$ and $\xi$ are the
eigenvalues. 

\index{thermal states}
A state that is in a thermal equilibrium with its environment at an
inverse temperature $\beta$ is a
mixed state described by its density matrix,
\begin{equation}
  \sigma=\frac1{\mathcal Z}\expe^{-\beta H}
\end{equation}
where $H$ is the system's Hamiltonian and $\mathcal
Z=\trace\expe^{-\beta H}$ is the partition function. Given a
basis in which the Hamiltonian is diagonal $H=\sum_k
E_k\ket{\psi_k}\bra{\psi_k}$, this becomes 
\[\sigma=\frac1{\mathcal Z}\sum_k\expe^{-\beta
  E_k}\ket{\psi_k}\bra{\psi_k}.\]

In \cite{SOStherm} we compute the time evolution of an excited pure
state in an Ising chain with a magnetic impurity, and show that the
eigenvalues of the reduced density matrix for a part of the chain
resembles those for a state in thermal equilibrium at some temperature
$\beta^{-1}$. To this end, we compute the classical fidelity between
the eigenvalues. If we denote the ordered eigenvalues of the reduced
density matrix $\lambda_k$ and the thermal density matrix
$\varphi_k=\expe^{-\beta E_k}/\mathcal Z$, the classical fidelity
becomes
\index{fidelity}
\[F_{\mathrm c}(\lambda,\varphi)=\sum_n\sqrt{\lambda_k\varphi_k}.\]
The classical fidelity is an adequate measure of the distance between
two probability distributions, being 1 if the distributions are
equal, and less than one else. Hence the fidelity is not a metric, but
is intimately related to the trace distance, which is, and thus can be
considered equivalent to a metric \cite{Nielsen&Chuang}.

For two arbitrary quantum states described by density matrices $\rho$
and $\sigma$, the fidelity between them is defined as
\begin{equation}
  F(\rho,\sigma)=\trace\sqrt{\rho^{1/2}\sigma\rho^{1/2}},
\end{equation}
which reduces to the classical fidelity of their eigenvalues in the
case where $\rho$ and $\sigma$ commute. However, for the non-commuting
case, it gets more complicated. Given that the two
states can be fermionized, and thus written in terms of the Majorana fermions
\begin{align*}
  \rho&=N_\rho\exp\left(-\sum_{kl}\Omega_{kl}\check\gamma_k\check\gamma_l\right)&
  N_\rho&=\frac1{\prod_k\left(1+\expe^{-\Omega_{kk}}\right)}\\
  \sigma&=N_\sigma\exp\left(-\sum_{kl}\beta E_{kl}\check\gamma_k\check\gamma_l\right)&
  N_\sigma&=\frac1{\prod_k\left(1+\expe^{-\beta E_{kk}}\right)},
\end{align*}
we need to compute the matrix product
$\sigma^{1/2}\rho\sigma^{1/2}$. The Campbell-Baker-Hausdorff formula
for exponentials of non-commuting variables $\mathcal A$ and $\mathcal
B$ reads\index{Campbell-Baker-Hausdorff formula}
\begin{align}
  \expe^{\mathcal A}\expe^{\mathcal B}&=\expe^{\mathcal D}\qquad\text{where}\nonumber\\
  \mathcal D&=
  \mathcal A+\mathcal B+\frac12[\mathcal A,\mathcal
  B]+\frac1{12}\left[\mathcal A-\mathcal B,[\mathcal A,\mathcal
    B]\right]+\cdots,
  \label{eq:CBH}
\end{align}
given that the series converges (which is not obvious). If we have
operators $\Theta=\frac12\theta_{ij}\check\gamma_i\check\gamma_j$ and
$\Xi=\frac12\xi_{ij}\check\gamma_i\check\gamma_j$ for some
anti-symmetric 
matrices $\theta$ and $\xi$, we can use the commutator--anti-commutator relation
\begin{align*}
  [\check\gamma_i\check\gamma_j,\check\gamma_i\check\gamma_j]&=\check\gamma_i\{\check\gamma_j,\check\gamma_k\}\check\gamma_l-\check\gamma_i\check\gamma_k\{\check\gamma_j,\check\gamma_l\}+\{\check\gamma_i,\check\gamma_k\}\check\gamma_l\check\gamma_j-\check\gamma_k\{\check\gamma_i,\check\gamma_l\}\check\gamma_j\\
  &=\delta_{jk}\check\gamma_i\check\gamma_l-\delta_{jl}\check\gamma_i\check\gamma_k+\delta_{ik}\check\gamma_l\check\gamma_j-\delta_{il}\check\gamma_k\check\gamma_j
\end{align*}
to find that 
\begin{equation}
  [\Theta,\Xi]=\frac12\varphi_{ij}\check\gamma_i\check\gamma_j=\Phi
\end{equation}
where $\varphi=[\theta,\xi]$. This relation holds for all iterated
commutators involved in (\ref{eq:CBH}), and we may therefore conclude
that  
\[\expe^{\mathcal A}\expe^{\mathcal B}=\expe^{\mathcal D},\]
where $\mathcal A=\frac12A_{ij}\check\gamma_i\check\gamma_j$, 
$\mathcal B=\frac12B_{ij}\check\gamma_i\check\gamma_j$, and 
$\mathcal D=\frac12D_{ij}\check\gamma_i\check\gamma_j$ with $D$ given
by the expansion (\ref{eq:CBH}). The generalization to more than two
factors is simple, and we can compute the matrix product
$\sigma^{1/2}\rho\sigma^{1/2}$ and (block) diagonalize the result into the
matrix $\exp\left(-K_{ij}\hat\gamma_i\hat\gamma_j\right)$ with new
Majorana fermions $\hat\gamma$. Thus the fidelity becomes
\[F(\sigma,\rho)=\trace\sqrt{N_\sigma N_\rho}
\expe^{-\frac12K_{ij}\hat\gamma_i\hat\gamma_j},\]
which is straightforward to compute.

Hence it is possible to find the full fidelity rather than just the
classical fidelity of the eigenvalues, with the possible catch that
the Campbell-Baker-Haudorff series (\ref{eq:CBH}) may not
converge. However, it is not expected that the difference between the
classical and quantum versions will be
large, since the diagonalization of the Hamiltonian and the density
matrix will be ``almost'' the same. That is, the diagonalization
matrices $O_\sigma$ and $O_\rho$ respectively fulfill
$O_\sigma^{-1}O_\rho\approx\mathds 1$. However, discrepancies in this
equality may map back to small eigenvalues that are of little
importance to the fidelity.

\section{The quantum Hall effects}
\label{sec:QHE}
\begin{figure}[b]
  \centering
  \begin{pspicture}(15,10)
    \rput*(8,5){\includegraphics[width=.7\textwidth]{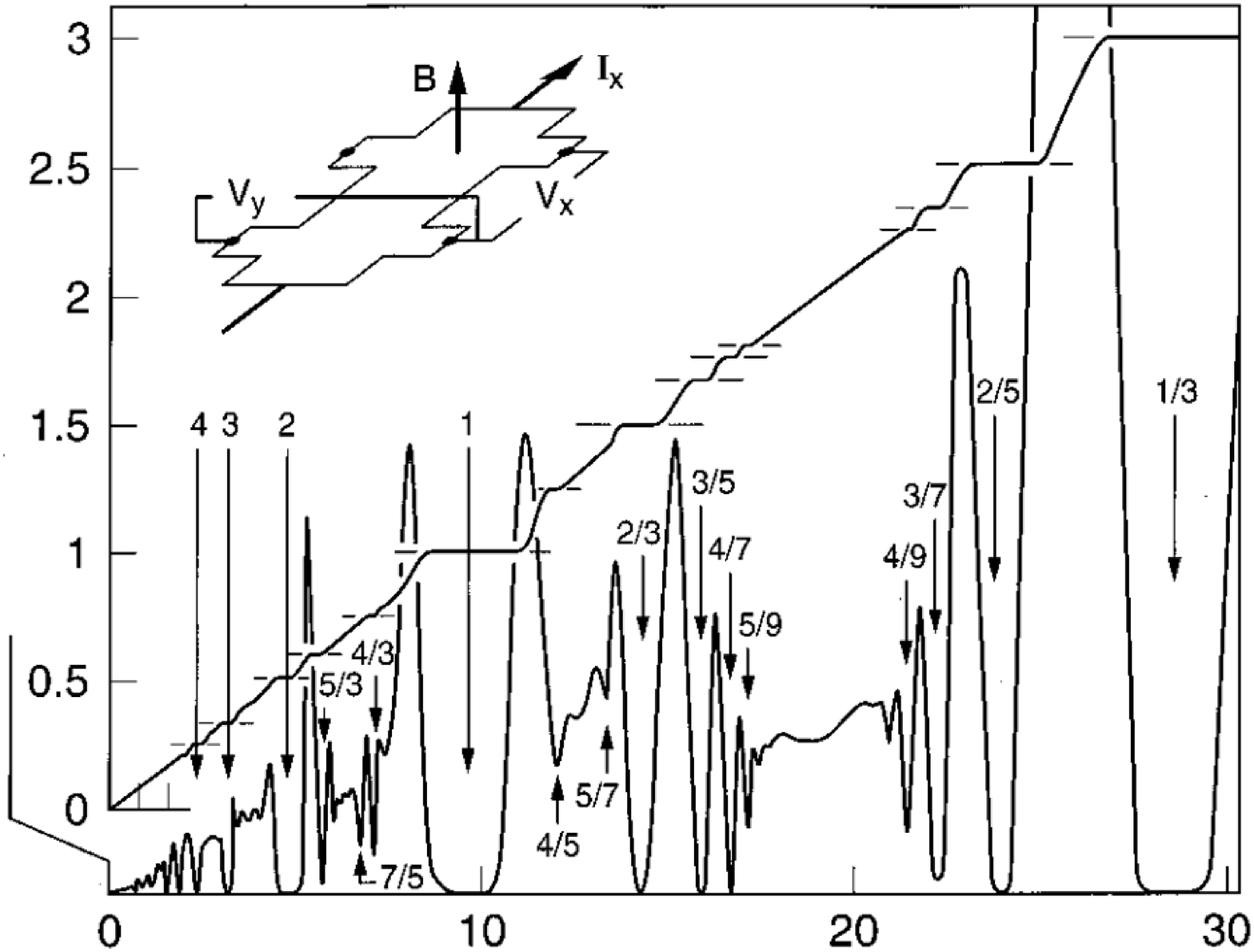}}
    \rput(2,2.5){$R_{xx}$}
    \rput(2.2,6){$R_{xy}$}
    \rput(8,.3){B [$T$]}
  \end{pspicture}
  \caption[Experimental setup and data for the fractional quantum Hall
  effect.]%
  {The resistivity of a quantum Hall system, sketched in the upper
    left. Both resistivities in orthogonal $R_{xx}=V_x/I_x$ and
    transverse $R_{xy}=V_y/I_x$ are shown and the most important
    fractional and integer quantum Hall states are indicated. The
    experiment uses a 2D system with electron density
    $n=2.33\times10^{15}\unit{m}^{-2}$ cooled to $85\unit{mK}$. 
    Image from \cite{Stormer99}.} 
  \label{fig:QHE}
\end{figure}

\index{quantum Hall effect|(}
The exposure of an essentially two dimensional conductor to an magnetic
field gives rise to a voltage perpendicular to the current flowing
through the slab, which has long been recognized as the Hall
effect. A schematic experimental setup is shown in Figure
\ref{fig:QHE}. 
The Hall resistance $R_{xy}$ is the Hall voltage across the
slab $V_y$ by the exposed current $I_x$, and in the classical case
the Hall voltage is proportional to the imposed magnetic
field. Indeed, straightforward calculation gives the law
$R_{xy}=B/ne$, $B$ being the perpendicular magnetic field, $n$ the
density of conduction electrons%
\footnote{In a typical setup on uses
  \chem{GaAs}/\chem{AlGaAs} heterostructures with
  $n\sim10^{15}\unit{m^{-2}}$. \cite{Stormer99}} 
and $e$ the elementary charge. When the system is cooled down, and the
magnetic field becomes large, the Hall resistance becomes quantized at
exact levels of
\[\nu=\frac{N_eh}{eB}=\frac{N_e}{N_s};\qquad\qquad N_s=\frac{eB}h,\]
with $\nu$ an integer called the filling factor\index{filling factor},
as first reported in 1980 by von Klitzing \etal \cite{Klitzing80}. $h$
is Planck's constant, and the phrase {\it filling factor} refers to
the fact that $N_s$ is the number of available states in each
Landau level and $N_e$ the number of available electrons. As the
magnetic field is increased, 
$N_s$ increases proportionally, and when $\nu$ becomes integer, the
Fermi level resides 
between Landau levels, thus all available states are occupied and the
electron fluid is essentially incompressible until the magnetic field
increases to another integer. This integer quantum Hall effect (IQHE)
is explained as a problem of non-interacting (apart from the Fermi
statistics) electron problem first explained by Laughlin 
in 1981 \cite{Laughlin81}.

In 1982 Tsui \etal \cite{Tsui82} identified levels of constant
resistance also at levels of fractional $\nu$, a result completely
unexpected at the time, and a matter much harder to explain than
IQHE. The fractional quantum Hall effect (FQHE) occurs at filling
factors $\nu=p/q$ for certain fractions, where the most prominent
levels are those with $\nu=1/q$ and $\nu=1-1/q$ with $q$ odd (called
primary states), and $\nu=p/(2p\pm1)$ (higher order states),
cf. Figure \ref{fig:QHE}. However,
many more fractions appear, even fractions with even denominator. FQHE
is, in contrast to the integer variant a complex many-particle
problem, and a number of explanations exist. Perhaps the most
prominent of these are Laughlin's trial wave function
\cite{Laughlin83}, which for the $\nu=1/q$ state of $n$ particles
reads  \index{Laughlin wave function}
\[\psi_{1/q}(z_1,\cdots,z_n)=\prod_{i<j}^n\left(z_i-z_j\right)^q\exp\left(-\frac1{4l_0^2}\sum_{k=1}^n|z_k|^2\right),\]
where $z_k$ is the 2D complex position of particle $k$, and
normalization is omitted. $l_0=\sqrt{\hbar/eB}$ is the magnetic
length. This reflects attaching $q$ flux quanta to each electron,
since the charge concentration of electrons and charge deficit of a
vortex associated with a flux quantum attract each other, thus
creating fractionally charged excitations.

More recently, Jain's composite fermion picture \cite{Jain89}
has become a more\index{composite fermion} 
fashionable explanation, particularly since it also  explains many
higher-order states. Essentially, the system of strongly interacting
electrons in a strong magnetic field is mapped onto a system with
a reduced magnetic field and a series of non-interacting composite
fermions consisting of one electron and two flux quanta
\cite{Jain00}. Thus, the fractional quantum Hall effect becomes a
integer quantum Hall problem of fractionally charged fermions.

\subsection{FQHE on a torus}
Most of the literature has so far focused on explaining the inner
workings of the FQHE, and little is known about the entanglement
properties of the quantum Hall states. Nevertheless, it has been
argued that
on the IQHE plateaus the entanglement is zero, since this is
essentially a non-interacting system \cite{Shi04}. 
As
the magnetic field is increased (or decreased) and the plateau changes
from one FQHE state to another, we have a quantum phase transition,
and possibly the criticality technique described in \cite{SOS1} and
Section \ref{ssec:determining_criticality} could be applied here.

Even though the QHE system is a generic two dimensional problem, it
can be mapped onto a 1D problem \cite{Chakraborty98,Yoshioka83}. That
is, with 
periodic boundary conditions, one have $N$ different single-particle
wave functions in each Landau level, which in a rectangular geometry
with sides $L_x$ and $L_y$ are
\begin{equation}
  \psi_j(\vkt r)=\left(\frac1{L_y\sqrt\pi l_o}\right)^{1/2}\sum_{k=-\infty}^\infty
  \exp\left[\frac\imi{l_0^2}(X_j+kL_x)y-\frac1{2l_0^2}(X_j+kL_x-x)^2\right].
\end{equation}
Here, $1\leq j\leq N_s$ and $X_j=jL_x/N_s$ are localizations along the
$x$-direction\footnote{This example is in the Landau gauge $\vkt
  A=Bx\hat{\vkt y}$.}. This amounts to $N_s$ localizations along the $x$ 
direction with Gaussian with of the order of the magnetic length. The
probability distribution is independent of $y$.

In a second quantized version, the Hamiltonian is
\begin{equation}
  H=\sum_j\mathcal Wa_j^\dag a_j+
  \sum_{j_1}\sum_{j_2}\sum_{j_3}\sum_{j_4}\mathcal A_{j_1,j_2,j_3,j_4}
  a_{j_1}^\dag a_{j_2}^\dag a_{j_3}a_{j_4}
  \label{eq:FQHE_H}
\end{equation}
where $a_j$ is the destruction operator at site $j$. The
single electron energy $\mathcal W$ is a known constant
\cite{Yoshioka83} and the coupling term is
\begin{align*}
  \mathcal A_{j_1,j_2,j_3,j_4}&=\frac12\int\dd^2\vkt r_1\dd^2\vkt r_2
  \,\psi^*_{j_1}(\vkt r_1)\psi^*_{j_2}(\vkt r_2)V(\vkt r_1-\vkt r_2)
  \psi_{j_3}(\vkt r_2)\psi_{j_4}(\vkt r_1)\\
  &=\frac{\pi e^2}{\varepsilon_0L_xL_y}\sum_{\vkt q\not=0}
  \frac1q\,\exp\left[-\frac12l_0^2q^2-\imi q_xL_x(j_1-j_3)/N_s\right]
  \delta'_{j_1-j_4,q_yL_y/2\pi}\delta'_{j_1+j_2,j_3+j_4}
\end{align*}
where  $\delta'$ is a Kronecker delta modulus $N_s$ and $V(\vkt
r)$ is the real-space Coulomb interaction. The sum over $\vkt q$ is a
sum over all allowed $\vkt q$-vectors (except the zero vector), that
is with integers $s$ and $t$ such that $\vkt q=\left(\frac{2\pi
    s}{L_x},\frac{2\pi t}{L_y}\right)$. Thus a basis of the Hilbert 
space is defined by the $N_e$ occupation numbers,
$\ket{j_1,\cdots,j_{N_a}}$. For a given filling fraction $\nu$ this
means that the Hilbert space has dimension ${N_s}\choose{N_e}$, which
obviously quickly becomes far too large to handle. Indeed, the Hilbert
space's dimension grows exponentially with the number of electrons for
a fixed $\nu$, and faster than $2^{N_e}$. Again, utilizing symmetries
makes it possible to go slightly further. With the most na\"{\i}ve
approach however, investigating six particles in a $\nu=1/3$ system
involves matrices of size ${18 \choose 6}=18564$, which is
computationally time consuming. Larger systems are essentially
inaccessible. 

\subsection{Entanglement in the FQHE}
The entanglement in the FQHE case is conceptually different from
the spin chains setting. This is mainly due to the
complications involved in dealing with identical
particles. In spin chains we consider each particle fixed with a
non-overlapping wave function, and the particles can
hence be thought of as separate, or non-identical. This is not the
case in the FQHE where the overlap of the wave function may be
considerable. Entanglement is ill-defined in this case, mainly since
the anti-symmetrized wave function defined by the Slater determinant
already carries quantum correlations while they particles involved
cannot be said to be entangled, in the sense that there exists no less
correlated state. It has been suggested that entropy of
reduced density matrices larger than that of a Slater determinant
state is an entangled fermion state \cite{Ghirardi04}. For bosons
the argument is likewise, but since the least correlated bosonic state
is a product state whose entropy is zero, positive entropy indicates
an entangled state here. However, it is not obvious that the entropy
(possibly subtracted the Slater determinant entropy) is a genuine
entanglement measure. The entropy of a single particle in a Slater
determinant with $N$ particles is $\log_2N$.

For the FQHE case the reduced density matrix of  $N_e'<N_e$ electrons is
of size ${N_s \choose N_e'}$ and thus the reduced density matrix may
be {\it larger} than the original matrix, if $N_e'>N_e/2$. Despite
this apparent 
paradox, it is possible to compute the entropy of the reduced density
matrix. To this end, consider the density matrix of the ${3\choose 2}$
system in the basis $\{\ket{110},\ket{101},\ket{011}\}$,
$\rho=[\rho_{ij}]$ in this basis.
Tracing out
one particle to remain with one amounts to tracing over all
wave functions with one particle, and in the new basis
$\{\ket{100},\ket{010},\ket{001}\}$ this becomes
\[\rho'=\bra{1\cdot\cdot}\rho\ket{1\cdot\cdot}+\bra{\cdot1\cdot}\rho\ket{\cdot1\cdot}+\bra{\cdot\cdot1}\rho\ket{\cdot\cdot1}
=\frac12
\begin{matrise}
\rho_{11}+\rho_{22} & \rho_{23} & \rho_{13} \\
\rho_{32} & \rho_{11}+\rho_{33} & \rho_{12} \\
\rho_{31} & \rho_{21} & \rho_{22}+\rho_{33}
\end{matrise}.\]

Numerical calculation of the entropy for the ground state of
(\ref{eq:FQHE_H}) by the prescription above, shows that the integer
quantum Hall effect has entropy $S_n=\log_2{N_s\choose n}$ when the
system is traced out to leave $n$ particles, as predicted in
\cite{Iblisdir06}. Tracing out to a single particle, the entropy is
simply $S_1=\log_2N_e$. The values for the entropy in the IQHE states
are independent of the nature of the interaction, since IQHE is an
essentially non-interacting phenomenon.

For the FQHE the case is more complicated, and exact analytical
results are rare. Using the prescription above, some basic results for
very few particles can be found. In particular, we find that in the
limit where $L_y\ll L_x$, the entropy is exactly that of the IQHE,
$\log_2{N_s\choose n}$ when $n$ particles are traced out. In this
limit, the localizations are so far apart that the particles are
essentially non-interacting, and the IQHE regime is
recovered. However, as the aspect ratio $L_y/L_x$ is increased, at
some value the entropy for a single particle jumps discontinuously to
a higher value, and after this increases slowly with the aspect
ratio until saturating.
\index{quantum Hall effect|)}

\fancyhead[LE]{\thepage}
\fancyhead[RO]{\thepage}

\bibliographystyle{mybst}

\bibliography{thesis}

\newcommand{\etalchar}[1]{$^{#1}$}
\begin{thebibliography}{VWPGC06}
\providecommand{\url}[1]{\texttt{#1}}
\providecommand{\urlprefix}{URL }
\providecommand{\eprint}[2][]{\url{#2}}

\bibitem[ABV01]{Arnesen01}
\textsc{M.~C. Arnesen, S.~Bose, and V.~Vedral}, \emph{Natural Thermal and
  Magnetic Entanglement in the 1D Heisenberg Model}, Phys. Rev. Lett.,
  \textbf{87}, 017901 (2001).

\bibitem[ADD{\etalchar{+}}06]{Andre06}
\textsc{A.~Andr\'{e}, D.~Demille, J.~M. Doyle, M.~D. Lukin, S.~E. Maxwell,
  P.~Rabl, R.~J. Schoelkopf, and P.~Zoller}, \emph{A coherent all-electrical
  interface between polar molecules and mesoscopic superconducting resonators},
  Nature Phys., \textbf{2}, 636 (2006).

\bibitem[ADR82]{Aspect82}
\textsc{A.~Aspect, J.~Dalibard, and G.~Roger}, \emph{Experimental Test of
  {B}ell's Inequalities Using Time-Varying Analyzers}, Phys. Rev. Lett.,
  \textbf{49}, 1804 (1982).

\bibitem[BB84]{BB84}
\textsc{C.~H. Bennett and G.~Brassard}, \emph{Quantum cryptography: Public key
  distribution and coin tossing}, in \emph{Proceedings of the IEEE
  International Conference on Computers, Systems and Signal Processing,
  Bangalore, India}, p. 175 (IEEE Press, New York, 1984).

\bibitem[BBC{\etalchar{+}}93]{Bennett93}
\textsc{C.~H. Bennett, G.~Brassard, C.~Cr\'epeau, R.~Jozsa, A.~Peres, and W.~K.
  Wootters}, \emph{Teleporting an unknown quantum state via dual classical and
  Einstein-Podolsky-Rosen channels}, Phys. Rev. Lett., \textbf{70}, 1895
  (1993).

\bibitem[Bel64]{Bell64}
\textsc{J.~S. Bell}, \emph{On the {E}instein {P}odolsky {R}osen Paradox},
  Physics, \textbf{1}, 195 (1964).

\bibitem[BM71]{BarouchMccoy71}
\textsc{E.~Barouch and B.~M. McCoy}, \emph{Statistical Mechanics of the {XY
  model. II. S}pin-Correlation Functions}, Phys. Rev. A, \textbf{3}, 786
  (1971).

\bibitem[BPZ84]{BPZ84b}
\textsc{A.~A. Belavin, A.~M. Polyakov, and A.~B. Zamolodchikov}, \emph{Infinite
  conformal symmetry in two-dimensional quantum field theory}, Nucl. Phys. B,
  \textbf{241}, 333 (1984).

\bibitem[BW92]{Bennett92}
\textsc{C.~H. Bennett and S.~J. Wiesner}, \emph{Communication via one- and
  two-particle operators on Einstein-Podolsky-Rosen states}, Phys. Rev. Lett.,
  \textbf{69}, 2881 (1992).

\bibitem[CC04]{Calabrese04}
\textsc{P.~Calabrese and J.~Cardy}, \emph{Entanglement Entropy and Quantum
  Field Theory}, J. Stat. Mech., p. P06002 (2004).

\bibitem[CC05]{calabrese-2005-0504}
---{}---{}---, \emph{Evolution of Entanglement Entropy in One-Dimensional
  Systems}, J. Stat. Mech., p. P04010 (2005).

\bibitem[Cha02]{Chakravarty02}
\textsc{S.~Chakravarty}, \emph{Theory of the d-density wave from a vertex model
  and its implications}, Phys. Rev. B, \textbf{66}, 224505 (2002).

\bibitem[Chr06]{Christandl06}
\textsc{M.~Christandl}, \emph{The Structure of Bipartite Quantum States -
  Insights from Group Theory and Cryptography}, Ph.D. thesis, University of
  Cambridge (2006), \eprint{quant-ph/0604183}.

\bibitem[CP95]{Chakraborty98}
\textsc{T.~Chakraborty and P.~Pietil\"{a}inen}, \emph{The Quantum Hall Effects:
  Fractional and Integral} (Springer, Berlin, 1995).

\bibitem[CW94]{Callan94}
\textsc{C.~Callan and F.~Wilczek}, \emph{On Geometric Entropy}, Phys. Lett. B,
  \textbf{333}, 55 (1994).

\bibitem[CW04]{Christandl04}
\textsc{M.~Christandl and A.~Winter}, \emph{``Squashed Entanglement'' - An
  Additive Entanglement Measure}, J. Math. Phys., \textbf{45}, 829 (2004).

\bibitem[CZ95]{Cirac95}
\textsc{J.~I. Cirac and P.~Zoller}, \emph{Quantum Computations with Cold
  Trapped Ions}, Phys. Rev. Lett., \textbf{74}, 4091 (1995).

\bibitem[EC05]{Eisert05}
\textsc{J.~Eisert and M.~Cramer}, \emph{Single-copy entanglement in critical
  quantum spin chains}, Phys. Rev. A, \textbf{72}, 042112 (2005).

\bibitem[Eke91]{Ekert91}
\textsc{A.~K. Ekert}, \emph{Quantum cryptography based on Bell\char39{}s
  theorem}, Phys. Rev. Lett., \textbf{67}, 661 (1991).

\bibitem[EPR35]{EPR35}
\textsc{A.~Einstein, B.~Podolsky, and N.~Rosen}, \emph{Can Quantum-Mechanical
  Description of Physical Reality Be Considered Complete?}, Phys. Rev.,
  \textbf{47}, 777 (1935).

\bibitem[Fey82]{Feynman82}
\textsc{R.~P. Feynman}, \emph{Simulating Physics with Computers}, Int. J.
  Theor. Phys., \textbf{21}, 467 (1982).

\bibitem[FMS97]{Francesco97}
\textsc{P.~D. Francesco, P.~Mathieu, and D.~Sénéchal}, \emph{Conformal Field
  Theory} (Springer, New York, 1997).

\bibitem[FQS84]{Friedan84}
\textsc{D.~Friedan, Z.~Qiu, and S.~Shenker}, \emph{Conformal Invariance,
  Unitarity, and Critical Exponents in Two Dimensions}, Phys. Rev. Lett.,
  \textbf{52}, 1575 (1984).

\bibitem[Gin89]{Ginsparg1988}
\textsc{P.~Ginsparg}, \emph{Applied conformal field theory}, in
  \textsc{E.~Brézin and J.~Z. Justin} (editors), \emph{Fields, Strings and
  Critical Phenomena} (Les Houches, Session XLIX, 1989).

\bibitem[GKLC01]{Giedke01}
\textsc{G.~Giedke, B.~Kraus, M.~Lewenstein, and J.~I. Cirac},
  \emph{Entanglement Criteria for all Bipartite Gaussian States}, Phys. Rev.
  Lett., \textbf{87}, 167904 (2001).

\bibitem[GM04]{Ghirardi04}
\textsc{G.~Ghirardi and L.~Marinatto}, \emph{General criterion for the
  entanglement of two indistinguishable particles}, Phys. Rev. A, \textbf{70},
  012109 (2004).

\bibitem[GWK{\etalchar{+}}03]{Giedke03}
\textsc{G.~Giedke, M.~M. Wolf, O.~Kr\"uger, R.~F. Werner, and J.~I. Cirac},
  \emph{Entanglement of Formation for Symmetric {G}aussian States}, Phys. Rev.
  Lett., \textbf{91}, 107901 (2003).

\bibitem[Haw75]{Hawking75}
\textsc{S.~Hawking}, Comm. Math. Phys., \textbf{43}, 199 (1975).

\bibitem[HHH00]{Horodecki00}
\textsc{M.~Horodecki, P.~Horodecki, and R.~Horodecki}, \emph{Limits for
  Entanglement Measures}, Phys. Rev. Lett., \textbf{84}, 2014 (2000).

\bibitem[HLW94]{Holzhey94}
\textsc{C.~Holzhey, F.~Larsen, and F.~Wilczek}, \emph{Geometric and
  Renormalized Entropy in Conformal Field Theory}, Nucl.Phys. B, \textbf{424},
  443 (1994).

\bibitem[HW97]{Hill97}
\textsc{S.~Hill and W.~K. Wootters}, \emph{Entanglement of a Pair of Quantum
  Bits}, Phys. Rev. Lett., \textbf{78}, 5022 (1997).

\bibitem[ILO06]{Iblisdir06}
\textsc{S.~Iblisdir, J.~I. Latorre, and R.~Or\'us}, \emph{Entropy and Exact
  Matrix Product Representation of the Laughlin Wave Function} (2006),
  \eprint{cond-mat/0609088}.

\bibitem[Jai89]{Jain89}
\textsc{J.~K. Jain}, \emph{Composite-fermion approach for the fractional
  quantum {H}all effect}, Phys. Rev. Lett., \textbf{63}, 199 (1989).

\bibitem[Jai00]{Jain00}
---{}---{}---, \emph{The composite fermion: A quantum particle and its quantum
  fluids}, Physics Today, \textbf{53}, 39 (2000).

\bibitem[Jen05]{Jenssen05}
\textsc{R.~Jenssen}, \emph{An Information Theoretic Approach to Machine
  Learning}, Ph.D. thesis, The University of Troms\o, Norway (2005).

\bibitem[Kad66]{Kadanoff66}
\textsc{L.~P. Kadanoff}, \emph{Scaling laws for Ising models near $T_c$},
  Physics, \textbf{2}, 263 (1966).

\bibitem[KDP80]{Klitzing80}
\textsc{K.~v. Klitzing, G.~Dorda, and M.~Pepper}, \emph{New Method for
  High-Accuracy Determination of the Fine-Structure Constant Based on Quantized
  Hall Resistance}, Phys. Rev. Lett., \textbf{45}, 494 (1980).

\bibitem[KE04]{Kimble04}
\textsc{H.~J. Kimble and S.~J. van Enk}, \emph{Push-button teleportation},
  Nature, \textbf{429}, 712 (2004).

\bibitem[Kor04]{Korepin2004}
\textsc{V.~E. Korepin}, \emph{Universality of Entropy Scaling in One
  Dimensional Gapless Models}, Phys. Rev. Lett., \textbf{92}, 096402 (2004).

\bibitem[KT73]{KosterlitzThouless73}
\textsc{J.~M. Kosterlitz and D.~J. Thouless}, \emph{Ordering, metastability and
  phase transitions in two-dimensional systems}, Journal of Physics C: Solid
  State Physics, \textbf{6}, 1181 (1973).

\bibitem[Lau81]{Laughlin81}
\textsc{R.~B. Laughlin}, \emph{Quantized Hall conductivity in two dimensions},
  Phys. Rev. B, \textbf{23}, 5632 (1981).

\bibitem[Lau83]{Laughlin83}
---{}---{}---, \emph{Anomalous Quantum Hall Effect: An Incompressible Quantum
  Fluid with Fractionally Charged Excitations}, Phys. Rev. Lett., \textbf{50},
  1395 (1983).

\bibitem[Lie67]{Lieb67}
\textsc{E.~H. Lieb}, \emph{Residual Entropy of Square Ice}, Phys. Rev.,
  \textbf{162}, 162 (1967).

\bibitem[Llo99]{Lloyd99}
\textsc{S.~Lloyd}, \emph{Quantum search without entanglement}, Phys. Rev. A,
  \textbf{61}, 010301 (1999).

\bibitem[LLRV05]{latorre05}
\textsc{J.~I. Latorre, C.~A. L\"{u}tken, E.~Rico, and G.~Vidal},
  \emph{Fine-grained entanglement loss along renormalization-group flows},
  Phys. Rev. A, \textbf{71}, 034301 (2005).

\bibitem[LRV04]{Latorre:2003kg}
\textsc{J.~I. Latorre, E.~Rico, and G.~Vidal}, \emph{Ground state entanglement
  in quantum spin chains}, Quant. Inf. and Comp., \textbf{4}, 48 (2004),
  \eprint{quant-ph/0304098}.

\bibitem[Mer85]{Mermin85}
\textsc{N.~D. Mermin}, \emph{Is the moon there when nobody looks? Reality and
  the quantum theory}, Physics Today, \textbf{38}, 38 (1985).

\bibitem[MW66]{MerminWagner66}
\textsc{N.~D. Mermin and H.~Wagner}, \emph{Absence of Ferromagnetism or
  Antiferromagnetism in One- or Two-Dimensional Isotropic Heisenberg Models},
  Phys. Rev. Lett., \textbf{17}, 1133 (1966).

\bibitem[Myh04]{Myhr04}
\textsc{G.~O. Myhr}, \emph{Measures of entanglement in quantum mechanics},
  Master's thesis, NTNU, Norway (2004), \eprint{quant-ph/0408094}.

\bibitem[NC00]{Nielsen&Chuang}
\textsc{M.~A. Nielsen and I.~L. Chuang}, \emph{Quantum Computation and Quantum
  Information} (Cambridge University Press, Cambridge, UK, 2000).

\bibitem[OAFF02]{Osterloh:2002}
\textsc{A.~Osterloh, L.~Amico, G.~Falci, and R.~Fazio}, \emph{Scaling of
  entanglement close to a quantum phase transition}, Nature, \textbf{416}, 608
  (2002).

\bibitem[ON02a]{Osborne:2002}
\textsc{T.~J. Osborne and M.~A. Nielsen}, \emph{Entanglement in simple quantum
  phase transitions}, Phys. Rev. A, \textbf{66}, 032110 (2002).

\bibitem[ON02b]{OsborneNielsen01}
---{}---{}---, \emph{Entanglement, Quantum Phase Transitions, and Density
  Matrix Renormalization}, Quant. Inf. Proc., \textbf{1}, 45 (2002).

\bibitem[Ons44]{Onsager44}
\textsc{L.~Onsager}, \emph{Crystal Statistics. I. A Two-Dimensional Model with
  an Order-Disorder Transition}, Phys. Rev., \textbf{65}, 117 (1944).

\bibitem[Ort05]{Ortega05}
\textsc{E.~R. Ortega}, \emph{Quantum correlations in $(1+1)$-dimensional
  systems}, Ph.D. thesis, Universitat de Barcelona, Spain (2005),
  \eprint{quant-ph/0509037}.

\bibitem[Pol70]{Polyakov70}
\textsc{A.~M. Polyakov}, \emph{Conformal symmetry of critical fluctuations},
  JETP Lett., \textbf{12}, 381 (1970).

\bibitem[SAB{\etalchar{+}}06]{Steffen06}
\textsc{M.~Steffen, M.~Ansmann, R.~C. Bialczak, N.~Katz, E.~Lucero,
  R.~McDermott, M.~Neeley, E.~M. Weig, A.~N. Cleland, and J.~M. Martinis},
  \emph{Measurement of the Entanglement of Two Superconducting Qubits via State
  Tomography}, Science, \textbf{313}, 1423 (2006).

\bibitem[Sac99]{Sachdev}
\textsc{S.~Sachdev}, \emph{Quantum Phase Transitions} (Cambridge University
  Press, Cambridge, UK, 1999).

\bibitem[SC06]{Syljuasen06}
\textsc{O.~F. Sylju{\aa}sen and S.~Chakravarty}, \emph{Resonating Plaquette
  Phase of a Quantum Six-Vertex Model}, Phys. Rev. Lett., \textbf{96}, 147004
  (2006).

\bibitem[Sch05]{Schollwoeck2005}
\textsc{U.~Schollw\"{o}ck}, \emph{The density-matrix renormalization group},
  Rev. Mod. Phys., \textbf{77}, 259 (2005).

\bibitem[Shi04]{Shi04}
\textsc{Y.~Shi}, \emph{Quantum entanglement in second-quantized condensed
  matter systems}, J. Phys. A, \textbf{37}, 6807 (2004).

\bibitem[Sho94]{Shor94}
\textsc{P.~Shor}, \emph{Algorithms for Quantum Computation: Discrete Logarithms
  and Factoring}, in \textsc{S.~Goldwasser} (editor), \emph{Proceedings of the
  35\textsuperscript{th} Annual Symposium on Foundations of Computer Science},
  p. 124 (IEEE Press, Los Alamitos, California, 1994).

\bibitem[Skr05a]{SOSbose}
\textsc{S.~O. Skr{\o}vseth}, \emph{Entanglement in bosonic systems}, Phys. Rev.
  A, \textbf{72}, 062305 (2005).

\bibitem[Skr05b]{SOSposter}
---{}---{}---, \emph{Entanglement signatures in critical quantum systems}, in
  \emph{ERATO conference on Quantum Information Science}, p. 177 (Quantum
  Computation and Information Project, ERATO, JST, Tokyo, Japan, 2005).

\bibitem[Skr06a]{SOSent}
---{}---{}---, \emph{Entanglement properties of quantum spin chains}, Phys.
  Rev. A, \textbf{74}, 022327 (2006).

\bibitem[Skr06b]{SOStherm}
---{}---{}---, \emph{Thermalization through unitary evolution of pure states},
  Europhys. Lett., \textbf{76}, 1179 (2006).

\bibitem[Sla41]{Slater41}
\textsc{J.~C. Slater}, \emph{Theory of the Transition in \chem{KH_2PO_4}}, J.
  Chem. Phys., \textbf{9}, 16 (1941).

\bibitem[SO05]{SOS1}
\textsc{S.~O. Skr{\o}vseth and K.~Olaussen}, \emph{Entanglement used to
  identify critical systems}, Phys. Rev. A, \textbf{72}, 022318 (2005).

\bibitem[Sre93]{Srednicki93}
\textsc{M.~Srednicki}, \emph{Entropy and Area}, Phys. Rev. Lett., \textbf{71},
  666 (1993).

\bibitem[STG99]{Stormer99}
\textsc{H.~L. Stormer, D.~C. Tsui, and A.~C. Gossard}, \emph{The fractional
  quantum Hall effect}, Rev. Mod. Phys., \textbf{71}, S298 (1999).

\bibitem[Sut70]{Sutherland70}
\textsc{B.~Sutherland}, \emph{Two-Dimensional Hydrogen Bonded Crystals without
  the Ice Rule}, J. Math. Phys., \textbf{11}, 3183 (1970).

\bibitem[TSG82]{Tsui82}
\textsc{D.~C. Tsui, H.~L. Stormer, and A.~C. Gossard}, \emph{Two-Dimensional
  Magnetotransport in the Extreme Quantum Limit}, Phys. Rev. Lett.,
  \textbf{48}, 1559 (1982).

\bibitem[VC04]{Verstraete04}
\textsc{F.~Verstraete and J.~I. Cirac}, \emph{Renormalization algorithms for
  Quantum-Many Body Systems in two and higher dimensions} (2004),
  \eprint{cond-mat/0407066}.

\bibitem[Vid03]{Vidal03b}
\textsc{G.~Vidal}, \emph{Efficient Classical Simulation of Slightly Entangled
  Quantum Computations}, Phys. Rev. Lett., \textbf{91}, 147902 (2003).

\bibitem[Vid05]{Vidal2005}
---{}---{}---, \emph{Entanglement renormalization} (2005),
  \eprint{cond-mat/0512165}.

\bibitem[VLRK02]{Vidal:2002rm}
\textsc{G.~Vidal, J.~I. Latorre, E.~Rico, and A.~Kitaev}, \emph{Entanglement in
  quantum critical phenomena}, Phys. Rev. Lett., \textbf{90}, 227902 (2002).

\bibitem[VPRK97]{Vedral97}
\textsc{V.~Vedral, M.~B. Plenio, M.~A. Rippin, and P.~L. Knight},
  \emph{Quantifying Entanglement}, Phys. Rev. Lett., \textbf{78}, 2275 (1997).

\bibitem[VSB{\etalchar{+}}01]{Vandersypen01}
\textsc{L.~M.~K. Vandersypen, M.~Steffen, G.~Breyta, C.~S. Yannoni, M.~H.
  Sherwood, and I.~L. Chuang}, \emph{Experimental realization of Shor's quantum
  factoring algorithm using nuclear magnetic resonance}, Nature, \textbf{414},
  883 (2001).

\bibitem[VW02]{Vidal02}
\textsc{G.~Vidal and R.~F. Werner}, \emph{Computable measure of entanglement},
  Phys. Rev. A, \textbf{65}, 032314 (2002).

\bibitem[VWPGC06]{Verstraete06}
\textsc{F.~Verstraete, M.~M. Wolf, D.~Perez-Garcia, and J.~I. Cirac},
  \emph{Criticality, the Area Law, and the Computational Power of Projected
  Entangled Pair States}, Phys. Rev. Lett., \textbf{96}, 220601 (2006).

\bibitem[Wan01a]{Wang01b}
\textsc{X.~Wang}, \emph{Effects of anisotropy on thermal entanglement}, Phys.
  Lett. A, \textbf{281}, 101 (2001).

\bibitem[Wan01b]{Wang01}
---{}---{}---, \emph{Entanglement in the quantum {H}eisenberg {$XY$} model},
  Phys. Rev. A, \textbf{64}, 012313 (2001).

\bibitem[WDM{\etalchar{+}}05]{Wei05}
\textsc{T.-C. Wei, D.~Das, S.~Mukhopadyay, S.~Vishveshwara, and P.~M.
  Goldbart}, \emph{Global entanglement and quantum criticality in spin chains},
  Phys. Rev. A, \textbf{71}, 060305(R) (2005).

\bibitem[Wer89]{Werner89}
\textsc{R.~F. Werner}, \emph{Quantum states with {E}instein-{P}odolsky-{R}osen
  correlations admitting a hidden-variable model}, Phys. Rev. A, \textbf{40},
  4277 (1989).

\bibitem[WGK{\etalchar{+}}04]{Wolf04}
\textsc{M.~M. Wolf, G.~Giedke, O.~Kr\"uger, R.~F. Werner, and J.~I. Cirac},
  \emph{{G}aussian entanglement of formation}, Phys. Rev. A, \textbf{69},
  052320 (2004).

\bibitem[Whi92]{White1992}
\textsc{S.~R. White}, \emph{Density matrix formulation for quantum
  renormalization groups}, Phys. Rev. Lett., \textbf{69}, 2863 (1992).

\bibitem[Wil75]{Wilson75}
\textsc{K.~G. Wilson}, \emph{The renormalization group: critical phenomena and
  the Kondo problem}, Rev. Mod. Phys., \textbf{47}, 773 (1975).

\bibitem[WN92]{WhiteNoack92}
\textsc{S.~R. White and R.~M. Noack}, \emph{Real-space quantum renormalization
  groups}, Phys. Rev. Lett., \textbf{68}, 3487 (1992).

\bibitem[Woo98]{Wooters98}
\textsc{W.~K. Wootters}, \emph{Entanglement of Formation of an Arbitrary State
  of Two Qubits}, Phys. Rev. Lett., \textbf{80}, 2245 (1998).

\bibitem[YHL83]{Yoshioka83}
\textsc{D.~Yoshioka, B.~I. Halperin, and P.~A. Lee}, \emph{Ground State of
  Two-Dimensional Electrons in Strong Magnetic Fields and $\frac{1}{3}$
  Quantized Hall Effect}, Phys. Rev. Lett., \textbf{50}, 1219 (1983).

\end{thebibliography}

\printindex 


\end{document}